\documentclass[review]{elsarticle}

\usepackage[colorlinks,citecolor=blue,linktoc=all,linkcolor=cyan]{hyperref}
\usepackage{graphicx}

\usepackage[T1]{fontenc}
\usepackage{dsfont}               
\usepackage{mathrsfs}             
\usepackage{slashed}              
\usepackage{amsmath}
\usepackage{amssymb}
\usepackage{amsbsy}
\usepackage{amsfonts}
\usepackage{physics}
\usepackage{orcidlink}

\usepackage{tikz}
\usepackage{pgfplots}
\usepackage{pgfplotstable}

\graphicspath{{fig/}}

\numberwithin{equation}{section}
\numberwithin{table}{section}
\numberwithin{figure}{section}

\journal{Progress in Particle and Nuclear Physics}

\usepackage{titlesec}
\usepackage{sectsty}
\titleformat{\section}{\normalfont\Large\bfseries}{\thesection}{1em}{}
\titleformat{\subsection}{\normalfont\large\bfseries}{\thesubsection}{1em}{}
\titleformat{\subsubsection}{\normalfont\normalsize\bfseries}{\thesubsubsection}{1em}{}

\begin{document}
	
	\begin{frontmatter}
		
\title{Physics of the Electron-Ion Collider in China}
		
				
		\author[mymainaddress,mysecondaryaddress]{Bo-Wen Xiao\,\orcidlink{0000-0002-8738-3117}\corref{mycorrespondingauthor}}
		
		\cortext[mycorrespondingauthor]{Corresponding author}
		\ead{xiaobowen@cuhk.edu.cn}

        \author[thirdaddress,fourthaddress]{Yuxiang Zhao\,\orcidlink{0000-0001-8684-9766}}
        \ead{yxzhao@impcas.ac.cn}
      
       \author[fifthaddress,mysecondaryaddress]{Jian Zhou\,\orcidlink{0000-0001-5962-9004}}
       \ead{jzhou@sdu.edu.cn}
       
       \address[mymainaddress]{School of Science and Engineering, The Chinese University of Hong Kong (Shenzhen),\\ Longgang, Shenzhen, Guangdong, 518172, P.R. China}
		\address[mysecondaryaddress]{Southern Center for Nuclear-Science Theory (SCNT), Institute of Modern Physics, Chinese Academy of Sciences, Huizhou 516000, Guangdong Province, China}
	\address[thirdaddress]{Institute of Modern Physics, Chinese Academy of Sciences, Lanzhou 730000, China}
\address[fourthaddress]{University of Chinese Academy of Sciences, Beijing 100049, China}
\address[fifthaddress]{School of Physics and Key Laboratory of Particle Physics and Particle Irradiation (MOE), Shandong University, QingDao, Shandong, 266237, China}

\begin{abstract}
The Electron-Ion Collider in China (EicC), a cutting-edge facility under development, aims to unveil the internal structure of nucleons and nuclei by leveraging collisions of high-intensity polarized electrons and ions (polarized protons, polarized deuterons, polarized $^{3}$He, and unpolarized heavy ions up to Uranium) at center-of-mass energies of 15-20 GeV and luminosity of (2-4)$\times 10^{33}$cm$^{-2}$s$^{-1}$. Its primary physics objectives include 3D tomography of nucleon spin and momentum structure, fundamental questions regarding the origin of nucleon mass, partonic structure of nuclei and parton interactions with the nuclear environment, and exploration of exotic hadronic states. In this paper, we review the physics potential of the EicC and highlight its unique capabilities for advancing precision nucleon structure studies by combining its specialized kinematic coverage and high luminosity. Since traditional topics like 3D nucleon structure have already been well-covered by several extensive reviews, we have deliberately dedicated significant space to recent progress in nucleon mass decomposition, nucleon energy-energy correlation, quantum information, and artificial intelligence applications in high-energy nuclear physics, which have been emerging rapidly and attracted a tremendous amount of attention in the community. 
\end{abstract}

		
	\end{frontmatter}
	
	\newpage
	\thispagestyle{empty}
	\tableofcontents
	

	\newpage	
\section{Introduction}\label{intro}
\begin{center}
What is matter truly made of, and how is it held together? 
\end{center}

The quest to understand the fundamental structure of matter has been an enduring pursuit of natural philosophy for over two millennia. The ancient Greek philosophers Leucippus and Democritus first proposed around 400 BCE that nature consists of indivisible atoms existing in void, a remarkably prescient conceptual framework that, while remaining philosophical speculation for centuries, planted the seeds for modern atomic theory. This foundational concept evolved through rigorous scientific inquiry: from Dalton's empirically based atomic theory in the early 1800s, through Thomson's discovery of the electron (1897), Rutherford's discovery of the atomic nucleus and the proton (1911 to 1919), to Bohr's quantum mechanical description of the atom (1913).

The 20th century brought revolutionary insights as scientists developed increasingly sophisticated tools to probe the atomic nucleus. The discovery of the neutron by Chadwick in 1932 completed the basic picture of nuclear structure. Early hints that the proton might not be a point-like particle emerged from measurements of its anomalous magnetic moment by Stern and collaborators in the 1930s, which deviated from the value expected for a point-like Dirac particle. More direct evidence came from elastic electron-proton scattering experiments by Hofstadter and collaborators~\cite{Hofstadter:1956qs} in the 1950s, which revealed an extended charge distribution and indicated internal structure. In 1964, Gell-Mann~\cite{Gell-Mann:1964ewy} and independently Zweig~\cite{Zweig:1964ruk} proposed the quark model, suggesting that hadrons are composed of more fundamental constituents called quarks. This theoretical framework provided an elegant explanation for the observed patterns in hadron spectroscopy, though direct experimental evidence for quarks remained elusive initially.

The pivotal breakthrough came in the late 1960s with deep inelastic scattering (DIS) experiments at Stanford Linear Accelerator Center (SLAC). These landmark experiments~\cite{Bloom:1969kc,Breidenbach:1969kd}, employing deep-inelastic scattering of high-energy electrons off proton targets with detection of scattered electrons at large momentum transfer, provided the definitive evidence that protons and neutrons are composite hadronic systems with complex internal structure. The discovery of point-like constituents within the proton through DIS, later identified as the quarks predicted by Gell-Mann and Zweig, marked a paradigm shift in our understanding of matter and led directly to the development of Quantum Chromodynamics (QCD), the fundamental theory of the strong interaction.

Building on these experimental breakthroughs, QCD emerged as the cornerstone theory describing the strong interaction between quarks and gluons, the fundamental constituents of hadrons. QCD is a non-Abelian gauge theory based on the $SU(3)$ color symmetry, characterized by two remarkable properties: asymptotic freedom~\cite{Politzer:1973fx,Gross:1973ju} at short distances and confinement at long distances. Asymptotic freedom implies that the strong coupling constant $\alpha_s(Q^2)$ decreases with increasing momentum transfer $Q^2$, allowing perturbative calculations at high energies. Conversely, confinement, the phenomenon that quarks and gluons are permanently bound within hadrons and cannot be observed as free particles, dominates at low energies and remains one of the most profound unsolved problems in theoretical physics, intimately connected to the Yang-Mills mass gap problem~\cite{JaffeWitten2000}.

The study of hadron structure through high-energy scattering has revealed a rich, complex and dynamic picture. In the parton model framework, refined by QCD, the proton's structure is described by PDFs $f_i(x, Q^2)$, which give the probability of finding a parton of type $i$ (quark or gluon) carrying a longitudinal momentum fraction $x$ of the proton's momentum, probed at a resolution scale $Q^2$. The Dokshitzer–Gribov–Lipatov–Altarelli–Parisi (DGLAP) evolution equations~\cite{Gribov:1972ri,Altarelli:1977zs,Dokshitzer:1977sg} govern how these distributions change with $Q^2$, encoding the quantum dynamics of parton splitting. DIS experiments at facilities such as SLAC, CERN, and later Hadron-Electron Ring Accelerator (HERA) have mapped these distributions with increasing precision, revealing a complex internal structure: at typical experimental scales $Q^2\sim 10\, \text{GeV}^2$, valence quarks, which define the proton's quantum numbers, carry only about 40\% of the proton's longitudinal momentum in the infinite momentum frame, while gluons carry approximately 45\%, with sea quarks (quark-antiquark pairs arising from gluon splitting) contributing the remaining fraction.

The late 20th and early 21st centuries witnessed the construction of revolutionary particle accelerators that dramatically advanced hadron structure studies. The HERA at DESY, operational from 1992 to 2007, provided the high-energy electron-proton collider, achieving center-of-mass energies up to 320 GeV. HERA's high-energy collisions enabled remarkable precision in mapping the proton's quark and gluon distributions, particularly at small Bjorken $x$ ($x < 0.01$), where the rapid growth of the gluon density suggested the onset of novel QCD dynamics. The discovery of geometric scaling~\cite{Golec-Biernat:1998zce,Stasto:2000er,Iancu:2003ge} and other experimental signatures~\cite{Golec-Biernat:1999qor,Kowalski:2006hc} hinted at gluon saturation effects, a regime where gluon densities become so high that nonlinear QCD dynamics, including gluon recombination, become important. In this high-density regime, the gluon field is predicted to saturate into a universal state of matter known as the Color Glass Condensate 
(CGC)~\cite{Gelis:2010nm}.

Complementing HERA's detailed proton structure studies, the Large Hadron Collider (LHC) at CERN, which began operations in 2008, opened new frontiers through proton-proton and heavy-ion collisions at unprecedented energies (up to $\sqrt{s} = 13$ TeV for $pp$ collisions). While primarily designed for discovering new particles and testing the Standard Model at the energy frontier, the LHC has also contributed to our understanding of proton structure through precision measurements of electroweak boson and jet production, providing valuable constraints on PDFs at medium-to-high $x$ and high $Q^2$. The LHC heavy-ion program has explored the properties of the quark-gluon plasma, a deconfined state of matter created in ultra-relativistic heavy-ion collisions, offering complementary insights into QCD dynamics under extreme conditions.

The Relativistic Heavy Ion Collider (RHIC) at Brookhaven National Laboratory, which began operations in 2000, has played a complementary and crucial role in advancing our understanding of QCD. Operating primarily as a heavy-ion collider with collision energies up to $\sqrt{s_{NN}} = 200$ GeV for gold-gold collisions, RHIC discovered and characterized the strongly coupled quark-gluon plasma, demonstrating that this deconfined state of matter behaves as a nearly perfect fluid with extremely low viscosity. Beyond heavy-ion physics, RHIC's unique capability to collide polarized protons has opened a new frontier in spin physics. The RHIC spin program, with experiments such as STAR and PHENIX, has provided crucial constraints on the gluon helicity distribution $\Delta g(x)$ through measurements of longitudinal double-spin asymmetries in jet, pion, and direct photon production. These measurements have significantly advanced our understanding of the gluon contribution to the proton spin, particularly in the intermediate $x$ region ($0.05 < x < 0.2$) from the $\sqrt{s} = 200$ GeV data~\cite{STAR:2009vxb,PHENIX:2010aru,STAR:2014wox,PHENIX:2014axc}, complementing the information obtained from polarized deep inelastic scattering experiments. More recently, the $\sqrt{s} = 510$ GeV polarized proton-proton collision data \cite{STAR:2019yqm,PHENIX:2022lgn} have extended the reach to lower $x$ values (down to $x \sim 0.01$), providing the first direct constraints on the gluon helicity in the small $x$ region and offering critical input for understanding the evolution of spin structure functions toward the kinematic regime that will be explored at the future EICs.

\subsection{Physics at an electron-ion collider}

Despite these monumental achievements, fundamental questions about hadron structure remain at the forefront of nuclear physics. Three interconnected challenges stand out:

\textbf{1. The origin of nucleon mass:} While the Higgs mechanism accounts for the masses of fundamental fermions and bosons, it contributes about 1\% to the proton's mass ($M_p \approx 938$ MeV). The remaining 99\% arises from the dynamics of the strong interaction, the kinetic and potential energies of quarks and gluons, and the gluon field energy encoded in the trace anomaly of the energy-momentum tensor. Ji's mass decomposition~\cite{Ji:1994av,Ji:1995sv} provides a framework for partitioning the proton mass into contributions from quark masses ($M_m$), quark energy ($M_q$), gluon energy ($M_g$), and the trace anomaly ($M_a$), but experimental determination of these individual components remains a major challenge.

\textbf{2. The proton spin puzzle:} The discovery in 1988 by the European Muon Collaboration that quarks carry only about 30\% of the proton's spin~\cite{EuropeanMuon:1987isl} sparked the ``proton spin puzzle''. The remaining 70\% must come from gluon spin ($\Delta G$) and the orbital angular momentum (OAM) of quarks ($L_q$) and gluons ($L_g$), according to the Jaffe-Manohar sum rule~\cite{Jaffe:1989jz}: 
\begin{equation}
\frac{1}{2} = \frac{1}{2}\Delta\Sigma + L_q + \Delta G + L_g.
\end{equation}
Recent polarized scattering experiments have made remarkable strides in constraining the gluon helicity $\Delta G$, bringing us closer to solving the proton spin puzzle. The quest now focuses on mapping the OAM contributions $L_q$ and $L_g$, which will complete our understanding of how quark spin, gluon spin, and the orbital motions of both quarks and gluons conspire to generate the proton's total spin of $1/2$.

\textbf{3. The three-dimensional structure and gluon saturation:} Traditional PDFs provide a one-dimensional picture, describing only the longitudinal momentum distribution of partons. A complete understanding requires three-dimensional imaging through Generalized Parton Distributions (GPDs) and Transverse Momentum Dependent distributions (TMDs), which encode correlations between parton momentum and spatial position or transverse momentum, respectively. GPDs, accessed through deeply virtual Compton scattering (DVCS) and exclusive meson production, provide spatial maps of quarks and gluons inside the nucleon and connect to the nucleon's mechanical properties, such as pressure and shear forces. TMDs, measured in semi-inclusive deep inelastic scattering (SIDIS), reveal the intrinsic transverse momentum of partons and spin-orbit correlations. Furthermore, at small $x$ and in heavy nuclei, the gluon density is expected to reach a saturation regime where nonlinear QCD dynamics dominate, and CGC emerges. Experimental confirmation of gluon saturation and emergent properties of CGC remains a key goal.

\subsection{Development of US-EIC and EicC}

Currently under construction at Brookhaven National Laboratory in the United States, the US-EIC is designed to collide high-energy electrons with protons or nuclei, featuring polarization capabilities for electrons, protons, and light ions. This unique experimental platform will provide unprecedented precision in probing the three-dimensional structure of hadrons, uncovering the spatial and momentum distributions of quarks and gluons, and studying their dynamical evolution. It will enable scientists to explore the role of gluons in hadron structure and dynamics, investigate the origins of proton mass and spin, and probe the emergent properties of high-density gluonic matter, including the potential discovery of the CGC.

The EIC project in the United States has a long history of development, spanning more than a decade of planning and scientific evaluation~\cite{Boer:2011fh}. The initial concept for the EIC emerged in the early 2000s, driven by the success of the RHIC and the growing interest in exploring the initial-state physics of heavy-ion collisions. During this period, the study of hadron structure, particularly the origins of proton spin and mass, gluon saturation physics, TMDs, and GPDs, became prominent research areas, laying the groundwork for the EIC proposal.

In 2010, the publication of the EIC White Paper marked a significant milestone in the development of the project. The White Paper~\cite{Accardi:2012qut}, titled ``Electron Ion Collider: The Next QCD Frontier, Understanding the Glue That Binds Us All'', outlined the scientific goals and experimental design of the EIC, emphasizing its potential to address fundamental questions in QCD. In 2018, the National Academy of Sciences released a report highlighting ``the scientific case for EIC compelling, unique, and timely''~\cite{NAS2018}. These developments reached a significant milestone in January 2020, when the U.S. Department of Energy officially approved the construction of the EIC. In 2021, the EIC Yellow Report~\cite{AbdulKhalek:2021gbh} was published, detailing the physics objectives and detector requirements. This report was the result of a collaborative effort by an international team of over 400 scientists representing 151 research institutions. For recent developments in the EIC program, with an emphasis on the energy dependence of key measurements, see Ref.~\cite{Aschenauer:2017jsk}; for precision QCD studies and recent advances in the EIC science program, see Refs.~\cite{Burkert:2022hjz,Abir:2023fpo,Alexandrou:2026cnj}. 
A roadmap for the first years of operation has been established in the EIC's Early Science Report~\cite{Abbott:2026eqp}.
With construction underway since 2021, the EIC is expected to begin operations sometime in the 2030s, replacing RHIC as the flagship facility for high-energy nuclear physics research in the United States.

While the US-EIC represents a major experimental facility in the global effort to explore hadron structure, complementary initiatives are also being pursued internationally. In China, the proposal for an Electron-Ion Collider in China (EicC) has emerged as a key component of the country's strategic plan for nuclear physics research. The EicC, envisioned as a next-generation collider, will be based on the High Intensity Heavy-Ion Accelerator Facility (HIAF), a newly completed cutting-edge accelerator complex, now operational in Huizhou, Guangdong Province. The EicC aims to focus on the relatively low-energy regime, complementing the higher-energy capabilities of the US-EIC. By leveraging its unique energy range and precision measurement capabilities, the EicC will provide critical insights into the three-dimensional structure of protons and nuclei, with particular emphasis on sea-quark dynamics, the spin and mass structure of the proton, polarized gluon distributions in the large-$x$ region, cold nuclear matter (CNM) effects, and exotic states.

The EicC represents a major scientific and international initiative in the field of nuclear and particle physics in China. Its development has followed a systematic trajectory similar to that of the US-EIC, beginning with the publication of its White Paper in 2020. This document~\cite{CAO:2024fdz,Anderle:2021wcy} outlined the scientific goals, design concepts, and technical feasibility of the EicC, building the foundation for the conceptual design phase. The Conceptual Design Report (CDR) for the EicC, almost completed and set to be published soon, marks a significant milestone in its development, providing a detailed blueprint for its construction and operation. By leveraging the existing HIAF infrastructure and integrating state-of-the-art technologies for polarized electron and ion beams, the EicC will establish a world-class experimental platform for investigating the structure and dynamics of hadrons and nuclei.

The scientific agenda of the EicC encompasses a wide range of topics that address fundamental questions in QCD and hadron physics. One of the primary goals is to explore the one-dimensional spin structure of the proton. Building on the foundational EMC result~\cite{EuropeanMuon:1987isl} discussed earlier, precision measurements of spin-dependent structure functions at the EicC can provide critical data to determine the quark and gluon spin contributions to the proton spin.
This indirectly constrains the OAM contribution, thereby advancing our understanding of QCD dynamics and the origin of nucleon spin.

Another key area of research at the EicC is the three-dimensional imaging of hadrons, which involves mapping the spatial and momentum distributions of quarks and gluons within protons and nuclei. This can be achieved through the study of GPDs and TMDs, which provide complementary information about the internal structure of hadrons. GPDs encode the spatial distribution of partons as a function of their longitudinal momentum, offering insights into the spatial structure and mechanical properties of hadrons, such as pressure and shear forces. TMDs, on the other hand, describe the transverse momentum distributions of partons, providing dynamical information about quarks and gluons in the transverse momentum space. By combining measurements of GPDs and TMDs, the EicC will enable a comprehensive three-dimensional reconstruction of hadron structure, revealing a nuanced and intricate internal landscape shaped by QCD dynamics.

The EicC will also play a crucial role in studying cold nuclear matter effects, which refer to the modifications of parton distributions in nuclei relative to free protons. These effects, which include nuclear shadowing, anti-shadowing, and the EMC effect, provide valuable insights into the behavior of quarks and gluons in nuclear environments. By investigating these phenomena across a wide range of nuclear targets and energy scales, the EicC will enhance our understanding of the interplay between partonic and nuclear degrees of freedom, contributing to a more complete picture of QCD in nuclear systems.

In addition, the EicC will explore exotic states of matter, such as hybrid mesons and other exotic states, which are predicted by QCD but have yet to be conclusively observed. These states provide a unique window into the non-perturbative regime of QCD. By searching for these exotic states in exclusive and semi-inclusive processes, the EicC will test theoretical predictions and expand our knowledge of the QCD spectrum.

Another fundamental question that the EicC aims to address is the origin of proton mass. While the proton's mass is significantly larger than the sum of its constituent quark masses, the precise contributions from quark and gluon dynamics, as well as QCD's anomalous energy contributions, remain an open question. By studying the energy-momentum distributions of quarks and gluons within the proton, the EicC will provide critical data for quantifying the various contributions to proton mass and advancing our understanding of its origins.

Beyond these established research areas, the EicC will serve as a platform for exploring new ideas and miscellaneous topics in nuclear and particle physics. For example, the EicC could investigate novel observables for probing quantum entanglement, develop innovative techniques for studying parton correlations, and explore connections between hadron physics and other fields, such as quantum information theory. The flexibility and versatility of the EicC will enable it to address emerging scientific questions and adapt to new developments in the field, ensuring its long-term relevance and impact.

In summary, the US-EIC and the EicC represent complementary efforts to push the boundaries of high-energy nuclear physics, offering unique and synergistic approaches to unraveling the mysteries of hadron structure and dynamics. While the US-EIC is designed to operate at higher energies, enabling detailed investigations of gluon saturation and the CGC as well as other critical measurements at higher energies, the EicC will focus on the low-to-intermediate energy regime, providing valuable insights into sea quark dynamics, three-dimensional structure of hadrons, and CNM effects. Together, these facilities will provide a comprehensive understanding of hadron structure and QCD dynamics, addressing key scientific questions from multiple perspectives. By advancing the frontiers of QCD research, the US-EIC and EicC will shed light on fundamental questions such as the origins of proton mass and spin, the properties of high-density gluonic matter. Through international collaboration and interdisciplinary innovation, these world-class facilities will transform our understanding of the strong interaction and open new horizons in nuclear physics research.

\subsection{EicC physics highlights}

Complementary to the US-EIC, the EicC physics program addresses several interconnected research frontiers that together will provide a comprehensive understanding of nucleon and nuclear structure. The primary areas of investigation include:

\textbf{One-dimensional spin structure:} Through SIDIS measurements with both longitudinally polarized electron-proton and electron-$^3$He collisions, the EicC will achieve flavor separation of quark helicity distributions. The combination of excellent particle identification capabilities for pions and kaons, together with the effective polarized neutron source provided by $^3$He, will enable precise determination of the helicity distributions for up, down, and strange quarks in the kinematic region $x > 0.005$. Additionally, measurements of charm hadron production through the photon-gluon fusion process will provide unique constraints on the polarized gluon distribution $\Delta g(x)$ in the large $x$ region ($x > 0.1$), where the ratio $\Delta g/g$ is expected to be significantly larger than at small $x$.

\textbf{Three-dimensional nucleon tomography:} The EicC's kinematic coverage, focusing on the intermediate $x$ region ($0.005 < x < 0.3$), provides ideal conditions for comprehensive three-dimensional imaging of the nucleon through both TMDs and GPDs. For TMDs, high-statistics SIDIS measurements will dramatically improve our knowledge of the Sivers function $f_{1T}^\perp$, the transversity distribution $h_1$, and the worm-gear function $g_{1T}^\perp$, with uncertainties reduced by factors of 5 to 10 compared to current extractions. The flavor separation capabilities will enable the first significant constraints on strange quark TMDs. For GPDs, measurements of DVCS and deeply virtual meson production (DVMP) with transversely and longitudinally polarized beams will constrain the GPDs $H$ and $E$, providing access to quark OAM through the Ji sum rule and to the nucleon's mechanical properties through gravitational form factors.

\textbf{Cold nuclear matter effects:} The EicC's electron-nucleus collision program enables the study of parton distributions and dynamics in the nuclear environment. Measurements of nuclear modification of structure functions, transverse momentum broadening, and hadron multiplicities will provide precision constraints on nuclear PDFs and the QCD energy loss in CNM. Thus, it helps elucidate how the nuclear medium modifies parton distributions (nuclear shadowing, anti-shadowing, EMC effect) and how energetic partons interact with and propagate through nuclear matter.

\textbf{Nucleon mass decomposition:} Through measurements of near-threshold heavy quarkonium photoproduction, particularly $J/\psi$ and $\Upsilon$ production~\cite{Strakovsky:2021vyk}, the EicC will probe the gluonic gravitational form factors that encode information about the gluon contributions to the proton mass. Combined with precision measurements of unpolarized structure functions that constrain the second moments of parton distributions, these studies will provide crucial input for determining the various contributions to the proton mass according to Ji's decomposition: quark mass ($M_m$), quark energy ($M_q$), gluon energy ($M_g$), and trace anomaly ($M_a$).

\textbf{Exotic states and spectroscopy:} Beyond conventional hadrons composed of two or three valence quarks ($q\bar{q}$ mesons and $qqq$ baryons), QCD permits a richer spectrum of exotic hadrons: multiquark states such as tetraquarks ($qq\bar{q}\bar{q}$) and pentaquarks ($qqqq\bar{q}$), hybrid mesons containing explicit valence gluonic excitations, and glueballs composed purely of gluons. These exotic configurations directly probe the non-Abelian nature of QCD, the self-interaction of gluons, and confinement dynamics inaccessible through conventional hadrons alone.

The 21st century has witnessed remarkable experimental progress: the LHC has discovered 83 new hadrons, 24 of which are exotic candidates~\cite{Koppenburg:2025web}, while BESIII has discovered 31 new hadrons~\cite{BESIII:2025web} as of April 2026. Notable examples include the charged $Z_c$ state requiring at least four valence quarks, and the $P_c$ pentaquark states discovered by LHCb. However, fundamental questions persist regarding their internal structure: are they compact multiquark states, loosely bound hadronic molecules, or hybrid states with gluonic excitations?

EicC's unique capabilities for exotic hadron spectroscopy through electro-production mechanisms, which avoid kinematic triangle singularities prevalent in $e^+e^-$ annihilations, enable cleaner probes of resonant structures. Polarized beams facilitate crucial spin-parity ($J^{PC}$) determinations. The facility will systematically study charmonium-like $XYZ$ states and search for hidden-charm pentaquarks, as well as explore the less-studied bottom sector for hidden-bottom pentaquarks ($P_b$). Detailed discussions of exotic hadron physics at EicC, including production mechanisms, cross-section estimates, and discovery potential, have been presented in previous comprehensive reviews~\cite{Cao:2023oef,Anderle:2021wcy,Chen:2020ijn} and will not be repeated here. The clean $e$-$p$ environment and precision spectroscopy capabilities will provide unprecedented insights into non-perturbative QCD and confinement mechanisms.

\textbf{Emerging frontiers:} Beyond these established research areas, the EIC will serve as a platform for exploring novel physics concepts at the intersection of QCD and quantum information science. Recent theoretical developments suggest that electron-ion collisions can probe quantum entanglement between produced quark-antiquark pairs, with the potential to observe violations of Bell inequalities and test fundamental aspects of quantum mechanics in the context of strong interactions. Additionally, it is possible to develop new tools to study nuclear effects and hadronic structure through the interplay with quantum information theory.

Furthermore, the application of AI and ML techniques to detector optimization, event reconstruction, and data analysis will enhance the EIC's discovery potential and enable new approaches to extracting physics from complex multi-dimensional data. For example, neural networks can be used for solving the difficult inverse problems of extracting PDFs, GPDs, and TMDs from experimental data.

In the end, the complementary kinematic coverage and capabilities of the EicC, the 12 GeV upgraded Jefferson Lab, and the future US-EIC will together provide a complete picture of nucleon structure spanning from the valence quark dominated large $x$ region through the sea quark region to the gluon saturation regime at small $x$. This comprehensive program will address fundamental questions about the origin of mass, the decomposition of nucleon spin, the three-dimensional structure of hadrons, and the behavior of quarks and gluons in the nuclear environment, ushering in a precision era for QCD and nuclear physics.

\subsection{Expected data sets at the EicC}
The EicC machine parameters have been carefully optimized to access the sea-quark dominated regime at intermediate Bjorken-$x$ while maintaining good resolution in the valence quark region at high-$x$, thereby bridging the gap between fixed-target experiments and high-energy US-EIC.
The EicC design features a center-of-mass energy of $\sqrt{s}=15-20$ GeV, achieved through the collision of a 3-5 GeV electron beam with a 20-25 GeV proton beam or ion beam. The peak luminosity is designed to reach L $\sim 10^{33}-10^{34}$ cm$^{-2}$ s$^{-1}$. Both electron and proton/deuteron/$^3$He beams can be polarized with polarization exceeding 70\%. The facilicy is capable of accelerating protons and a range of ion species from light nuclei such as helium to heavy nuclei including calcium and lead.
Representative avaialbe beam particles and their corresponding energy, polarization, luminosity, and integrated luminosity per year are listed in Table \ref{tab:expected_data}.
It is important to emphasize that these parameters remian under active development as the project design evolves.

\begin{table}[htbp]
\centering
\caption{Expected data and their corresponding energy, polarization, instantaneous luminosity at the nucleon level and integrated luminosity per year. Be noted that these parameters remian under active development as the project design evolves.}
\resizebox{\textwidth}{!}{
\begin{tabular}
{p{0.12\textwidth}p{0.15\textwidth}p{0.15\textwidth}p{0.15\textwidth}p{0.2\textwidth}p{0.15\textwidth}}
\hline
Particle &
Momentum (GeV/c/u) &
C. o. M. energy (GeV/u) &
Polarization&
Luminosity($\rm{cm^{-2}s^{-1}}$) &
Integrated luminosity ($\mathrm{{fb}^{-1}}$)
\\ \hline 
$\rm{e}$ &
3.5&
&
80\%&

\\

$\rm{p}$&
20&
16.76&
70\%&
$\rm{4.25\times 10^{33}}$&
107
\\ 

$\rm{d}$&
12.90&
13.48&
Yes&
$\rm{10.09\times 10^{32}}$&
25
\\

$\rm{^{3}He^{++}}$&
17.21&
15.55&
Yes&
$\rm{7.44\times 10^{32}}$&
18
\\

$\rm{^{7}Li^{3+}}$ &
11.05&
12.48&
No&
$\rm{11.81\times 10^{32}}$&
29
\\

$\rm{^{12}C^{6+}}$ &
12.90&
13.48&
No&
$\rm{9.92\times 10^{32}}$&
25
\\

$\rm{^{40}Ca^{20+}}$ &
12.90&
13.48&
No&
$\rm{9.92\times 10^{32}}$&
25
\\

$\rm{^{197}Au^{79+}}$ &
10.35&
12.09&
No&
$\rm{11.40\times 10^{32}}$&
28
\\

$\rm{^{208}Pb^{82+}}$ &
10.17&
11.98&
No&
$\rm{11.18\times 10^{32}}$&
28
\\

$\rm{^{238}U^{92+}}$&
9.98&
11.87&
No&
$\rm{10.82\times 10^{32}}$&
27
\\ \hline
\end{tabular}}%
\label{tab:expected_data}%
\end{table}%

\begin{figure}[htbp]
\begin{center}
\includegraphics[width=0.7\textwidth]{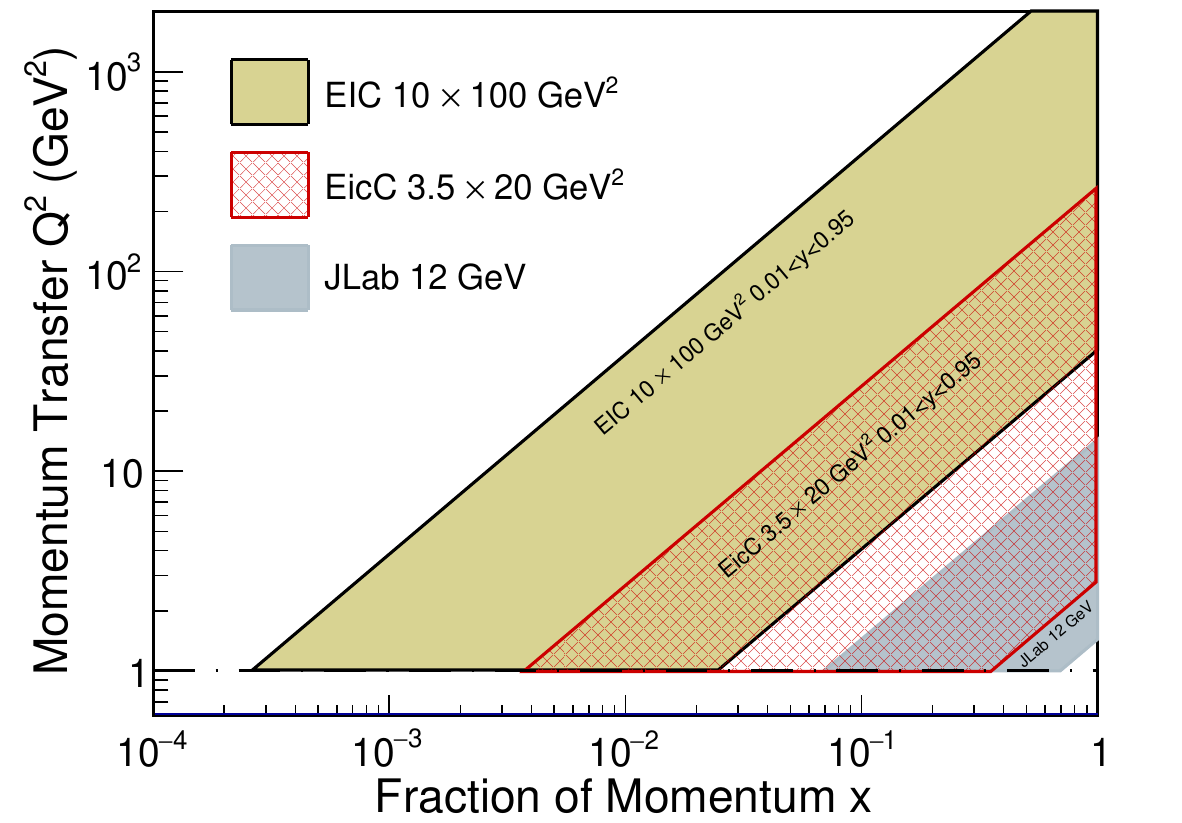}
\caption{\label{fig:Q2} Kinematic coverage of deep inelastic scattering process for different beam energy 
configurations at two proposed electron-ion colliders (EicC and US-EIC) as well as JLab.
Note that there are other energy configurations for both electron-ion colliders. The figure is adapted from Ref.~\cite{Anderle:2021wcy} (\href{https://creativecommons.org/licenses/by/4.0/}{CC BY 4.0}).
}
\end{center}
\end{figure}

The kinematic coverage of EicC in the $(x,Q^2)$ plane is shown in Fig.~\ref{fig:Q2}. The accessible Bjorken-$x$ range is particularly important for investigating both sea-quark and valence-quark dynamics. The $Q^2$ coverage extends from approximately 1 to 100~GeV$^2$, spanning the transition from the nonperturbative to the perturbative QCD regime.
Compared with the EIC at BNL, which operates over a center-of-mass energy range of 29--140~GeV, EicC provides higher luminosity in the moderate-$x$ region. This feature is especially advantageous for multidimensional studies of the nucleon spin structure, where both sizable spin-dependent signals and high statistical precision are essential. Moreover, for a given value of $x$, the corresponding $Q^2$ is generally much higher at the EIC than at EicC. At large $Q^2$, however, soft-gluon radiation encoded in the Sudakov factor becomes increasingly important and can substantially suppress spin-dependent signals. Therefore, EicC provides a particularly favorable kinematic regime for probing the nonperturbative origin of the proton spin structure.
Relative to the JLab fixed-target program, EicC also offers an extended reach in $Q^2$, thereby reducing the impact of higher-twist effects and the associated complications in the extraction of parton distribution functions.

	\newpage
	\section{One-dimensional spin structure}\label{second}

Within the collinear, leading-twist framework, the longitudinal spin of the nucleon is carried by quarks and gluons whose intrinsic spins are aligned (or anti-aligned) with the nucleon spin. Helicity distributions~\cite{Altarelli:1977zs,Jaffe:1989jz,deFlorian:2009vb,Aidala:2012mv}, which is the one-dimensional spin-dependent parton distribution functions, provide the most direct experimental access to these quark and gluon spin contributions. Precisely extracting these distributions~\cite{Aidala:2012mv} from polarized deep-inelastic scattering, semi-inclusive DIS, and polarized proton-proton collisions is essential for understanding the origin of the nucleon spin. Not only do these measurements test perturbative QCD evolution and the axial-anomaly mechanism~\cite{Altarelli:1988nr, Carlitz:1988ab}, but they also constrain the non-perturbative orbital angular momentum required to balance the nucleon's spin sum rule. Consequently, high-precision helicity distributions form the cornerstone of any comprehensive effort to unravel the full spin structure of the nucleon.

\subsection{Helicity formalism}
The helicity distribution~\cite{Altarelli:1977zs} of a parton species $f$ in a hadron is defined through the light-cone correlation function:
\begin{equation}
\Delta f(x, Q^2) = \int \frac{d\xi^-}{4\pi} e^{ix P^+ \xi^-} \langle P, S | \bar{\psi}(0) \gamma^+ \gamma_5 \psi(\xi^-) | P, S \rangle \Big|_{\xi^+ = \vec{\xi}_\perp = 0},
\end{equation}
where $|P, S\rangle$ denotes a hadron state with momentum $P$ and spin $S$, and the light-cone coordinates are defined by $a^\pm = (a^0 \pm a^3)/\sqrt{2}$. The $\gamma^+ \gamma_5$ structure projects out the helicity (chirality in the massless limit) of the parton along the hadron's direction of motion. Physically, $\Delta f(x, Q^2) = f_+(x, Q^2)- f_-(x, Q^2)$ represents the difference between the number densities of partons with helicity parallel (+) and antiparallel (-) to the parent hadron's helicity. While unpolarized PDFs $f(x, Q^2)$, defined with the operator $\bar{\psi} \gamma^+ \psi$, describe the probability density of finding a parton with momentum fraction $x$ regardless of spin and govern total cross sections in high-energy scattering, helicity distributions encode the spin-spin correlation between the parton and the hadron. The helicity distributions satisfy the sum rule $\int_0^1 dx \, \Delta f(x, Q^2) = \Delta \Sigma_f$ for quarks, where $\Delta \Sigma_f$ represents the contribution of quark flavor $f$ to the total proton spin.

The gluon helicity distribution $\Delta g(x, Q^2)$ can be similarly defined through the correlation function:
\begin{equation}
\Delta g(x, Q^2) = \int \frac{d\xi^-}{2\pi x  P^+} e^{ix P^+ \xi^-} \langle P, S | F^{+\alpha}(0) \tilde{F}_\alpha^{~+}(\xi^-) | P, S \rangle \Big|_{\xi^+ = \vec{\xi}_\perp = 0},
\end{equation}
where $F^{\mu\nu}$ is the gluon field strength tensor and $\tilde{F}^{\mu\nu} = \frac{1}{2}\epsilon^{\mu\nu\rho\sigma}F_{\rho\sigma}$ is its dual. The gluon helicity distribution plays a crucial role in resolving the proton spin puzzle, as it may account for a significant fraction of the proton's spin.  

Experimental measurements of both quark and gluon helicity distributions require longitudinally polarized beams and targets, making them more challenging than unpolarized PDF extractions but essential for a complete understanding of the spin structure of hadrons.

SIDIS extends the inclusive DIS framework by detecting both the scattered lepton and at least one hadron in the final state, typically denoted as $\ell N \to \ell' h X$, where $\ell$ is the incoming lepton, $N$ is the nucleon target, $\ell'$ is the scattered lepton, $h$ is the identified final-state hadron, and $X$ represents the undetected remnant. While inclusive DIS measures only the scattered lepton and integrates over all final-state hadrons, providing access primarily to quark distributions (sum over all flavors) through the leading-order photon-quark scattering $\gamma^* q \to q$ and gluon distribution through scaling violation over broad kinematic coverage beyond leading-order, SIDIS retains information about the struck parton in the initial state through the detected hadron. This additional measurement allows for several key advantages: 
\begin{enumerate}
    \item flavor tagging, as different quark flavors preferentially fragment into different hadrons (e.g., $u$ quarks into $\pi^+$, $s$ quarks into kaons)~\cite{Anderle:2021dpv};
    \item access to gluon-initiated processes such as photon-gluon fusion $\gamma^* g \to q\bar{q}$, where the produced quark-antiquark pair hadronizes into detectable particles~\cite{Hekhorn:2018ywm};
    \item sensitivity to the hadronization process itself through fragmentation functions~\cite{Metz:2016swz}.
\end{enumerate}
The kinematic variables in SIDIS include not only the standard DIS variables ($x$, $Q^2$, $y$) but also the hadron's momentum fraction $z = E_h/\nu$ and transverse momentum $p_T$ relative to the virtual photon direction.

\begin{figure}[htbp!]
\centering
\includegraphics[width=0.35\textwidth]{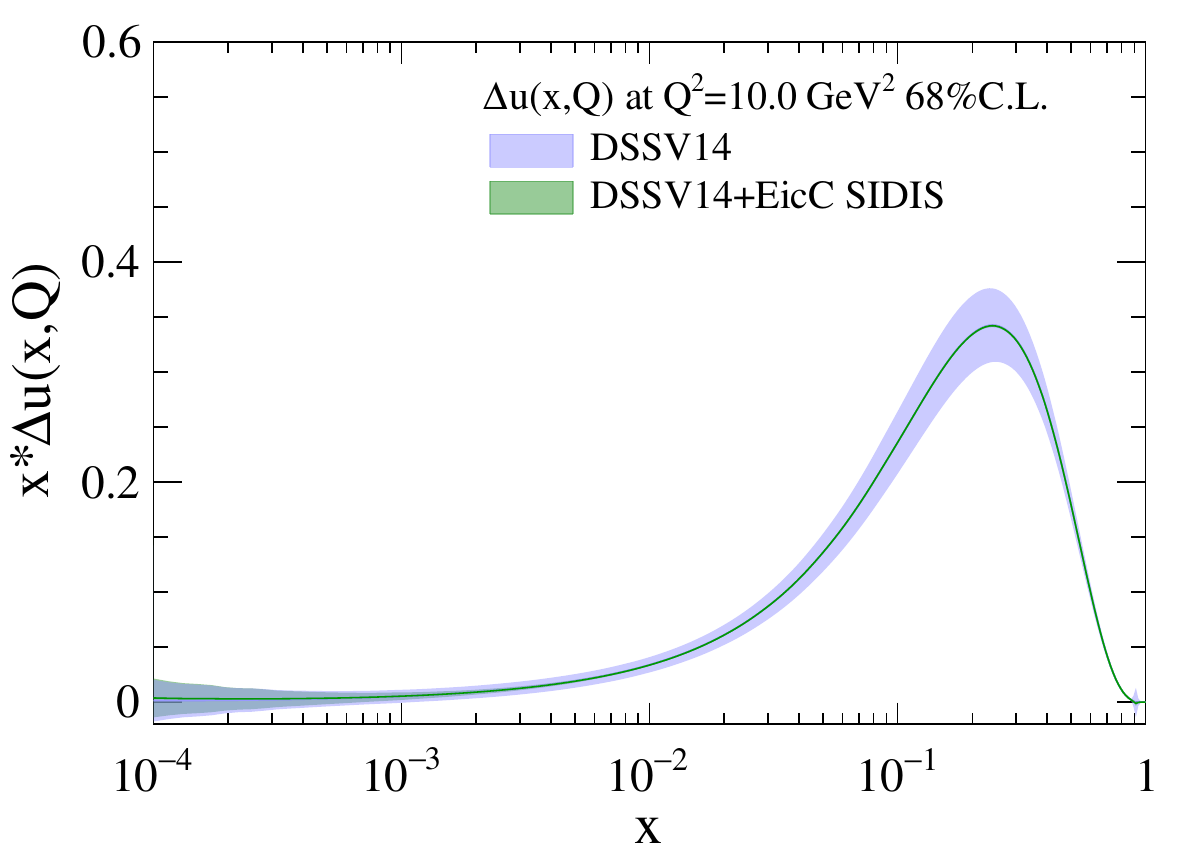}
\includegraphics[width=0.35\textwidth]{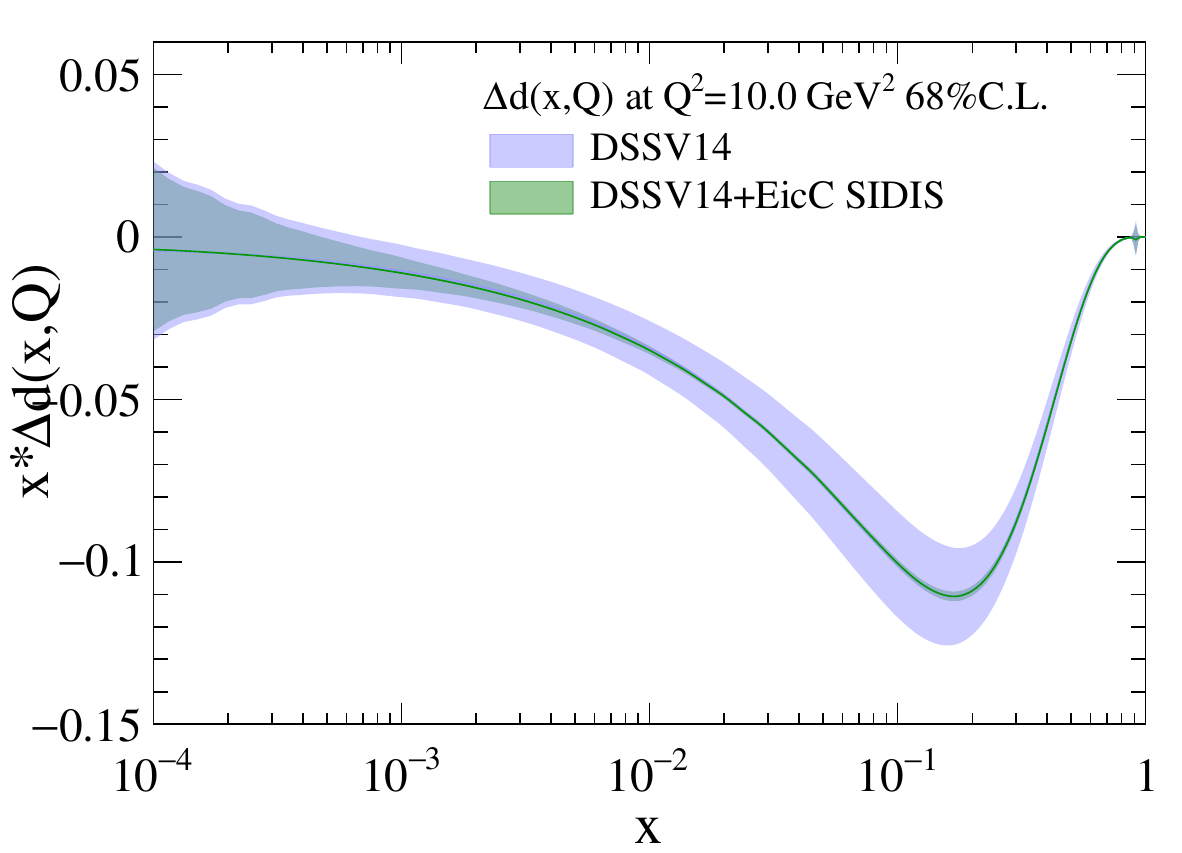}
\includegraphics[width=0.35\textwidth]{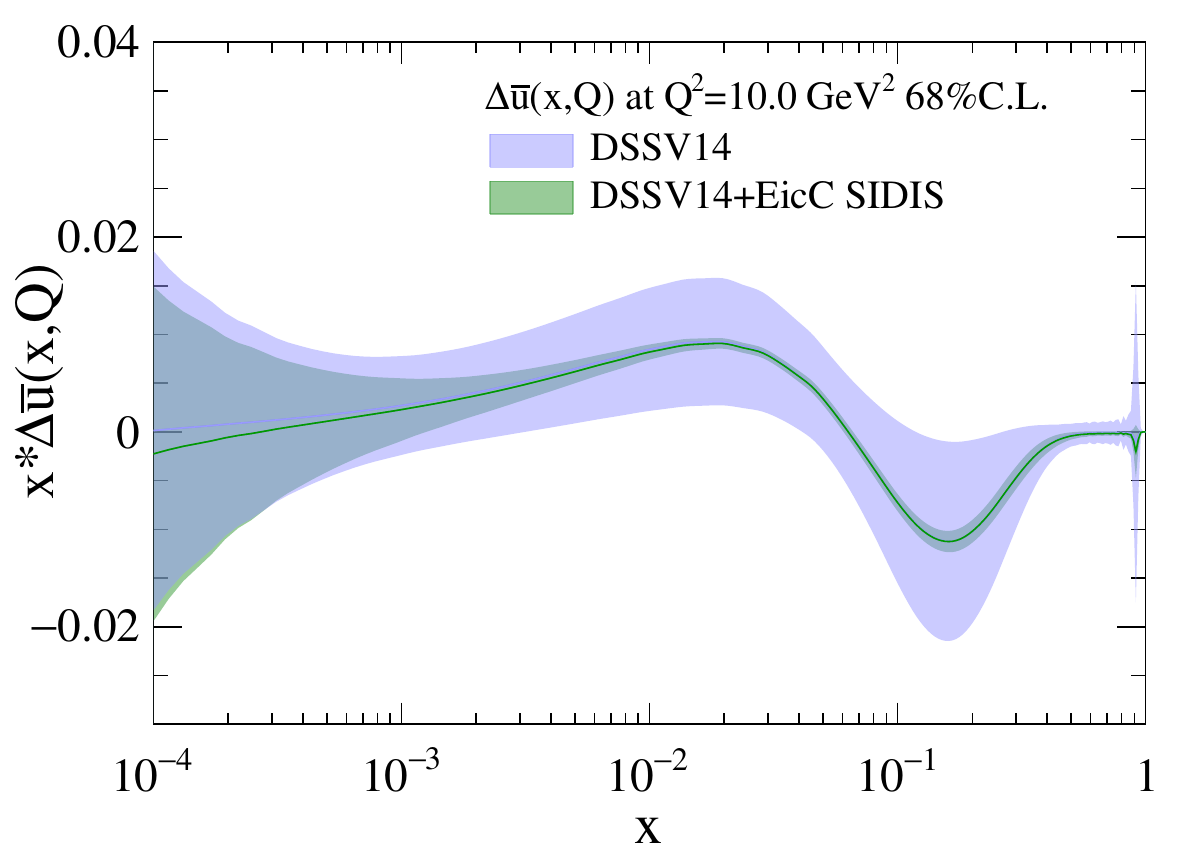}
\includegraphics[width=0.35\textwidth]{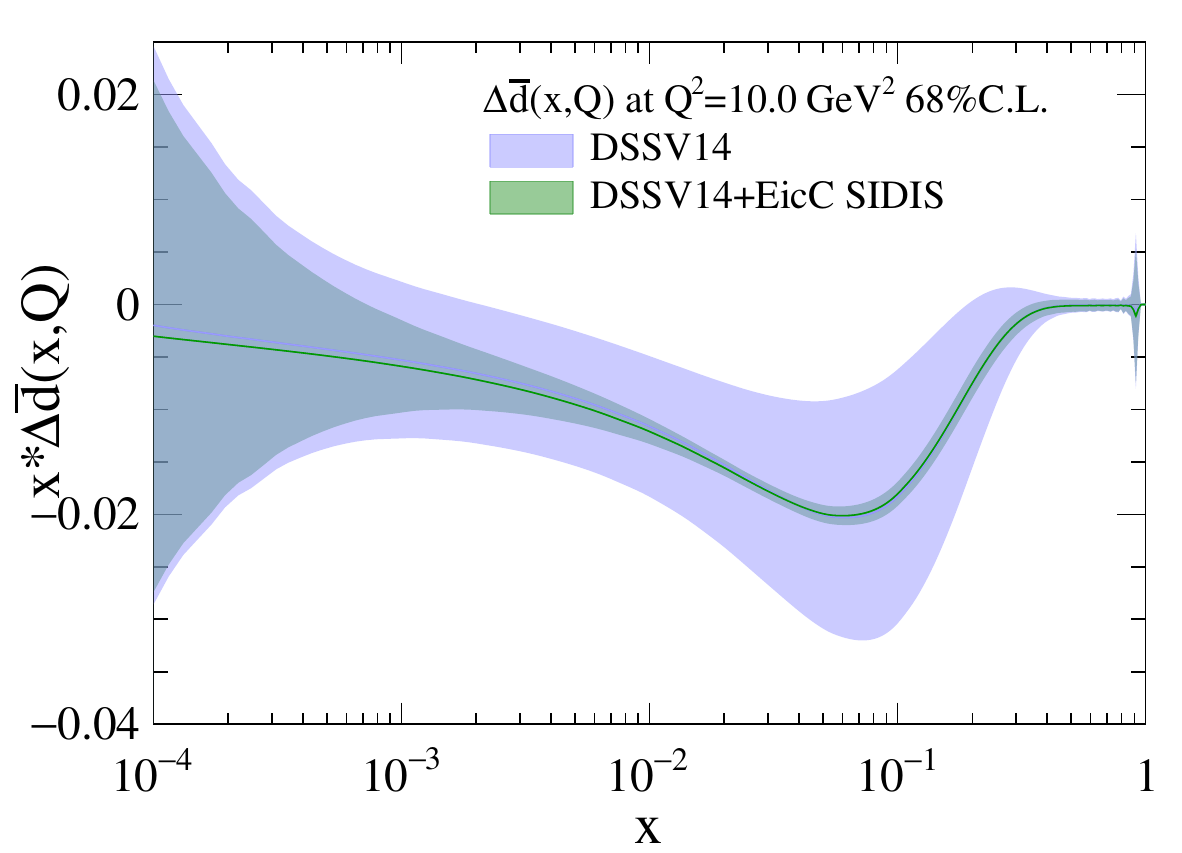}
\includegraphics[width=0.35\textwidth]{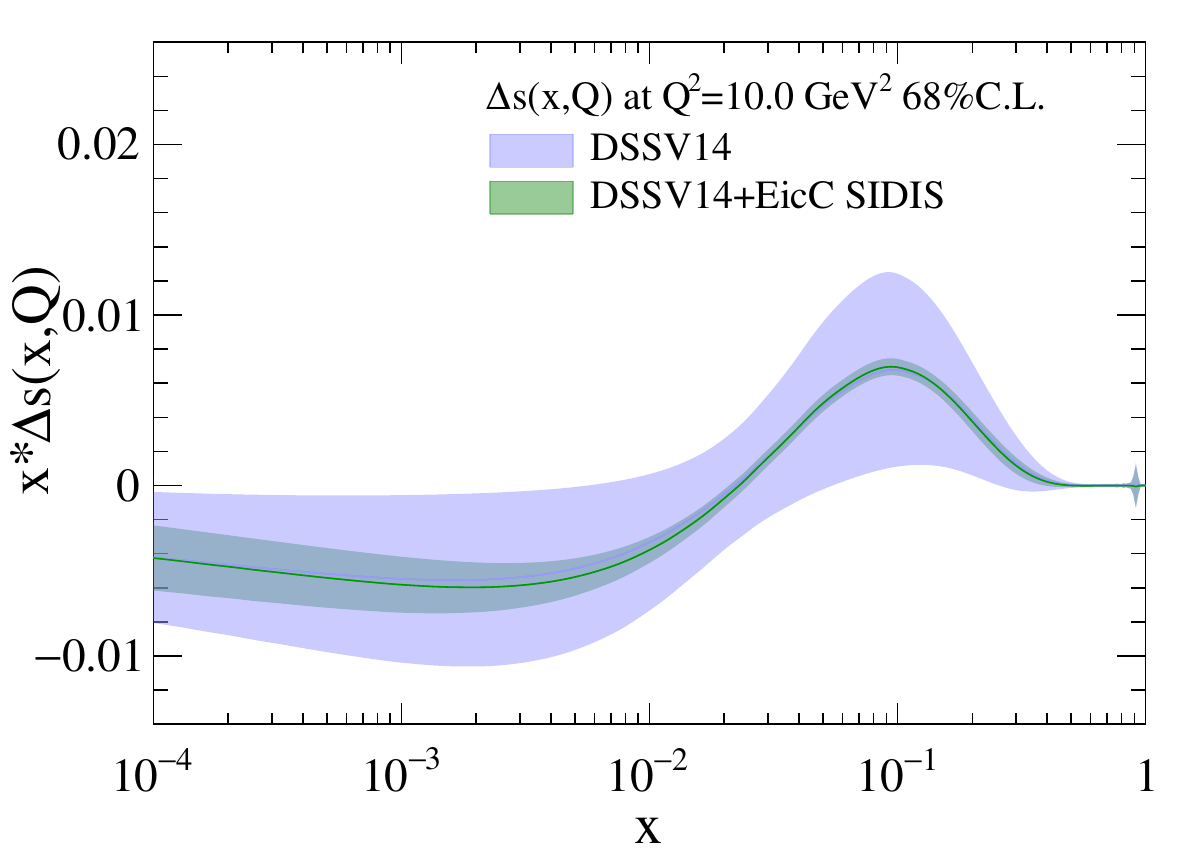}
\includegraphics[width=0.35\textwidth]{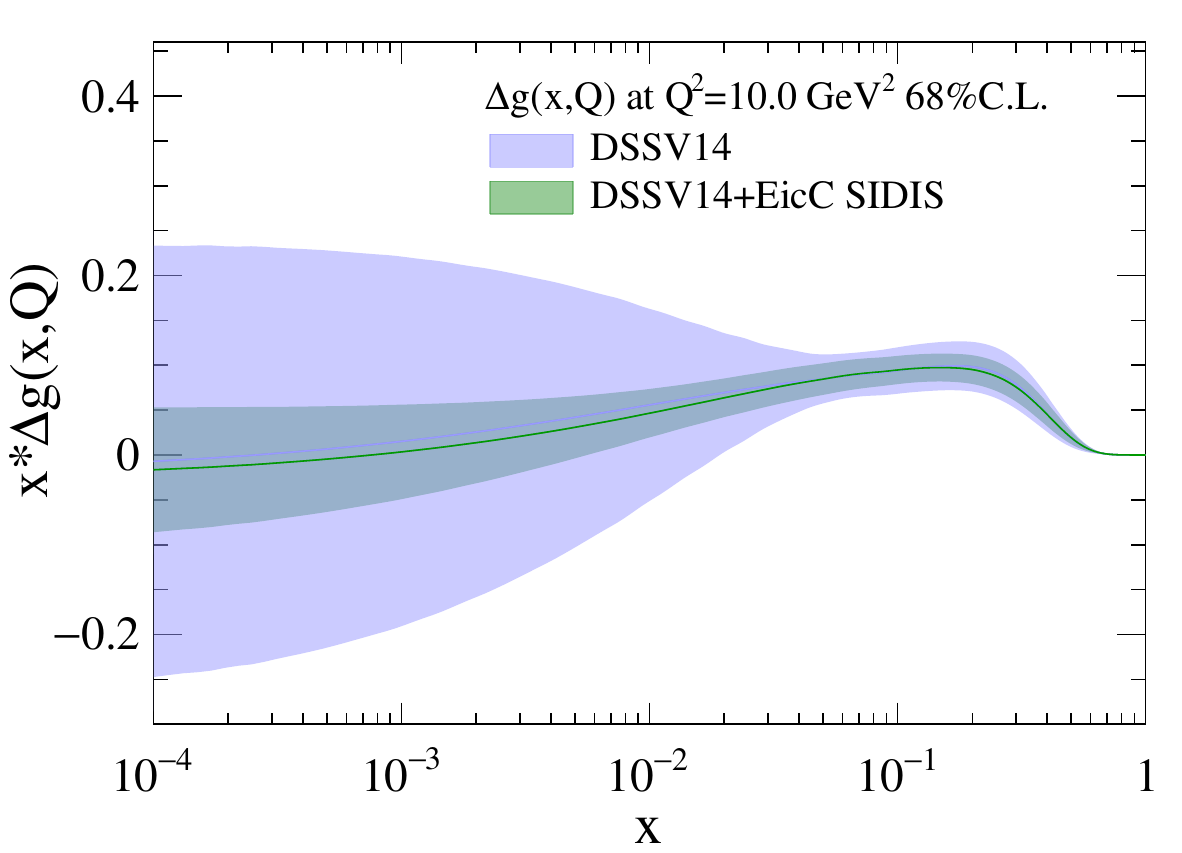}
\caption{Helicity distributions for different partons. Impact study shows that the precision of quark helicity distributions in the $x>0.01$ region can be significantly improved by including EicC SIDIS pseudo data from both e-p and e-$^3$He collisions. Details of the impact study can be referred to Ref. \cite{Anderle:2021dpv}.
Similar studies have also been performed for the US-EIC, one can find the impact in Refs.~\cite{AbdulKhalek:2021gbh,Khanpour:2026erj,Abbott:2026eqp}. Figure replotted from Ref.~\cite{Anderle:2021dpv}.}
\label{Fig:helicity}
\end{figure}

\subsection{Helicity studies at the EicC}
Understanding the helicity structure of the nucleon in terms of quark and gluon degrees of freedom is one of the major goals at the EicC. By combining high-intensity doubly polarized e-p and e-$^3$He collisions as well as good particle-identification capabilities for the SIDIS measurements, the EicC can precisely determine the helicity distributions of various quark flavors in the $x > 0.005$ regime. Compared to the double spin asymmetry measurements in the inclusive DIS process, SIDIS process can provide different weights for different quark flavors by involving fragmentation functions. Moreover, polarized 
$^3$He offers an effective polarized neutron source in the initial state, further enhancing the capability of
flavor separations.

Results of impact studies using EicC pseudodata are shown in Fig. \ref{Fig:helicity}. 
The impact study was performed by including 50 fb$^{-1}$ of e-p and 50 fb$^{-1}$ of e-$^3$He collision pseudo-data. The SIDIS process includes $\pi^{\pm}$ and $K^{\pm}$ final states.
More details can be found in~\cite{Anderle:2021dpv}.
Generally speaking, the uncertainties of
the quark helicity distributions will be significantly reduced by a factor of $\sim$10 in the $x>0.01$ regime once EicC SIDIS data are available, while inclusive DIS data have limited influence~\cite{Anderle:2021dpv}. More importantly, the sea 
quark distributions will be precisely determined at the EicC, offering novel insights into the spin structure
of the nucleon's sea. As for the gluon helicity distribution, the reduction of uncertainty in the $x$ region where EicC is not covered is mainly due to the assumptions for the parameterizations embedded in the DSSV14 PDF sets.

\begin{figure}[htb]
\centering
\includegraphics[width=0.45\textwidth,trim={0.5cm 0 0 2cm},clip]{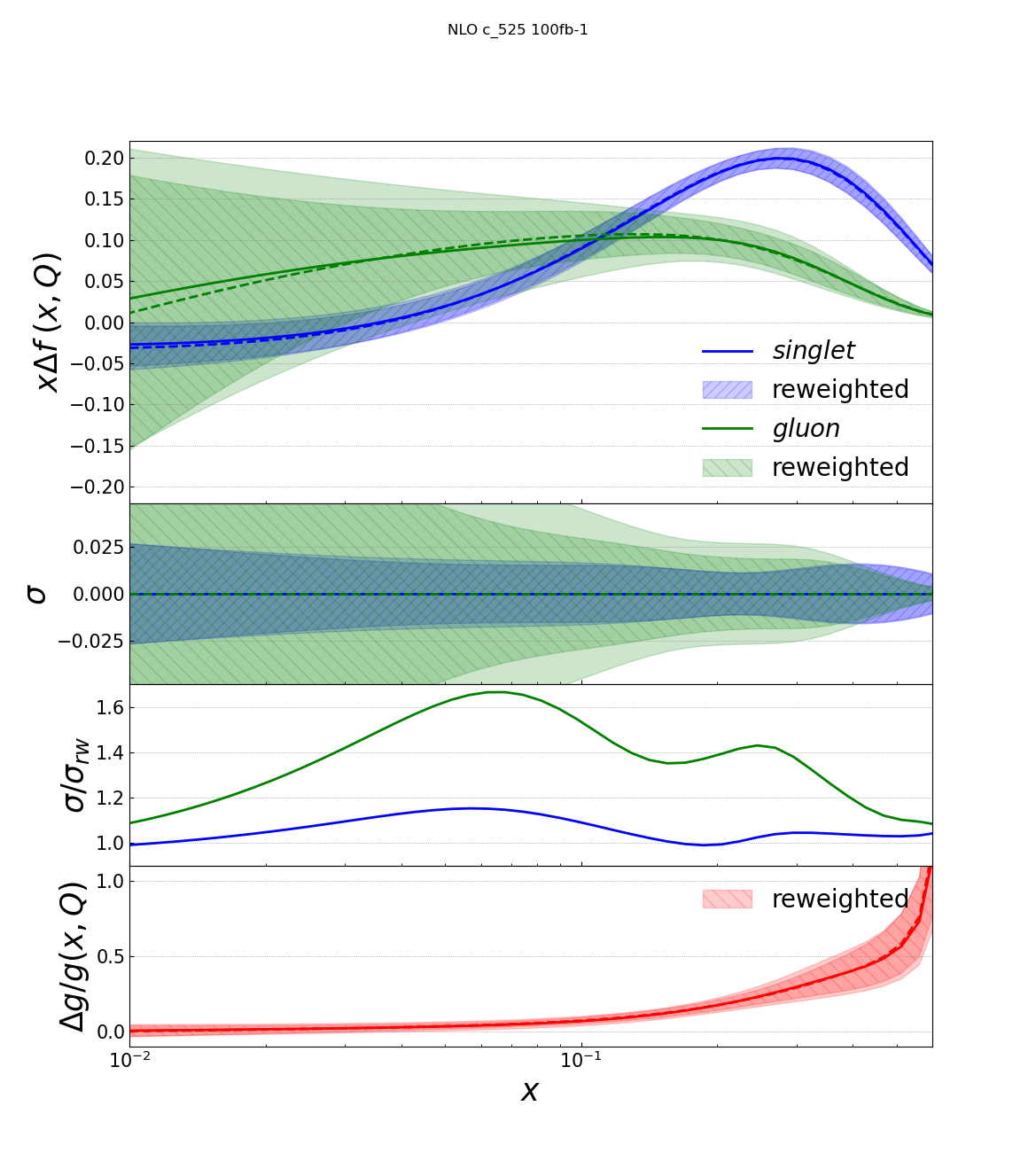} 
\includegraphics[width=0.45\textwidth,trim={0.5cm 0 0 2cm},clip]{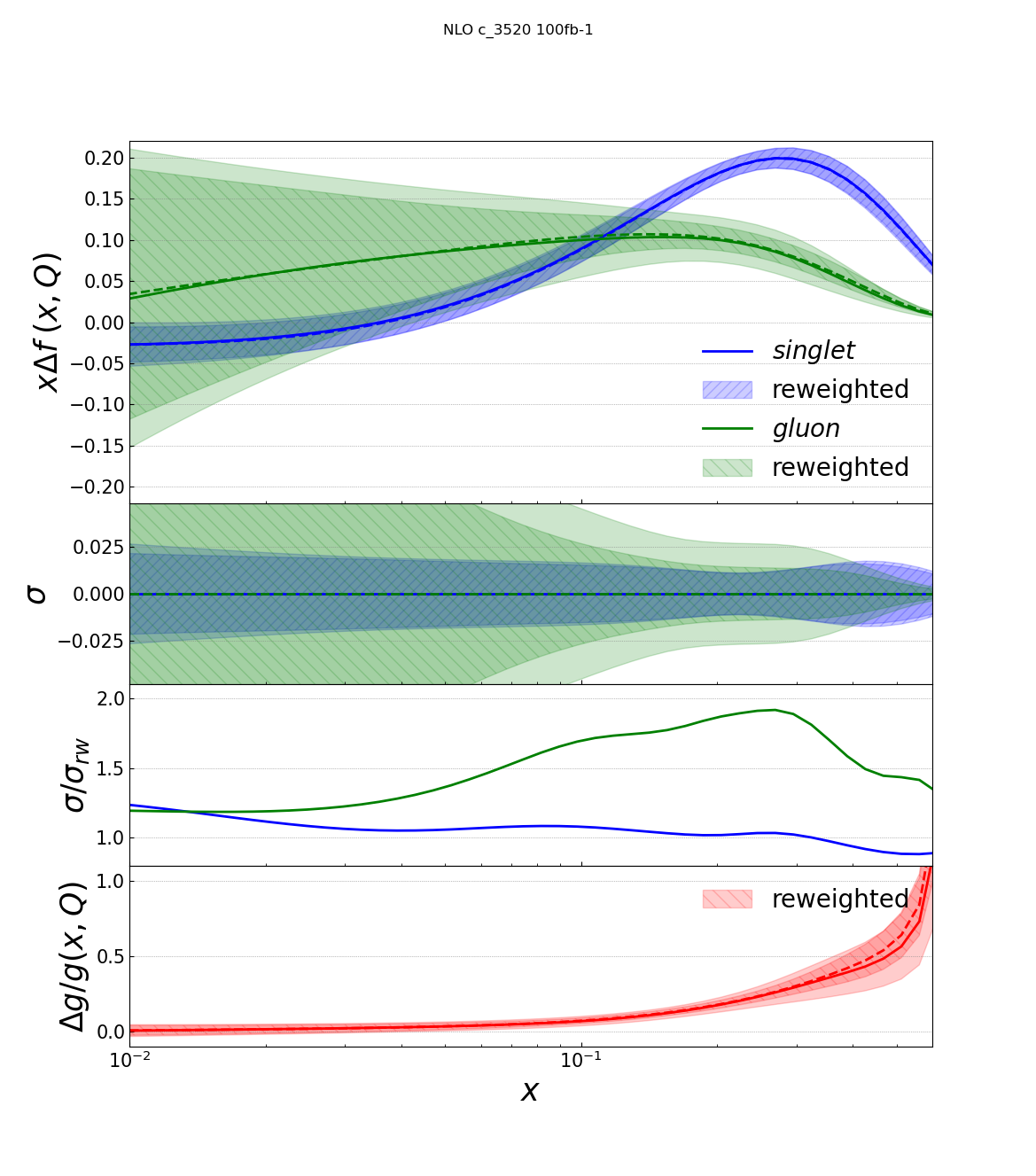}
\caption{Impact on the NNPDFpol1.1~~\cite{Nocera:2014gqa} singlet $\Delta \Sigma$ and gluon $\Delta g$ PDFs for the e-p collision energies 5 GeV $\times$ 25 GeV (left) and 3.5 GeV $\times$ 20 GeV (right) at EicC with integrated luminosity of 100 fb$^{-1}$ respectively. The hashed bands show the impact of the pseudo-data on the distributions' uncertainties, whereas the solid bands show the original uncertainty.
The bottom plots show the ratio of $\Delta g(x)/g(x)$ plotted as a function of the momentum fraction $x$, before and after including EicC pseudo-data. To produce the ratio, NNPDF2.3~\cite{Ball:2012cx} is used for the calculation of unpolarized gluon distributions. The figure is taken from Ref.~\cite{Anderle:2023uvi} (\href{https://creativecommons.org/licenses/by/4.0/}{CC BY 4.0}).}
\label{fig:NNPDFpolNLO}
\end{figure}

On the other hand, EicC has capability in the measurements of charm hadron production
in deep-inelastic scattering. It can provide constraints on the gluon distributions as the relevant unpolarized/polarized charm structure functions provide direct
access to the respective gluon distributions from the leading order, through the photon-gluon
fusion process. Especially in the large $x$ region ($x>0.1$), the ratio of polarized and unpolarized gluons, $\Delta g/g$, exhibits significantly larger values compared to the small $x$ region \cite{Khan:2022vot,Xu:2022yxb}. An impact study \cite{Anderle:2023uvi}, by doing double spin asymmetry $A_{\rm LL}$ measurements in 
the $\vec{e}+\vec{p}\to e' + D^0 + X$ process, shows that EicC will have sizable improvement for the determination
of $\Delta g/g$ in the high x region, as shown in Fig.~\ref{fig:NNPDFpolNLO}.

It is important to note that, in SIDIS, the measured hadron yields are convolutions of the nucleon's PDFs with non-perturbative fragmentation functions, which describe how the struck quarks hadronize into the detected hadrons. Consequently, the precision with which we can determine PDFs, especially for individual quark flavors, depends critically on the accuracy of the fragmentation functions used in the QCD global analysis. In practice, the fragmentation functions are extracted from complementary measurements in $e^+e^-$ annihilation, p-p collisions, and the SIDIS processes. One open issue is the difference in energy scales among different data samples. For example, previous $e^+e^-$ data were primarily taken in the $Q > 10~\mathrm{GeV}$ region, whereas SIDIS data were mainly taken in the low-$Q$ region. The consistency between $e^+e^-$ and SIDIS data should be checked in the low $Q$ region under QCD factorization.
Recent efforts have been made in the measurements of normalized differential cross sections of inclusive hadron productions at BESIII, taking advantage of the beam energy scan data at center-of-mass energy from 2 to 5 GeV \cite{BESIII:2022zit,BESIII:2024hcs,BESIII:2025mbc}.
New state-of-the-art global data fit for FFs has been performed by the ``non-perturbative collaboration'' (NPC collaboration) at full Next-to-Next-to-Leading Order, in which the collinear factorization has been tested with low-momentum-transfer data \cite{Gao:2025hlm}. 

\begin{figure}[htbp]
\begin{center}
\includegraphics[width=0.55\textwidth]{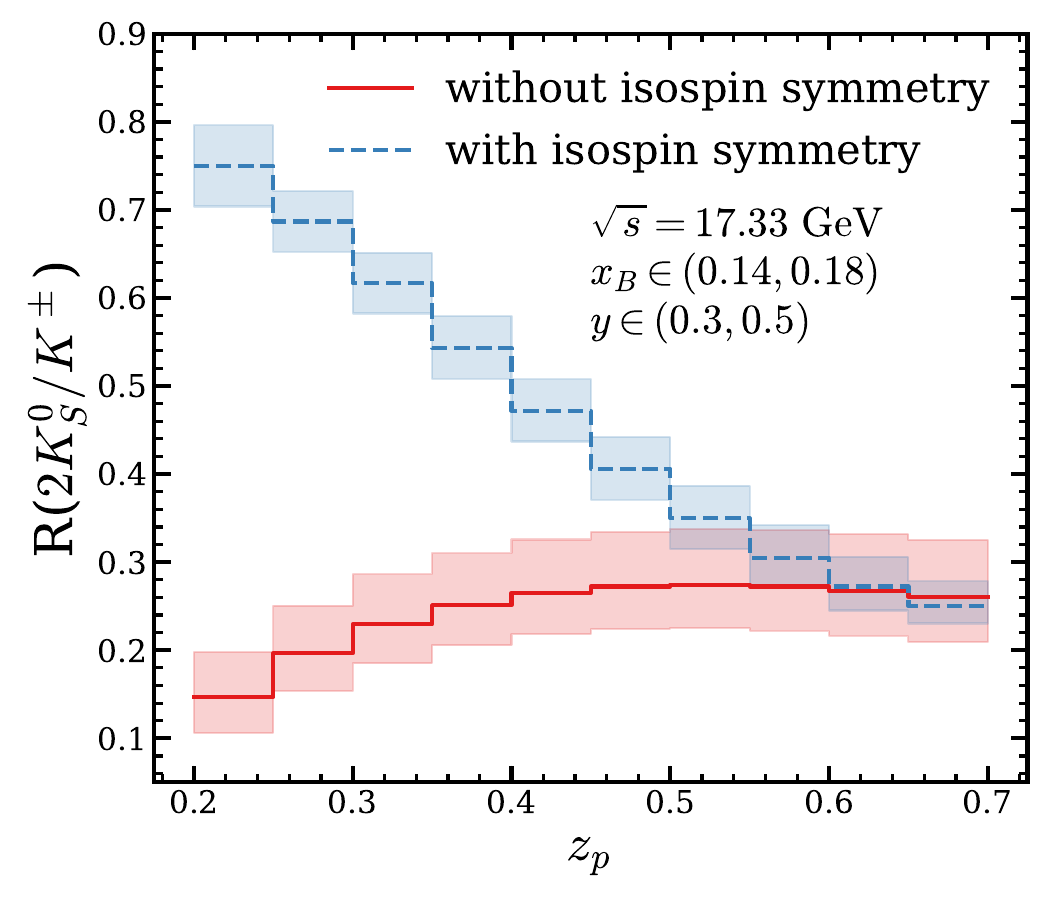}
\caption{\label{fig:FFs} Theoretical predictions for the $K_S^0$ and $K^{\pm}$ yield ratio in SIDIS process on a proton target with two scenarios: 1. $K_S^0$ yield is calculated by the $K_S^0$ fragmentation functions extracted via global data fits to $K_S^0$ productions; 2. $K_S^0$ yield is calculated by using $K^{\pm}$ fragmentation functions assuming isospin symmetry, namely, 
$D_q^{K_S^0} = \frac{1}{2} (D_{q'}^{K^+} + D_{q'}^{K^-})$, where $q(q') = u(d)$ or $d(u)$. 
More details can be found in Ref.~\cite{Gao:2025bko}. The figure is adapted from Ref.~\cite{Gao:2025bko} (\href{https://creativecommons.org/licenses/by/4.0/}{CC BY 4.0}).
}
\end{center}
\end{figure}

Unpolarized high-precision SIDIS data will be a by-product at EicC by averaging the polarization of beams in the initial state, therefore, fragmentation functions can be
systematically studied.
Moreover, isospin symmetry in the fragmentation process has been investigated recently \cite{BESIII:2025mbc,Gao:2025bko}, due to the fact that $K^{\pm}$ yield is observed to be systematically higher than that of $K_S^0$ yield in $e^+e^-$ collisions. 
Global data analysis shows that the SIDIS process is highly sensitive to isospin symmetry, as illustrated in 
Fig.~\ref{fig:FFs} with a center-of-mass energy of around 17 GeV, in the kinematic region $0.14<x<0.18$ and $0.3<y<0.5$, the predicted yield differences between $K_S^0$ and $K^{\pm}$ could be significant in the $z<0.5$ region, depending on whether isospin symmetry is assumed or not.

	\newpage
\section{Three-dimensional nucleon tomography}\label{sec:3dspin}

For decades, the internal structure of the nucleon has been primarily described within a one-dimensional  framework using collinear PDFs.  While this 1D picture has successfully predicted cross sections in inclusive deep inelastic scattering and hadron colliders, validating perturbative QCD, it is inherently incomplete. By integrating over transverse degrees of freedom, collinear PDFs average out the spatial distributions, intrinsic transverse motions, and spin-orbit correlations that are essential for a complete description of nucleon internal dynamics.

To achieve a more comprehensive understanding of nucleon structure, the field has moved toward three-dimensional imaging, often referred to as nucleon tomography. This effort relies on two complementary frameworks: TMDs (see Ref.~\cite{Boussarie:2023izj} and reference therein) and GPDs~\cite{Mueller:1998fv,Ji:1996ek,Diehl:2003ny}, both of which can be connected to the fully unintegrated Wigner distributions \cite{Ji:2003ak}. TMDs describe the 3D momentum structure, incorporating the intrinsic transverse momentum of partons and its correlation with the spins of the nucleon and the partons. GPDs provide a joint distribution of longitudinal momentum and transverse spatial coordinates (impact parameter space), allowing for the extraction of parton OAM and the mapping of mechanical properties, such as pressure and shear force distributions\cite{Burkert:2018bqq,Kumericki:2019ddg}. Together, these distributions are required to address unresolved questions, including the proton spin decomposition~\cite{EuropeanMuon:1987isl,Jaffe:1989jz,Ji:1996ek}.

Experimental extraction of these 3D distributions requires high-luminosity facilities with polarized beams and the capability to measure exclusive or semi-inclusive final states over a wide kinematic range. The proposed EicC is designed to meet these requirements. Operating at a center-of-mass energy of 15-20 GeV, the EicC will bridge the kinematic gap between the high-$x$ valence quark region studied at the 12 GeV Jefferson Lab and the low-$x$ gluon-dominated regime targeted by the future US-EIC. With highly polarized electron, proton, and light ion ($^3$He) beams, as well as comprehensive particle identification, the EicC will provide precise measurements of the sea-quark 3D structure, offering essential data to construct the spatial and momentum distributions of partons within the nucleon.

\subsection{Transverse momentum dependent parton distributions}

\subsubsection{TMD physics and formalism}
The theoretical foundation for mapping the transverse momentum structure of nucleons was established in the early 1980s through the seminal work of Collins and Soper~\cite{Collins:1981uk}, and subsequently extended by Collins, Soper, and Sterman~\cite{Collins:1984kg} into what is universally known as the Collins-Soper-Sterman (CSS) formalism. This rigorous factorization theorem demonstrated that in high-energy scattering processes involving a small measured transverse momentum $q_\perp \ll Q$ (where $Q$ is the large hard-scattering scale), the cross section can be systematically factorized into perturbatively calculable hard parts and non-perturbative TMD parton distributions and fragmentation functions. Formally, TMD PDFs are defined via the quark-quark correlation function on the light-cone:
\begin{equation}
\Phi(x,k_\perp,S)
=\int \frac{d y^- d^2 {y_\perp}}{(2\pi)^3} e^{-ixP^+ y^- + i{ k}_\perp\cdot {y}_\perp}\langle P,S|\bar{\psi}(y){\cal W}(y,0)\psi(0)|P,S\rangle,
\label{eq:tmd_correlator}
\end{equation}
where the Wilson line ${\cal W}(y,0)$ ensures color gauge invariance.  A hallmark of the CSS formalism is its elegant treatment of large logarithmic enhancements, $\ln(Q^2/q_\perp^2)$, arising from multiple soft gluon emissions. These large logarithms are resummed to all orders into a Sudakov form factor, which contains both perturbative and non-perturbative components. The non-perturbative part, encoded in the Collins-Soper (CS) kernel, dictates the energy evolution of TMDs at large $b_\perp$. Constraining the universality and kinematic dependence of the CS kernel is currently one of the central focuses in the TMD community.  At asymptotically large transverse momenta $q_\perp \sim Q$, the pure TMD factorization framework naturally breaks down, necessitating a systematic matching procedure to collinear factorization via the $Y$-term to restore the full kinematic dependence.

\begin{figure}[htb]
\centering
\includegraphics[width=0.6\textwidth]{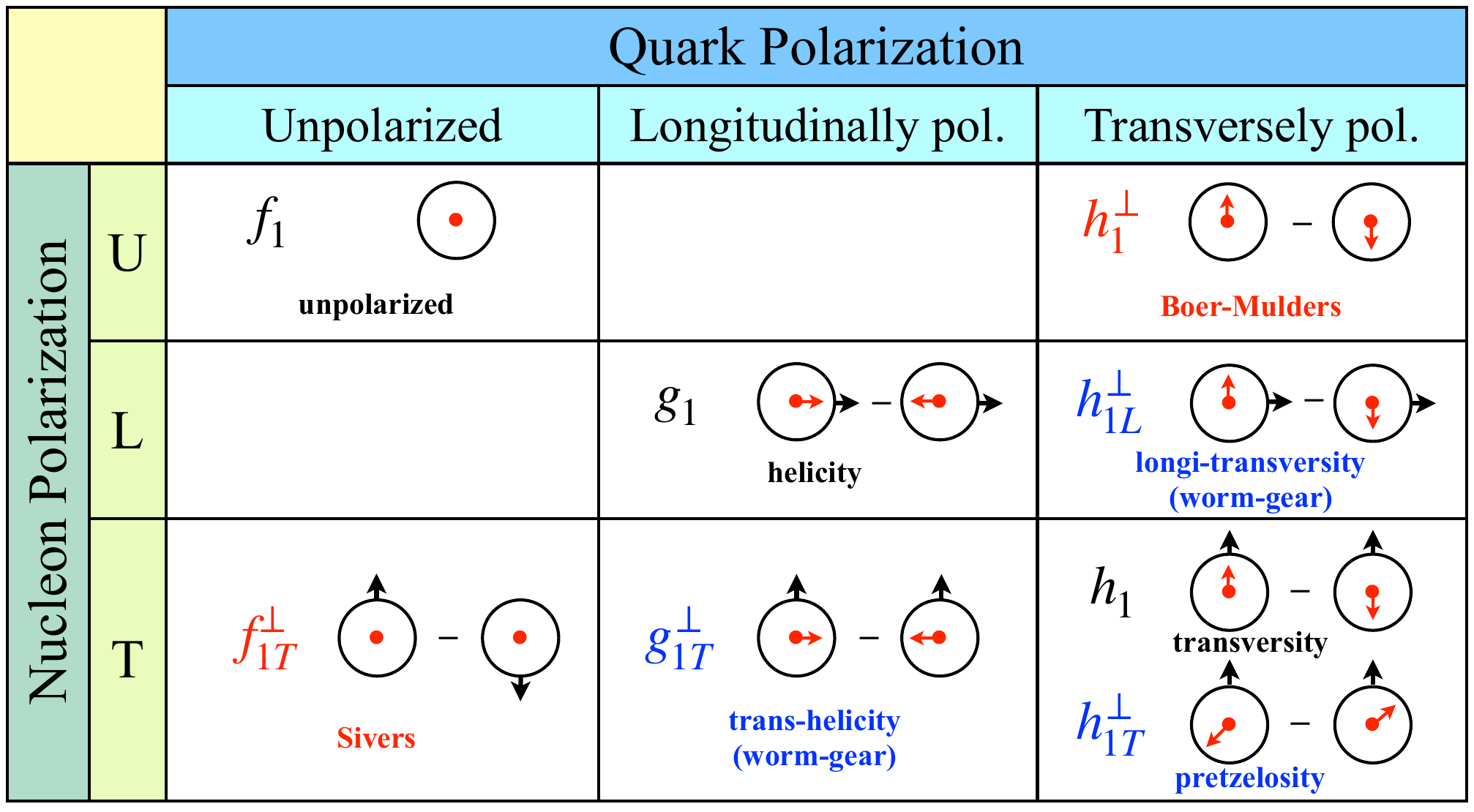} 
\caption{Eight quark TMDs at leading twist. }
\label{fig:tmds}
\end{figure}

Building upon this theoretical foundation, the phenomenological landscape of TMDs was vastly expanded in the late 1990s. Mulders and Tangerman provided a systematic decomposition of transverse-momentum-dependent quark correlators in terms of a complete set of Dirac structures, thereby classifying the polarization-dependent TMD distribution and fragmentation functions relevant for semi-inclusive scattering~\cite{Mulders:1995dh}. This framework was subsequently refined and extended by Boer and Mulders, who highlighted the phenomenological importance of time-reversal-odd TMDs, including the Boer-Mulders function~\cite{Boer:1997nt}. At leading twist, the quark correlator of a spin-$1/2$ nucleon contains eight independent TMDs, organized according to quark and nucleon polarizations as shown in Fig.~\ref{fig:tmds}. Rather than ordinary probabilistic densities in all cases, these functions encode spin-dependent momentum correlations and provide a three-dimensional, tomographic view of nucleon structure. 

Furthermore, TMDs are the key to mapping the ``spin-orbit'' effects. The Sivers function~\cite{Sivers:1990fh}, identified as a naive T-odd TMD, $f_{1T}^{\perp}$, quantifies the correlation between the transverse spin of the nucleon and the intrinsic transverse momentum of unpolarized quarks, and is sensitive to the initial- and final-state interactions encoded in the gauge link. Another leading-twist T-odd TMD, known as the Boer-Mulders function introduced in Ref.~\cite{Boer:1997nt}, $h_{1}^{\perp}$, describes the correlation between the transverse spin of quarks and their intrinsic transverse momentum inside an unpolarized nucleon. This distribution has attracted considerable attention in TMD phenomenology because it gives access to spin-momentum correlations even in unpolarized hadrons.

The introduction of T-odd TMDs posed a profound theoretical puzzle regarding time-reversal invariance. For years, these functions were widely believed to vanish due to naive time-reversal symmetry~\cite{Collins:1992kk}. This paradigm was shattered in 2002 when Brodsky, Hwang, and Schmidt demonstrated that final-state interactions (FSI) via one-gluon exchange between the struck quark and the target remnant naturally generate a non-zero single-spin asymmetry~\cite{Brodsky:2002cx}. Shortly after, Collins rigorously proved that~\cite{Collins:2002kn}: the non-vanishing T-odd functions are inextricably linked to the intricate path of the gauge link (Wilson line) ${\cal W}(y,0)$. Specifically, this gauge link takes the form of a staple-like path extending along the light-cone boundary, elegantly capturing the color interactions between the active parton and the spectating remnants. Because the Wilson line points to future infinity ($+\infty$) to account for final-state interactions in SIDIS, and to past infinity ($-\infty$) for initial-state interactions in the Drell-Yan (DY) process, the T-odd TMDs are forced to change sign. This leads to the celebrated modified universality prediction: $f_{1T}^\perp|_{\rm SIDIS} = -f_{1T}^\perp|_{\rm DY}$, a fundamental test of the gauge nature of QCD, which awaits to be tested with high precision in the future experiment.

The study of TMDs also sheds new light on our understanding of nucleon spin structure. It has been realized that the spin budget must be balanced among the spins of its quarks and gluons, and, most elusively, their OAM. Beyond providing indirect constraints on OAM by precisely measuring quark and gluon helicity distributions, EicC will provide more direct signature for parton OAM through TMDs. The ``worm-gear'' TMD, $g_{1T}^\perp$, for instance, describes the probability of finding a longitudinally polarized quark within a transversely polarized nucleon. A non-zero value for this function can only arise from a coherent interference between quantum mechanical states that differ by one unit of OAM, specifically between the S-wave and P-wave components of the nucleon's light-front wavefunctions, thus serving as a  model dependent signature of parton orbital motion. In addition to the aforementioned distributions, the TMD helicity distribution $g_{1L}(x,k_T)$ offers a crucial window into the correlation between the longitudinal polarization of partons and their transverse momentum. A recent global analysis achieved the first extraction of the TMD helicity distributions~\cite{Yang:2024drd}, revealing a profound physical mechanism: the Wigner rotation effect. When a proton is boosted to the infinite momentum frame, the intrinsic transverse momentum of quarks causes a relativistic Wigner rotation of their spin states, mixing longitudinal and transverse components. It was  further demonstrated that the extracted quark helicity TMDs are consistent with existing global fits of the collinear quark helicity PDFs, once $k_T$ is integrated out to a hard scale~\cite{Yang:2024drd}.

To map these TMDs, three complementary experimental processes are utilized. The primary tool is SIDIS, $e + N \to e' + h + X$, where the detection of a final-state hadron makes the cross section sensitive to the struck quark's transverse momentum. Under the one-photon-exchange approximation, the SIDIS differential cross section is expressed in terms of 18 structure functions, constructed from convolutions of TMD PDFs and TMD Fragmentation Functions (FFs). During the past two decades, great efforts have been made to measure the Sivers asymmetry and other TMD-related observables via the SIDIS process at many experimental facilities around the world, including HERMES, COMPASS, and JLab~\cite{JeffersonLabHallA:2011ayy,COMPASS:2014bze,COMPASS:2012dmt,HERMES:2020ifk,HERMES:2009lmz,JeffersonLabHallA:2011vwy}. The second crucial process is DY lepton pair production in hadron-hadron collisions ($p + p \to l^+ + l^- + X$)~\cite{COMPASS:2017jbv}, which involves the convolution of two TMD PDFs without fragmentation functions, serving as the critical testing ground for the T-odd sign change. Finally, electron-positron annihilation into two hadrons ($e^+ + e^- \to h_1 + h_2 + X$) is indispensable for directly probing the TMD FFs~\cite{Belle:2005dmx,Belle:2011cur}, such as the Collins fragmentation function and di-hadron fragmentation function, which are essential ingredients for cleanly extracting chiral-odd parton distributions like transversity from SIDIS data.

However, these measurements face various difficulties. The JLab experiments were carried out at relatively low energies, where higher-twist effects can be sizable, and target mass corrections must be evaluated to extract clean leading-twist signals~\cite{Accardi:2009md}. HERMES data were mostly collected in the valence quark region, providing limited sensitivity to sea quark distributions and leaving the flavor dependence of TMDs incompletely constrained. The separation of current fragmentation and target fragmentation also remains a challenging task at fixed target facilities, often heavily relying on the Berger criterion and making the isolation of pure current-region TMD fragmentation somewhat ambiguous~\cite{Boglione:2016bph}. Consequently, TMDs, especially the spin-dependent ones, remain poorly determined, demanding high-precision measurements from next-generation colliders.

\subsubsection{EicC's unique advantages}
The EicC provides specialized kinematic coverage for a comprehensive SIDIS program. Operating in the intermediate-$x$ region ($0.005 < x < 0.3$) and spanning $1\text{ GeV}^2 < Q^2 < 30\text{ GeV}^2$, the EicC uniquely probes the domain of sea quarks, where constraints are currently insufficient. Its phase space is critical for accurately extracting sea quark TMDs and evaluating power-suppressed corrections. Recent impact studies focusing on this exact kinematic region explicitly demonstrate EicC's capability to deliver precise, quantitative constraints on key TMDs \cite{Zeng:2022lbo,Zeng:2023nnb,Yang:2024bfz}.

The EicC's capability to operate with highly polarized proton and effective neutron (${}^{3}\mathrm{He}$) beams is critical for achieving comprehensive light-flavor separation. A combined analysis of $e$--$p$ and $e$--${}^{3}\mathrm{He}$ collisions under similar kinematic conditions can exploit isospin symmetry to disentangle the $u$- and $d$-quark distributions. In addition, comprehensive final-state hadron identification, particularly the high-precision detection of both $\pi^{\pm}$ and $K^{\pm}$ mesons, provides sensitivity to the separation of valence- and sea-quark contributions and to the less well-constrained strange-quark distributions. Charm-quark distributions require dedicated heavy-flavor measurements and are discussed separately. Together, these complementary measurements provide unprecedented constraints on global phenomenological analyses.

\begin{figure}[htb]
\centering
\includegraphics[width=0.7\textwidth,trim={0.5cm 0 0 2cm},clip]{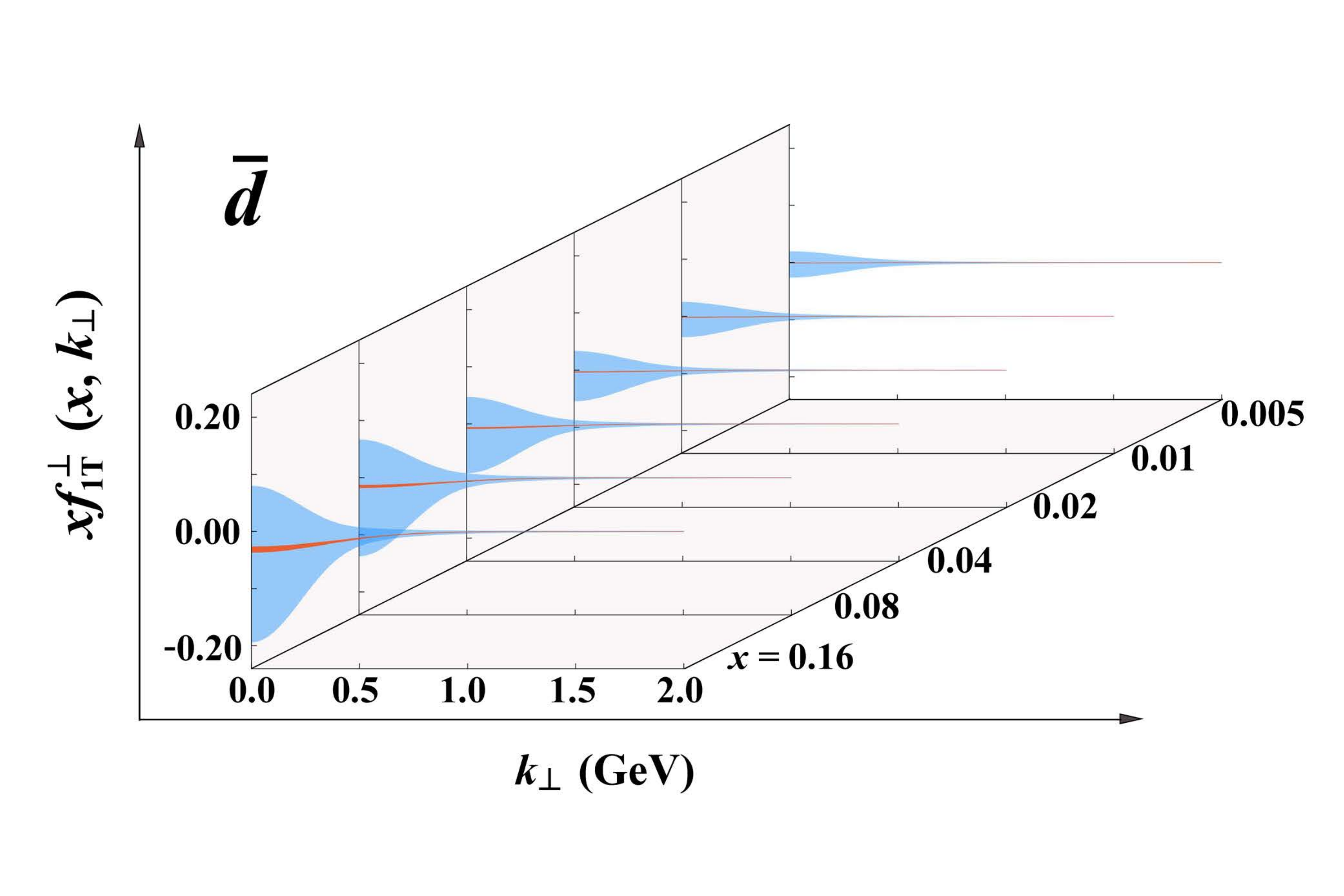} 
\caption{The transverse momentum distribution of the Sivers functions at different $x$ values for $\bar{d}$ quark. The light blue bands represent the
uncertainties of the fit to world SIDIS data, the red bands represent the EicC projections.
The figure is adapted from Ref.~\cite{Zeng:2022lbo} (\href{https://creativecommons.org/licenses/by/4.0/}{CC BY 4.0}).}
\label{sivers}
\end{figure}

For the Sivers function, recent impact studies utilize the latest global phenomenological parameterizations to generate pseudodata specifically tailored to the EicC's kinematic acceptance and expected statistical reach~\cite{Zeng:2022lbo}. A new global analysis was then performed using more flexible parameterizations, fewer assumptions to allow for extensive flavor separation, and stricter kinematic requirements for data selection to ensure the validity of TMD factorization. This analysis was designed to systematically quantify the reduction in uncertainties relative to those constrained by current world data. The resulting studies demonstrate that the $f_{1T}^\perp(x, k_\perp^2)$ distributions for $u, d, \bar{u}$, and $\bar{d}$ quarks will all be significantly improved, typically by one order of magnitude in the intermediate $x$-range of $0.005 \sim 0.2$, as shown in Fig.~\ref{sivers} for $\bar d$ Sivers function, for example. The unprecedented precision projected for the sea quarks ($\bar{u}$, $\bar{d}$), which currently suffer from large statistical uncertainties in global fits, is a direct consequence of the EicC's high luminosity combined with its optimized coverage in the sea-quark-dominant region. Furthermore, by heavily leveraging the high-purity kaon identification capabilities, the EicC simulations show the unique capability to constrain the $s$ and $\bar{s}$ Sivers distributions. This will provide a novel, quantitative perspective on the spin-orbit correlations within the nonperturbative strange sea, whose sign and magnitude remain largely undetermined in current phenomenological extractions. In addition, for access to the largely unexplored gluon Sivers function, detailed impact studies at EICs via the di-jet and di-hadron channels can be found in Refs.~\cite{Zheng:2018ssm,AbdulKhalek:2021gbh}.

In the small-$x$ region, the Sivers function uniquely intersects with the dynamics of the QCD odderon (the C-odd $t$-channel gluon exchange). Theoretical studies utilizing the CGC formalism demonstrate that both gluon and quark Sivers functions at small $x$ are dynamically generated by the spin-dependent odderon~\cite{Zhou:2013gsa,Boer:2015pni,Dong:2018wsp,Boer:2022njw}. A striking consequence of the odderon's C-odd nature is that the generated quark and anti-quark Sivers functions have equal magnitudes but opposite signs: $f_{1T,q}^\perp(x, k_\perp^2) = -f_{1T,\bar{q}}^\perp(x, k_\perp^2)$. 

\begin{figure}[htb]
\centering
\includegraphics[width=0.8\textwidth]{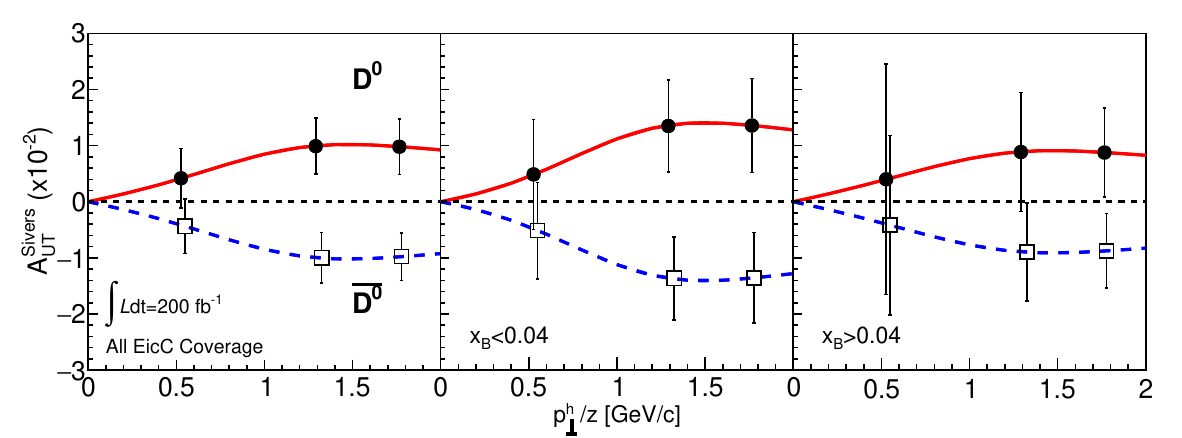} 
\caption{The projected single spin asymmetry for $D^0$ and $\bar  D^0$ production in SIDIS process at EicC. The figure is adapted from Ref.~\cite{Zhu:2024iwa} (\href{https://creativecommons.org/licenses/by/4.0/}{CC BY 4.0}).
 }
\label{d0}
\end{figure}

Experimental verification of this predicted sign change would serve as compelling evidence for the spin-dependent odderon. At EicC, this can be tested by measuring the transverse single-spin asymmetry (SSA) in open charm production ($ep \to e' D^0 X$ versus $ep \to e' \bar{D}^0 X$), assuming the intrinsic charm Sivers distribution is negligible. The signature of this mechanism is a distinct sign separation between the $A_{UT}$ asymmetries of $D^0$ and $\bar{D}^0$ mesons. 
Dedicated impact studies for EicC kinematics ($Q^2 > 2\text{ GeV}^2$, $x < 0.1$) have been performed using full detector simulations with a silicon and MPGD based tracking system for the decay topology reconstruction and the high performance particle-identification system for the pion and kaon separations~\cite{Zhu:2024iwa,Anderle:2023uvi}. The signals were cleanly reconstructed via the $D^0 \to K^- \pi^+$ decay channel, and the background was effectively handled using the side-band method across different $p_{h\perp}/z$ bins. Assuming an integrated luminosity of 200~fb$^{-1}$, the statistical projections demonstrate that EicC can clearly differentiate the $A_{UT}$ of $D^0$ and $\bar{D}^0$ mesons (Fig.~\ref{d0}). Specifically, the projected uncertainties show a separation significance of approximately $3\sigma$ in the $1.1 < p_{h\perp}/z < 2\text{ GeV}/c$ region. This highlights EicC's capability to measure the open charm SSA and decisively test the odderon-induced sign change in the Sivers distribution.

\begin{figure}[htb]
\centering
\includegraphics[width=0.55\textwidth]{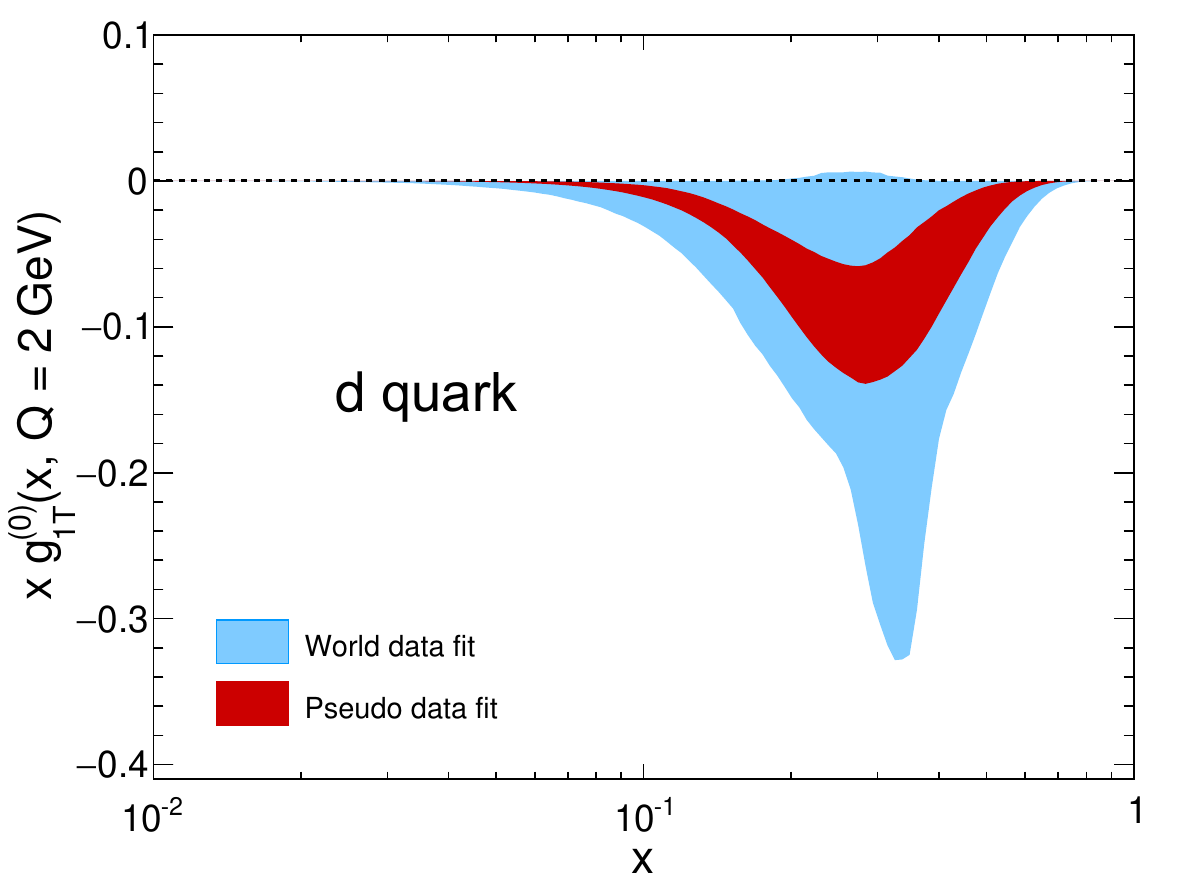} 
\caption{The impact on the constraint of d quark worm gear function with $Q=2$ GeV. The blue band represents the current uncertainty, while the red band shows the projected uncertainty after taking into account EicC pseudo-data. The impact study was assumed with 50 fb$^{-1}$ of $ep$ and $e^3$He collisions, separately. The figure is adapted from Ref.~\cite{Yang:2024bfz} (\href{https://creativecommons.org/licenses/by/4.0/}{CC BY 4.0}). 
 }
\label{wormgear}
\end{figure}

Similarly, extensive impact studies have been performed for the trans-helicity worm-gear function $g_{1T}^\perp$. Projections incorporating EicC's specific kinematics indicate that the precision of $g_{1T}^\perp(x, k_\perp^2)$ distributions for $u$ and $d$ quarks will be substantially improved in the $0.05 < x < 0.5$ region where the signal is sizable, as shown in Fig.~\ref{wormgear}. Crucially, these simulations establish that EicC will enable the first reliable extraction of sea quark worm-gear functions, which are very little known from existing data~\cite{Yang:2024bfz}, completing our leading-twist tomographic picture.

For the transversity distribution $h_1(x, k_\perp^2)$, global analyses currently show the $d$-quark distribution is less constrained than the $u$-quark due to fewer effective neutron target data. EicC will make significant improvements on the precision of both $u$ and $d$ transversity distributions by taking advantage of proton and effective neutron data~\cite{Zeng:2023nnb}. Besides, EicC will be able to constrain sea quark transversity distributions, providing valuable information on the intrinsic sea in nucleon spin structures. Complementary to this, EicC data will substantially improve the precision of both pion and kaon Collins fragmentation functions by a factor of two or more, facilitating accurate extractions via simultaneous fits to SIDIS and $e^+e^-$ annihilation (SIA) data. Projected measurements of di-hadron fragmentation functions (DiFFs) will also serve as an independent cross-verification.

\begin{figure}[htb]
\centering
\includegraphics[width=0.6\textwidth]{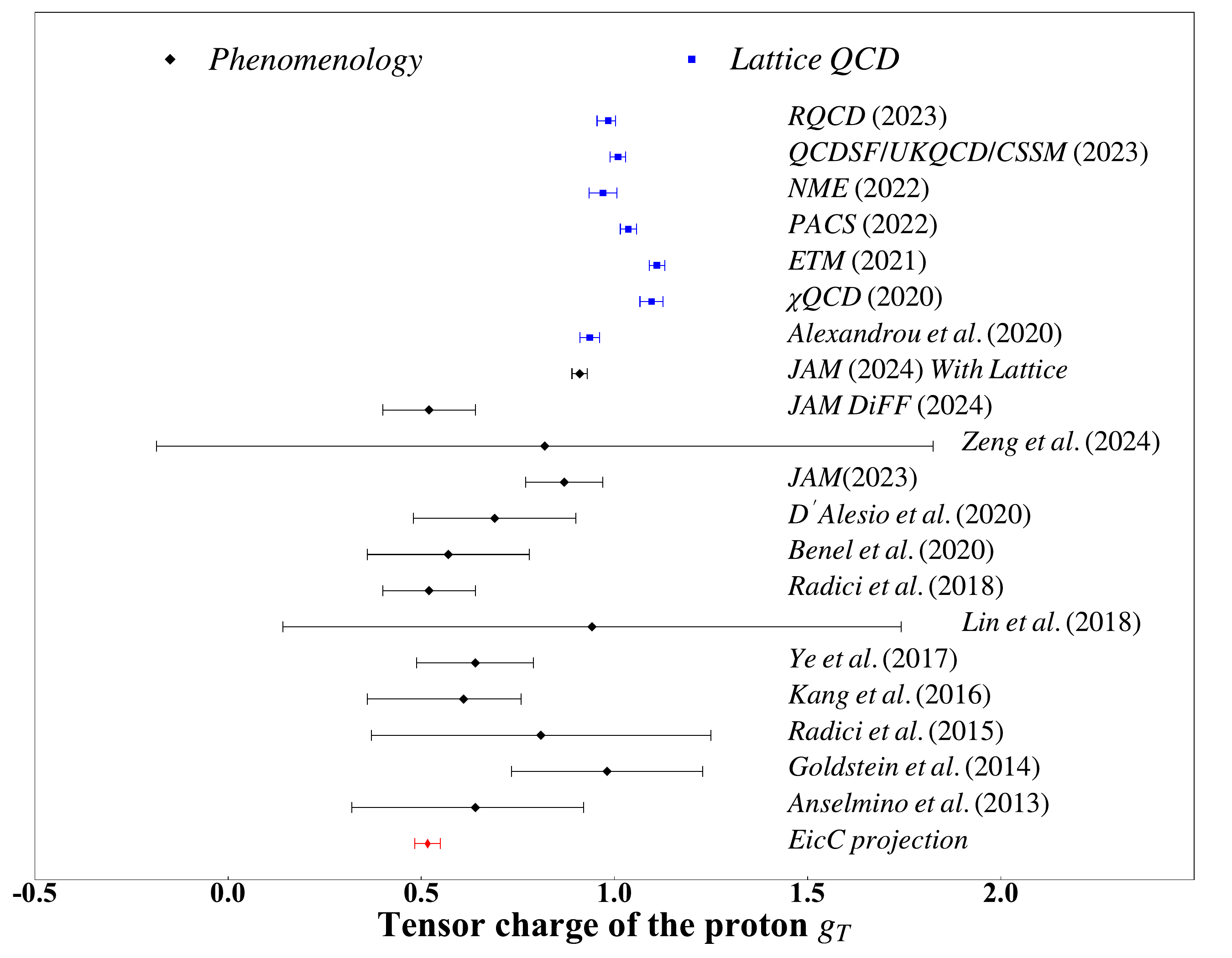} 
\caption{ EicC projection on the tensor charge $g_T$ along with results from lattice QCD calculations, and
phenomenological extractions. The figure is adapted from Ref.~\cite{Zeng:2023nnb} (\href{https://creativecommons.org/licenses/by/4.0/}{CC BY 4.0}). The impact study was assumed with 50 fb$^{-1}$ of $ep$ and $e^3$He collisions, separately.
 }
\label{gt}
\end{figure}

The integral of the transversity distribution gives the nucleon tensor charge, a fundamental quantity characterizing the coupling to a tensor current. The tensor charges are key inputs for beyond-Standard-Model (BSM) searches (e.g., electric dipole moments) and tests of QCD dynamics. A comprehensive phenomenological study on tensor charge can be found in Ref.~\cite{Cocuzza:2023oam}. While the tensor charge is often expected to be valence-dominated, whether the sea quark contribution is negligible remains to be definitively verified. The EicC SIDIS program will play an essential role in superseding this assumption by providing direct experimental constraints on the sea quark tensor charges~\cite{Zeng:2023nnb}.  It is shown in Fig.~\ref{gt} that  
the proton tensor charge can be determined at EicC with precision comparable to the lattice calculations.

Precision measurements of $\Lambda$ and $\bar\Lambda$ hyperon polarization further clarify parton fragmentation and spin transfer mechanisms. Based on the simulations of the $\Lambda \rightarrow p \pi^-$ topological decay structure, residual backgrounds can be strongly suppressed and effectively estimated using the side-band method on the invariant mass distributions~\cite{Ji:2023cdh}. Theoretical predictions and statistical projections based on just one month of expected EicC data show that the spontaneous transverse polarization in unpolarized collisions, driven by the polarizing fragmentation function $D_{1T}^\perp$, can be mapped with high fidelity. With transversely polarized proton beams, the induced $\Lambda$ polarization probes the convolution of $h_1$ and the transversity fragmentation function $H_1$. High-statistics samples will enable the separation of quark flavors in $\Lambda$ FFs, offering specific sensitivity to strange quark transversity. Furthermore, the wide coverage of $Q^2$ and $z$ provides an ideal platform to study contributions from both current and target fragmentation regions and the transition between them, alongside exploring decay contributions from heavier particles like $\Sigma$ and $\Lambda_c$~\cite{Ji:2023cdh}.  Moreover,
 in unpolarized collisions, EicC enables the study of the Boer-Mulders function, which requires careful separation from the Cahn effect in the $\cos 2\phi$ azimuthal asymmetry.  The intermediate-$Q^2$ reach is advantageous for characterizing higher-twist dynamics.

These experimental efforts will be highly complementary to rapid advancements in lattice QCD calculations of TMD-related quantities. Utilizing the Large-Momentum Effective Theory (LaMET), groundbreaking nonperturbative calculations have been achieved for the Collins-Soper kernel~\cite{Ebert:2018gzl}, the intrinsic soft function~\cite{LatticeParton:2020uhz}, the nucleon unpolarized TMDPDF~\cite{LatticePartonCollaborationLPC:2022myp}, and the pion TMD wave function~\cite{LatticeParton:2023xdl}. Determinations of TMD ratios, including generalized Sivers~\cite{Musch:2011er,Yoon:2017qzo}, Boer-Mulders~\cite{Engelhardt:2015xja}, and worm-gear shifts~\cite{Yoon:2017qzo}, are steadily increasing in precision. The high-precision EicC data will provide crucial experimental benchmarks to test and guide these first-principles calculations.

Finally, in electron-nucleus ($eA$) collisions, the SIDIS-type observables offer unique advantages for studying parton transport in CNM~\cite{Liang:2008vz,Schafer:2013mza}. By measuring the deflected electron, the kinematics of the initial hard parton are directly controlled, enabling sensitive probes of three-dimensional parton propagation dynamics. Multi-dimensional-differential measurements of flavor-dependent hadron multiplicities ($x, Q^2, z_h, p_T$) will robustly constrain the transport parameter $\hat{q}$, which characterizes transverse momentum broadening~\cite{Liang:2008vz,Schafer:2013mza,Ru:2019qvz,Ru:2023ars}. The EicC will deliver high-precision measurements in the intermediate-$x$ region where the broadening effect is sizable. Leveraging various nuclear species for path-length selection will drastically enhance sensitivity to medium modifications. The extensive EicC kinematics, spanning relatively large $x$ and moderate $Q^2$, are perfectly suited to disentangle partonic-level versus hadronic-level interactions and precisely map the transition from partonic transport to hadronic absorption in the nuclear environment. Furthermore, the EicC will likely reach the small-$x \sim 0.01$ region in eA collisions, providing a valuable probe into the initial state of the Color Glass Condensate.

\subsection{Generalized parton distributions}
\subsubsection{GPD physics and formalism}
GPDs characterize the 3D nucleon structure in the joint transverse position and longitudinal momentum phase space~\cite{Mueller:1998fv,Ji:1996ek,Ji:1996nm,Radyushkin:1997ki}, formally defined as off-forward hadronic matrix elements of non-local quark and gluon operators along the light-cone. They can be viewed as Wigner phase-space distributions integrated over the parton transverse momentum, inherently correlating the spatial distribution of partons in the transverse plane with their longitudinal momentum. Accessible via hard exclusive processes, GPDs depend on three kinematic variables: the average longitudinal momentum fraction $x$ of the active parton, the longitudinal momentum transfer (skewness) $\xi$ representing half the longitudinal momentum fraction transferred to the target, and the squared total four-momentum transfer $t$.

At leading twist, the off-forward matrix element of the nucleon's quark-quark correlation operator is parametrized by eight independent GPDs for each quark flavor. The four chiral-even GPDs conserve parton helicity and include the unpolarized $H(x,\xi,t)$ and $E(x,\xi,t)$, as well as the helicity-dependent $\tilde{H}(x,\xi,t)$ and $\tilde{E}(x,\xi,t)$. Physically, $H$ and $\tilde{H}$ conserve the nucleon helicity, whereas $E$ and $\tilde{E}$ are associated with a nucleon helicity flip. The other four are the chiral-odd (transversity) GPDs, $H_T$, $E_T$, $\tilde{H}_T$, and $\tilde{E}_T$, which involve a flip of the active parton's helicity. The kinematics of GPDs naturally separate their domain into distinct physical regimes. In the Dokshitzer-Gribov-Lipatov-Altarelli-Parisi (DGLAP) regions ($|x| > \xi$), GPDs describe the emission and reabsorption of a quark (or antiquark) with different momentum fractions, generalizing standard PDFs. In contrast, within the Efremov-Radyushkin-Brodsky-Lepage (ERBL) region ($|x| < \xi$), GPDs describe the emission of a quark-antiquark pair from the nucleon, sharing phenomenological features with meson distribution amplitudes.  The $Q^2$ dependence of GPDs exhibits distinct behaviors in the two different regions.

GPDs exhibit extensive connections to various fundamental observables that characterize different aspects of nucleon structure. In the forward limit ($t \to 0$, $\xi \to 0$), $H$ and $\tilde{H}$ reduce to the unpolarized and polarized PDFs, respectively. Integrating over $x$ yields the elastic electromagnetic and axial form factors, e.g., $\int_{-1}^1 dx H^q(x,\xi,t) = F_1^q(t)$.   At $\xi = 0$, a two-dimensional Fourier transform with respect to the transverse momentum transfer $\Delta_\perp$ yields impact-parameter dependent distributions, denoting the probability density of finding a parton with momentum fraction $x$ at transverse distance $b_\perp$ from the transverse center of momentum~\cite{Burkardt:2000za,Burkardt:2002hr,Ralston:2001xs}.

Lorentz invariance dictates the polynomiality of GPDs: their $n$-th Mellin moments are even polynomials in $\xi$ up to degree $n+1$. The highest power of $\xi$ isolates the D-term, which determines the spatial distribution of mechanical forces, such as pressure and shear stress, inside the nucleon~\cite{Polyakov:2002yz}. These mechanical properties are encoded in the gravitational form factors (GFFs) derived from the second Mellin moments of the unpolarized GPDs.
\begin{figure}[htb]
\centering
\includegraphics[width=0.75\textwidth]{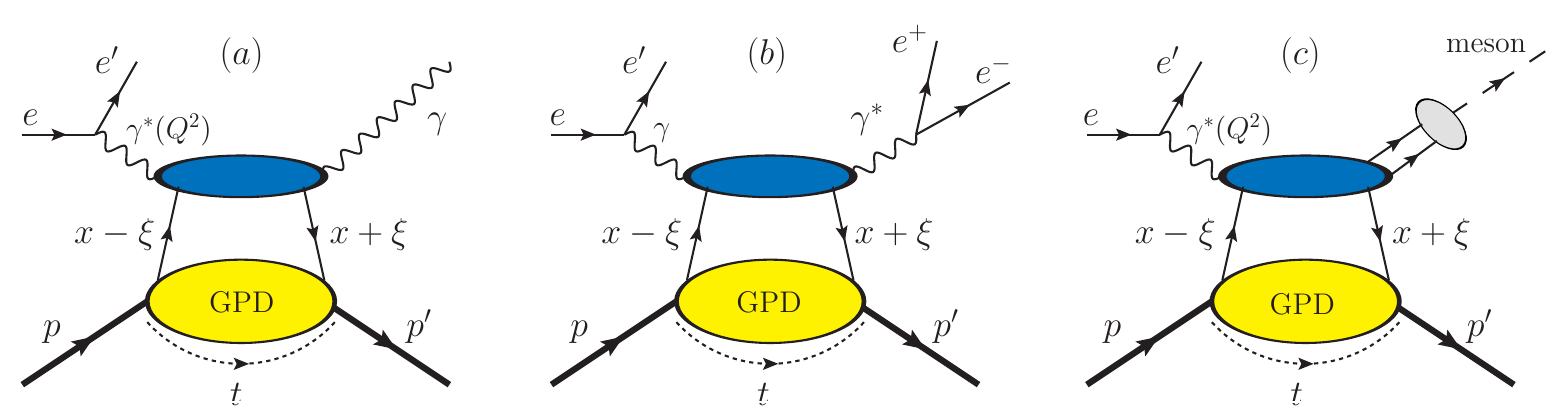} 
\caption{ DVCS process(left panel), TCS process(middle panel) and DVMP process(right panel).
 }
\label{DVCS}
\end{figure}

In electron-proton ($ep$) collisions, GPDs are primarily accessed through exclusive processes where the target nucleon remains intact. The ``golden channel'' is deeply virtual Compton scattering (DVCS, $ep \to e'p'\gamma$), characterized by a clean final state and well-understood higher-order QCD corrections. Complementary insights are gained from timelike Compton scattering (TCS, $\gamma p \to p' e^+ e^-$)~\cite{Berger:2001xd} and deeply virtual meson production (DVMP, $ep \to e'p'M$), where the production of specific vector (e.g., $\rho, \omega, J/\psi$) or pseudoscalar (e.g., $\pi, \eta$) mesons acts as a quantum number filter to separate different quark flavors and gluon contributions.  These processes are depicted in Fig.~\ref{DVCS}. Beyond TCS, DVMP and DVCS, double DVCS represents another important exclusive channel, where the virtualities of both photons can be varied independently, enabling a more complete mapping of GPDs in the $x-\xi$ plane~\cite{Deja:2023ahc}.

Previous global analyses performed within the PARTONS framework~\cite{Berthou:2015oaw} have extracted DVCS Compton Form Factors from world proton data using either physically constrained GPD-inspired parameterizations of border and skewness functions~\cite{Moutarde:2018kwr} or neural-network representations designed to reduce model bias~\cite{Moutarde:2019tqa}. 

In DVCS, GPDs do not represent the physical observable directly; instead, they enter the scattering amplitude through convolution integrals known as Compton form factors (CFFs). For instance, the CFF associated with the unpolarized GPD $H$ is given by
\begin{equation}
\mathcal{H}(\xi,t) = \int_{-1}^1 dx H(x,\xi,t) \left[ \frac{1}{x-\xi+i\epsilon} + \frac{1}{x+\xi-i\epsilon} \right].
\label{eq:CFF}
\end{equation}
Due to the propagator structure, the imaginary part of the hard-scattering amplitude at leading order is proportional to a $\delta$-function, $\delta(x \pm \xi)$. Consequently, the imaginary part of the CFFs probes the GPDs directly at the boundary points $x = \pm \xi$. Conversely, the real part involves a principal value integral over the entire $x$ domain and is uniquely sensitive to the D-term. Because high-energy scattering amplitudes are often dominated by their imaginary parts, many global fits of GPDs have historically been performed at leading order, capitalizing on this direct relationship. However, notable exceptions include the classic Kumeri\v{c}ki--M\"uller model~\cite{Kumericki:2009uq} and the recent global extraction incorporating lattice-QCD inputs~\cite{Guo:2025muf}, both of which are formulated at next-to-leading order accuracy.

Moreover, this structure implies that extracting the complete $x$-dependence of GPDs poses a severe mathematical challenge known as the ``de-convolution problem''~\cite{Bertone:2021yyz}. Specifically, there exists an infinite class of ``shadow GPDs''~\cite{Bertone:2021yyz,Moffat:2023svr} that integrate to zero in the real part and vanish at $x=\pm\xi$, completely decoupling from the CFFs at leading order. Overcoming this limitation requires advancing beyond leading order extractions, measuring observables across a wide kinematic phase space, analyzing multi-channel processes simultaneously. Additionally, neural networks offer a promising framework for solving this deconvolution problem, as discussed further in Sec.~\ref{AI-inverse-problem}.

Resolving the proton spin puzzle is a primary motivation for studying GPDs. The spin of the nucleon arises from the complex interplay of the intrinsic spin and OAM of its constituent quarks and gluons. Two principal frameworks are widely used to describe this decomposition: the Jaffe-Manohar decomposition~\cite{Jaffe:1989jz} and the Ji decomposition~\cite{Ji:1996ek}. The Jaffe-Manohar sum rule, derived from the canonical energy-momentum tensor often formulated in the light-cone gauge, expresses the proton spin as $\frac{1}{2} = \frac{1}{2}\Delta\Sigma + \Delta G + \mathcal{L}_q + \mathcal{L}_g$. This cleanly separates both quark and gluon contributions into intrinsic helicities ($\Delta\Sigma, \Delta G$) and canonical OAMs ($\mathcal{L}_q, \mathcal{L}_g$), though the individual canonical OAM terms are inherently gauge-dependent.
In contrast, the Ji decomposition relies on the Belinfante-Rosenfeld symmetric energy-momentum tensor, yielding terms that are manifestly gauge-invariant. It partitions the total nucleon spin into the quark helicity ($\frac{1}{2}\Delta\Sigma$), the quark kinetic (or mechanical) OAM ($L_q$), and the total gluon angular momentum ($J_g$). Crucially, Ji established a rigorous connection between the parton total angular momentum $J_{q,g}$ and GPDs via the second Mellin moments at the forward limit:
\begin{equation}
J_{q,g}=\frac{1}{2}\int_{-1}^{1}dx x \left[ H_{q,g}(x,\xi,0)+E_{q,g}(x,\xi,0) \right].
\label{eq:quark_angular_mom}
\end{equation}
By independently extracting the quark helicity $\Delta\Sigma$ from polarized inclusive or semi-inclusive scattering, one can isolate the elusive quark kinetic OAM through the relation $L_q = J_q - \frac{1}{2}\Delta\Sigma$. This unique capability to directly access $J_q$ establishes GPD measurements as an indispensable tool for achieving a complete, quantitative understanding of the proton spin structure.

First-principles lattice QCD calculations provide vital theoretical foundations complementary to experimental extractions. Breakthroughs using the large-momentum effective theory (LaMET) enabled the first computation of the nucleon helicity GPD $\tilde{H}(x, \xi, t)$ at the physical pion mass~\cite{Lin:2021brq}. By Fourier transforming the $\xi=0$ GPDs, lattice QCD delivers continuous determinations of the impact-parameter dependent polarized distribution $\Delta q(x,b)$. These constraints, especially in regions inaccessible to experiment (e.g., exactly $\xi=0$), are indispensable for regularizing global phenomenological fits of EicC data.

\subsubsection{EicC's unique advantages}
Once again, for the GPD measurements, the EicC kinematic coverage bridges the gap between JLab and the US-EIC, focusing on the sea quark regime. This intermediate range offers distinct advantages for GPD extraction through both DVCS and exclusive meson production.

For DVCS, the higher $Q^2$ reach at EicC compared to JLab suppresses higher-twist contamination, placing GPD extractions on firmer theoretical footing. At intermediate center-of-mass energies, the DVCS and Bethe-Heitler amplitudes are comparable in magnitude, which enhances their interference and facilitates the extraction of the scattering amplitude phase. Additionally, the ratio of longitudinal to transverse virtual photon fluxes ($\epsilon$) varies across the accessible phase space, enabling direct separation of $\sigma_L$ and $\sigma_T$---a measurement impractical at higher-energy colliders where $\epsilon \approx 1$.

Exclusive pseudoscalar meson production at EicC provides unique access to chiral-odd GPDs. The $ep \to e'p'M$ amplitude for $M = \pi^0, \eta$ decomposes into twist-2 contributions from the helicity GPDs $\tilde{H}$ and $\tilde{E}$ (longitudinal photon) and twist-3 contributions from the four transversity GPDs $H_T$, $E_T$, $\tilde{H}_T$, $\tilde{E}_T$ (transverse photon)~\cite{Goloskokov:2009ia,Goloskokov:2011rd}. At EicC energies, the transverse cross section $\sigma_T$ substantially exceeds $\sigma_L$, so that the twist-3 transversity GPD convolutions ($\langle \bar{E}_T \rangle$ and $\langle H_T \rangle$, with $\bar{E}_T = 2\tilde{H}_T + E_T$) dominate $\pi^0$ and $\eta$ electroproduction~\cite{Goloskokov:2022mdn}. 

Handbag-model calculations predict that, for representative EicC kinematics with (W=8--16~GeV) and ($Q^{2}$=2--7~$\mathrm{GeV}^{2}$), the transverse cross section is typically about one order of magnitude larger than the longitudinal contribution,
$
\frac{\sigma_{T}}{\sigma_{L}}
\sim
\mathcal{O}(10),
$
although the precise ratio depends on (t) and on the adopted GPD parametrization~\cite{Goloskokov:2022mdn}. This transverse-photon dominance makes EicC a particularly favorable kinematic region for enhancing sensitivity to the chiral-odd GPDs, with reduced contamination from leading-twist chiral-even contributions.

At sufficiently large ($Q^2$), the transverse amplitude is expected to be power suppressed relative to the longitudinal amplitude. In the Bjorken limit, this implies
$
\frac{\sigma_{T}}{\sigma_{L}}
\sim
\frac{1}{Q^{2}}
$. Consequently, the high-($Q^{2}$) region accessible at the US EIC is expected to become increasingly sensitive to the longitudinal, chiral-even mechanism~\cite{Collins:1996fb}. The transition between transverse and longitudinal dominance, however, remains dependent on the kinematics and the underlying GPD parametrization.

\begin{figure}[htb]
\centering
\includegraphics[width=0.8\textwidth]{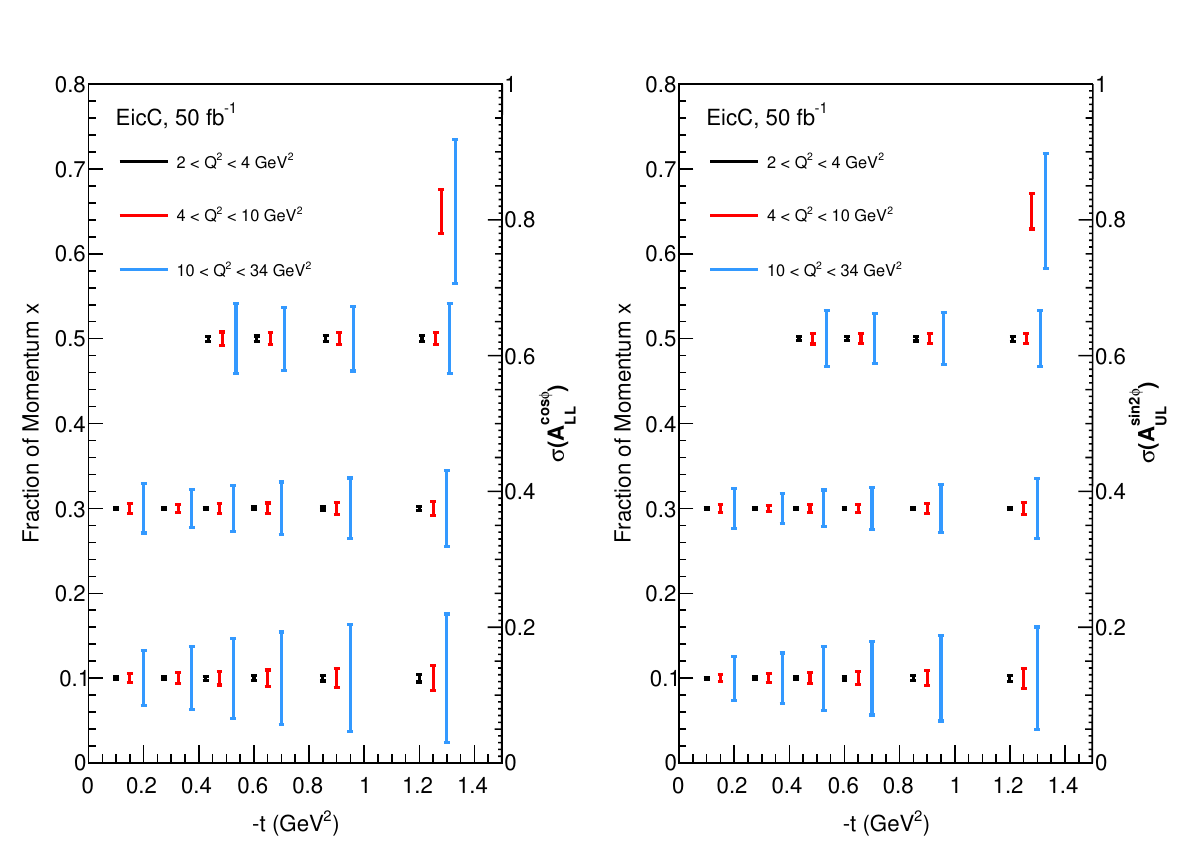} 
\caption{The statistics errors of the projected $A_{LL}^{\cos\phi}$ asymmetry (left) and $A_{UL}^{\sin 2\phi}$ asymmetry (right) for $\pi^{0}$ production in DVMP process at the EicC. The figure is adapted from Ref.~\cite{Anderle:2021wcy} (\href{https://creativecommons.org/licenses/by/4.0/}{CC BY 4.0}). 
}
\label{dvmp}
\end{figure}

This $\sigma_T$-dominated regime distinguishes EicC from higher-energy colliders where $\sigma_L$ prevails and chiral-even GPDs control the cross section, positioning EicC as the optimal facility for constraining chiral-odd GPDs. Measurements of target spin asymmetries---$A_{LL}^{\cos(\phi)}$, $A_{LT}^{\cos(\phi_s)}$, and $A_{LT}^{\cos(2\phi-\phi_s)}$---in both $\pi^0$ and $\eta$ channels enable robust $u$/$d$ flavor separation for transversity GPDs, exploiting the distinct isospin decompositions of the two meson channels~\cite{Goloskokov:2009ia,Goloskokov:2011rd,Goloskokov:2013mba,Goloskokov:2022rtb}. The projected $A_{LL}^{\cos(\phi)}$ and $A_{UL}^{\cos(\phi)}$ asymmetries at EicC are shown in Fig.~\ref{dvmp}. The combined extraction constrains the first Mellin moment of $H_T(x,0,0)$, which yields the tensor charge $\delta q$---a fundamental nucleon observable. With EicC’s high luminosity (up to $\sim 4 \times 10^{33}$ cm$^{-2}$ s$^{-1}$) and polarized beams, these asymmetries can be measured with percent-level statistical precision over $0.05 \lesssim x_B \lesssim 0.4$, providing stringent constraints on transversity GPD parametrizations complementary to the transversity PDF $h_1(x)$ extracted from SIDIS. In addition, the broad $W$ range accessible at EicC will permit detailed studies of the energy dependence of the transversity convolutions, further discriminating between GPD models and tightening constraints on the nucleon's transverse spin structure.

\begin{figure}[htb]
\centering
\includegraphics[width=0.75\textwidth]{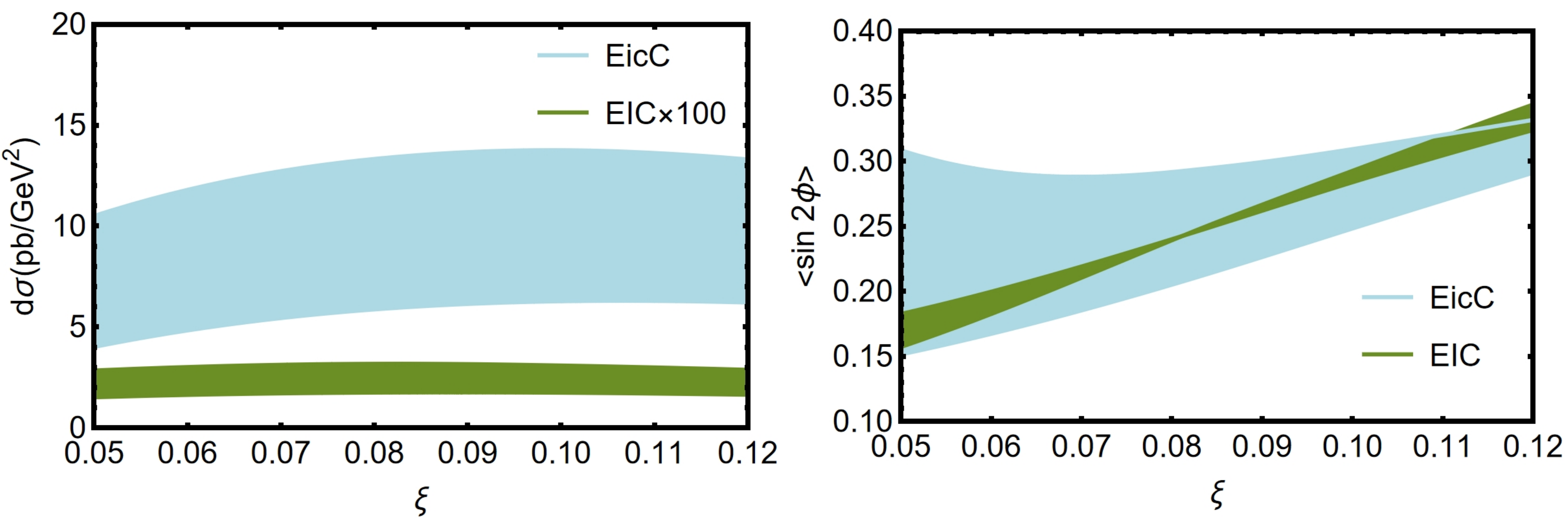} 
 \caption{(Left) The unpolarized cross section for US-EIC kinematics with $Q^2=10 \, \textrm{GeV}^2$ and $\sqrt{s_{ep}}=100 \, \textrm{GeV}$, as well as for EicC kinematics with $Q^2=3 \, \textrm{GeV}^2$ and $\sqrt{s_{ep}}=16 \, \textrm{GeV}$. The unpolarized cross section for the US-EIC case is re-scaled by a factor of 100. (Right) The average value of $\langle \sin(2\phi) \rangle$ in US-EIC and EicC kinematics. The variable $t$ is integrated over the range [$-0.5\, \textrm{GeV}^2$, $-\frac{4\xi^2 M^2}{1-\xi^2} $]. This figure is adapted from Ref.~\cite{Bhattacharya:2023hbq} (\href{https://creativecommons.org/licenses/by/4.0/}{CC BY 4.0}).
 }
\label{sin2phi}
\end{figure}

The $\pi^0$ channel further provides a direct probe of canonical quark OAM: the target longitudinal spin asymmetry in $ep \rightarrow e' p' \pi^0$ exhibits a $\sin 2\phi$ azimuthal modulation~\cite{Bhattacharya:2023hbq}. This observable is particularly powerful at the EicC, as shown in the left panel of Fig.~\ref{sin2phi}. With its moderate center-of-mass energy, the unpolarized cross section is substantially larger than at higher-energy facilities such as the US-EIC, yielding significantly higher event rates and enabling high-precision measurements of the asymmetry for a given $\zeta$. Moreover, as shown in the right panel of Fig.~\ref{sin2phi}, the asymmetry amplitude itself is enhanced at EicC kinematics, allowing clean extraction of the quark GTMD $F_{1,4}$ (and thus the canonical quark OAM)~\cite{Bhattacharya:2023hbq,Hatta:2021jcd,Hatta:2020bgy,Bhattacharya:2026qnd} directly in the DGLAP region ($\xi \sim 0.1$)---a regime complementary to the small-$x$ reach of higher-energy colliders~\cite{Bhattacharya:2023yvo}. Furthermore, both $F_{1,4}$ and $G_{1,1}$, which describe the strength of spin-orbit correlation of gluons, can be accessed in the vector meson production through various azimuthal dependent observables~\cite{Bhattacharya:2026qnd}. 

The EicC presents a remarkable opportunity to probe pion and kaon structures through Sullivan processes \cite{Chen:2020ijn, Anderle:2021wcy, Lu:2025bnm}. In these processes, the abundant ``meson cloud'' of a proton serves as a virtual target, while the baryon core (a neutron for the pion cloud, or a $\Lambda$ for the kaon cloud) acts as a spectator \cite{Sullivan:1971kd, Holtmann:1995qev}. Identifying these events requires the spectator to move into the far-forward region with minimal scattering angle, ensuring a smooth extrapolation to the on-shell meson structure ($-t \lesssim 0.6$ GeV$^2$ for pions, $-t \lesssim 0.9$ GeV$^2$ for kaons) \cite{Qin:2017lcd}.

Bridging the kinematic gap between JLab and US-EIC, EicC enables precision 3D meson tomography. Extracting pion and kaon form factors ($F_{\pi/K}$) via exclusive $ep \to en\pi^+$ and $ep \to e\Lambda K^+$ requires isolating the longitudinal cross section $\sigma_L$. At EicC's low-energy settings (e.g., 2.8 GeV $e$ on 12 GeV $p$), 
 an L-T separation may be feasible in principle using multiple beam-energy configurations. However, the limited $\epsilon$ lever arm, reduced luminosity at the lower-energy settings, and the amplification of statistical and systematic uncertainties require dedicated studies before the achievable precision and physics impact can be established.

\begin{figure}[t]
\centering
\includegraphics[width=.42\textwidth]{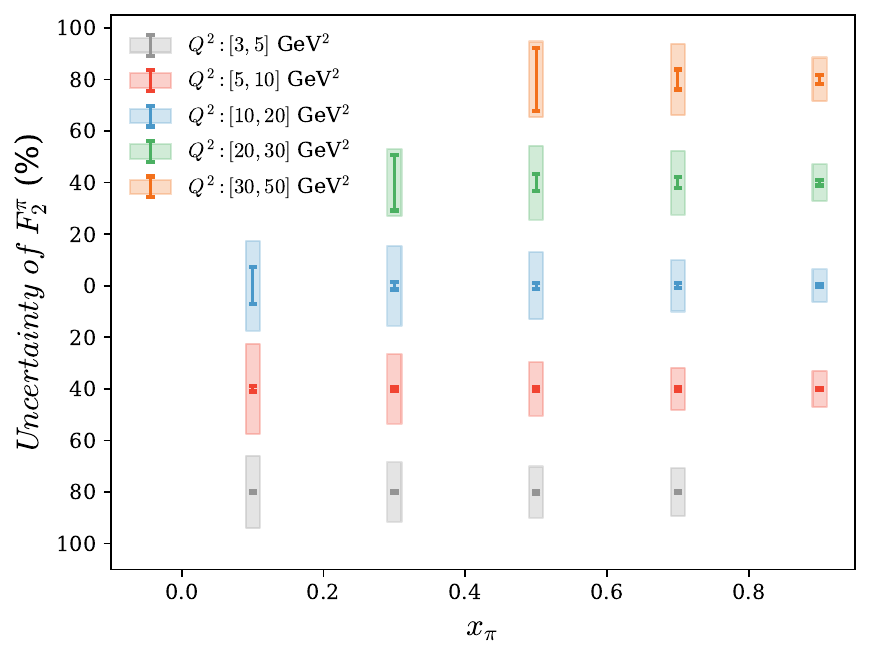}
\includegraphics[width=.42\textwidth]{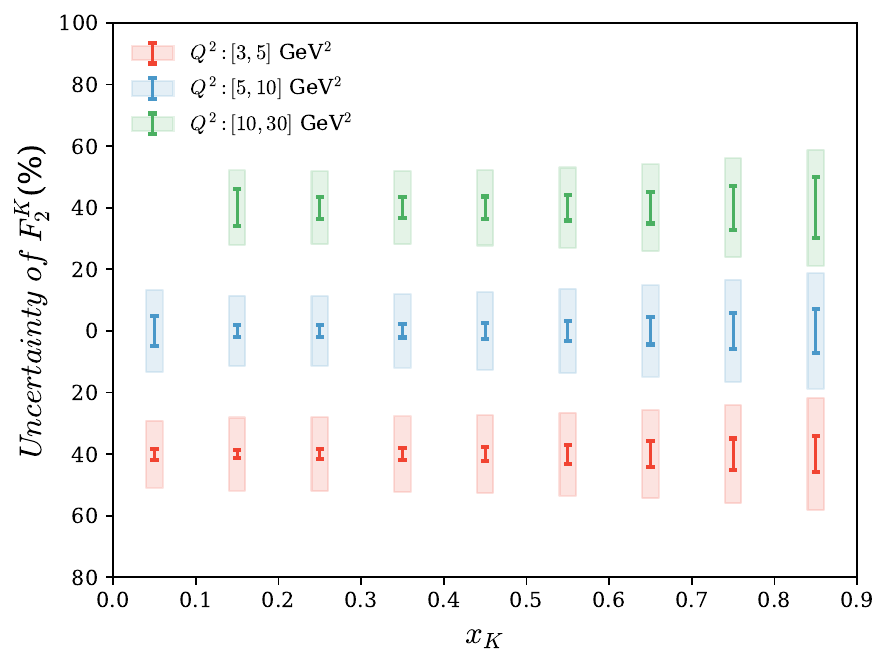}\hfill
\caption{The projected pion structure function $F_{2}^{\pi}$ (\textbf{left panel}), and the kaon structure function $F_{2}^{K}$ (\textbf{right panel}). Both cases assume an integrated luminosity of 50~fb$^{-1}$ and $3.5 \times 20$~GeV$^2$ energy setting. 
Error bars show the statistical uncertainty, while color boxes show the systematic uncertainties and do not include uncertainty from the flux factor. The figure is adapted from Ref.~\cite{Lu:2025bnm} (\href{https://creativecommons.org/licenses/by/4.0/}{CC BY 4.0}).
}
\label{pionstructure}
\end{figure}

Beyond form factors, the Sullivan process at EicC unlocks access to pion GPDs via Deeply Virtual Compton Scattering (DVCS) on a virtual pion target ($ep \to e'\gamma\pi^+n$). Unlike JLab, where events often fall in the $s$-channel resonance region, EicC's higher $\sqrt{s}$ safely places kinematics in the valence domain while preserving partonic interpretation \cite{Amrath:2008vx, Chavez:2021koz}. The projected pion structure function and the kaon structure function is shown in Fig.~\ref{pionstructure}. The interference between DVCS and the Bethe-Heitler (BH) process allows extraction of the pion GPD $H$. The golden observable is the beam-spin asymmetry (BSA):
\begin{equation}
\mathcal{A}(\varphi_{\text{Trento}}) = \frac{\sigma^{\uparrow}-\sigma^{\downarrow}}{\sigma^{\uparrow}+\sigma^{\downarrow}} \propto \text{Im}(\mathcal{H}_{\pi})\sin(\varphi_{\text{Trento}}).
\end{equation}
This asymmetry is exceptionally sensitive to the balance of quark and gluon contents in the pion, offering a direct probe into the QCD dynamics of a pseudo-Goldstone boson. EicC's high mid-rapidity acceptance yields an overall 27\% event acceptance for this fully exclusive channel, proving that pion 3D tomography is highly feasible. Furthermore, applying collinear factorization to the backward kinematic region allows the study of Transition Distribution Amplitudes (TDAs), free from significant BH background \cite{Lansberg:2006fv,Pire:2021hbl,Castro:2025rpx}. Ultimately, extracting these GPDs and TDAs will provide the first experimental constraints on the pion's Energy-Momentum Tensor (EMT), granting unprecedented insight into the pressure and shear forces holding this fundamental bound state together.

In summary, the combined kinematic coverage of EicC, JLab, and the US-EIC is essential for ultimately yielding complete 3D images of the proton from the large-$x$ valence region down to the low-$x$ saturation regime, and for a much more profound understanding of the proton spin puzzle.

\subsection{Nucleon energy energy correlation}
Energy-energy correlations(EEC) are among the cleanest QCD probes of energy flow. While in $e^{+}e^{-}$ annihilation the EEC is infrared and collinear safe and admits precise fixed--order and resummed predictions~\cite{Basham:1978bw}, the observable becomes significantly richer in deep inelastic scattering (DIS). Here, the incoming nucleon introduces a beam axis, a target remnant, and distinct current and target fragmentation regions. This DIS generalization retains the theoretical cleanliness of energy flow observables while converting the radiation pattern into valuable information about the parent nucleon. Consequently, nucleon energy--energy correlators (NEECs) are rapidly emerging as a novel class of event--shape observables for nucleon tomography. In this section, we briefly review how the classic EEC is generalized to DIS and related hadronic measurements, with a particular emphasis on forward nucleon energy correlators.

A central theoretical feature governing these generalizations is scale separation. In the short-distance regime, where $\theta Q \gg \Lambda_{\rm QCD}$ with $\theta$ being the angle between the two final-state hadrons in the energy-energy correlator, the correlator is controlled by perturbative splittings and can be described by fixed--order pQCD~\cite{Dixon:2019uzg}. However, in singular angular regions, logarithms of the angular variable and of the ratios between hard, collinear, and soft scales become large, necessitating resummation via renormalization--group (RG) methods. In the context of DIS, this intricate factorized structure must be matched onto both beam and target dynamics. As a result, the observable seamlessly connects hard scattering, perturbative branching, and the onset of confinement within a single unified framework.

A major step in realizing this unified framework was the formal definition of nucleon energy correlators (NECs) in DIS~\cite{Liu:2022wop}. Focusing on the target fragmentation region, NECs measure the angular distribution of forward energy in the beam remnant, allowing one to probe initial state structure without relying on fragmentation functions or jet algorithms. Crucially, they are predicted to interpolate smoothly between a perturbative Bjorken scaling regime ($\theta Q \gg \Lambda_{\rm QCD}$) and a non-perturbative confinement regime ($\theta Q \sim \Lambda_{\rm QCD}$)~\cite{Liu:2022wop}. Because the definition of the observable remains consistent across these two domains, NECs provide a direct, unambiguous way to track exactly where perturbative radiation gives way to confinement--dominated energy flow.

NEEC has now been rigorously tied to the operator description of target fragmentation. Through energy sum rules, a one-to-one mapping has been established between NECs and fracture functions, thereby organizing the broader class of DIS energy-flow observables in the target sector within the fracture-function formalism~\cite{Chen:2024bpj}. This relationship is especially important beyond leading twist, where spin and azimuthal asymmetries can be directly expressed in terms of NECs with a transparent conventional parton operator interpretation. Furthermore, semi-inclusive energy correlators (SIECs) have been introduced to correlate the target and current fragmentation regions, generating transverse-momentum moments in the process~\cite{Cao:2023oef}. These SIECs provide a unified language for extracting transverse--momentum--dependent (TMD) information from both the incoming nucleon and the fragmenting jet, all without requiring identified final states or a strictly back-to-back limit. In this sense, nucleon EECs complement standard tomographic techniques by revealing how intrinsic transverse motion and collinear radiation are dynamically redistributed over angle.

Incorporating spin dependence further broadens the potential of NEEC in exploring nucleon structure. Azimuthal--angle dependent EECs in SIDIS probe Collins--type EEC jet functions and hence transversity and related spin--orbit correlations \cite{Kang:2023gvg}. For longitudinal polarization, spin--resolved correlators provide a differential handle on helicity structure and its interplay with transverse motion, with joint N$^3$LL/NNLL predictions now available in both current and target fragmentation regions \cite{Gao:2025cwy}. Energy correlators therefore add a powerful event-shape layer to the usual TMD and GPD description of nucleon structure.

Moving to the small--$x$ regime, transverse EECs (TEECs) offer a distinct approach to spin observables specifically within the current fragmentation region. For the approximately back-to-back electroproduction of a hadron-electron pair, the TEEC is sensitive to the small-$x$ quark Sivers function and, in the CGC language, to a C-odd interaction associated with odderon exchange \cite{Bhattacharya:2025bqa}. The observable is constructed by summing over positively and negatively charged hadrons separately, with a predicted asymmetry at the $0.1\%$ level \cite{Bhattacharya:2025bqa}. Because it focuses on the current region, this back-to-back TEEC channel serves as a distinct, complementary probe to forward NEC measurements. By contrast, exploring odderon exchange in the forward sector requires a different strategy. In the target fragmentation region, the fully inclusive T--odd energy pattern gives a vanishing single--spin asymmetry, whereas track--based or charge--weighted NECs built from charged hadrons produce a nonzero asymmetry with opposite signs for positive and negative charges \cite{Mantysaari:2025mht}. Charge weighting then effectively provides a clean C--odd tag that suppresses contamination from C--even gluonic contributions. Looking beyond the odderon, these versatile energy correlators also open entirely new directions in gluon polarimetry and in light--ray operator product expansion(OPE) treatments of transverse spin \cite{Song:2025bdj,Chen:2021gdk}.

Although most work so far concerns $ep$ scattering, the same energy-correlator formalism can also probe CNM effects. For jet EECs in $eA$, multiple scattering and medium-induced radiation enhance the correlator at relatively large angles inside the jet cone \cite{Fu:2024pic}. Ratios of $eA$ to $ep$ measurements can therefore constrain the jet transport coefficient $\hat{q}$ and the path-length dependence of parton propagation in nuclei. Energy correlators are thus relevant not only for nucleon tomography but also for medium modification.

These developments make the EicC particularly relevant in the moderate-$x$ and spin-sensitive regime. Polarized beams and high luminosity support precise asymmetry measurements, while forward detectors and particle identification enable differential studies in $x$, $Q^2$, hadron charge, and angular variables. NECs require reliable reconstruction of the target fragmentation region, whereas SIECs and spin-dependent EECs also demand controlled acceptance in the current region. This reach is complementary to that of the higher-energy EIC: the US-EIC can push further into the small-$x$ and saturation domain, while the EicC is well placed to emphasize valence and sea-quark structure, flavor separation, and spin asymmetries in a cleaner kinematic window.

In summary, nucleon energy correlators turn energy flow into a controlled probe of hadron structure. Forward NECs, SIECs, and spin-dependent azimuthal or transverse EECs form a coherent set of observables connecting factorization, resummation, multiparton correlations, small-$x$ dynamics, and cold-nuclear-matter effects. With sufficient integrated luminosity, broad forward angular and energy acceptance, and polarized beams, the EicC could measure these observables differentially in $x$, $Q^2$, and the relevant correlation variables, thereby enabling quantitative comparisons with theoretical predictions and constraints on the underlying spin, small-$x$, and nuclear effects. Dedicated detector-level simulations will be required to determine the achievable precision, which is the beyond the scope this review.

	\newpage

\section{The origin of proton mass}\label{sec:mass}

Understanding the origin of proton mass is a fundamental objective in hadronic physics. The current light quark masses, generated via the Brout-Englert-Higgs mechanism \cite{Englert:1964et,Higgs:1964pj}, account for less than 1\% of the proton's total mass ($M_p \approx 938$ MeV). Since the sum of the valence quark masses ($2m_u + m_d \approx 9$ MeV) is negligible, the bulk of the proton's mass must be generated dynamically through the binding energy of the strong interaction.

This discrepancy is rooted in the classical properties of the theory. At the classical level, in the massless limit, the proton mass vanishes identically due to an exact cancellation between the quarks' kinetic energy and their negative potential energy. This is a consequence of the relativistic virial theorem~\cite{Brack:1983ht}, which, in a field theory formulation, implies that the trace of the QCD energy-momentum tensor (EMT) vanishes in the chiral limit. Consequently, the physical proton mass originates from quantum effects—specifically the trace anomaly of the energy-momentum tensor~\citep{Adler:1976zt,Nielsen:1977sy,Collins:1976yq}. The anomaly arises from the violation of conformal symmetry in the quantum theory~\cite{Collins:1976yq}, providing a mechanism through which a scale-invariant classical Lagrangian results in a massive bound state. This phenomenon is inextricably linked to the Yang-Mills mass gap problem, the confinement of quarks and gluons, and the dynamical generation of hadron masses.

As low-energy QCD is non-perturbative, analytical solutions for the proton mass remain a challenge. To gain insight into its internal structure, the mass is often decomposed into distinct physical components based on the QCD EMT~\citep{Ji:1994av,Ji:1995sv}. This has led to significant theoretical interest regarding the formal definitions of mass decomposition~\citep{Rodini:2020pis} and the renormalization properties of the individual EMT terms~\citep{Rodini:2020pis,Hatta:2018ina,Chen:2025iul,Ahmed:2022adh}. Ultimately, these studies seek to explain how a scale-invariant classical Lagrangian can result in a physical mass scale, a question that lies at the heart of one of the most significant challenges in mathematical physics:  The Yang-Mills mass gap problem, one of the seven Millennium Prize Problems identified by the Clay Mathematics Institute~\cite{JaffeWitten2000}, asks a deceptively simple question: does a quantum Yang-Mills theory necessarily exhibit a mass gap? That is, must there exist some minimal energy $\Delta > 0$ required to create any excitation above the vacuum state? Despite the apparent simplicity of the classical Yang-Mills Lagrangian with massless gluons, the quantum theory exhibits a dynamically generated mass scale $\Lambda_{\text{QCD}} \approx 200$ MeV. This emergent scale governs the spectrum of physical excitations and leads to the formation of bound hadronic states. Understanding how this mass gap arises from the massless gluon Lagrangian, and how it manifests in the rich spectrum of hadronic bound states, remains one of the central challenges in theoretical physics and mathematics.

This dynamical generation of the mass gap is inextricably linked to color confinement, ensuring both the short-range nature of the strong nuclear force and the permanent binding of quarks and gluons into color-neutral hadrons. Governed by these non-perturbative QCD dynamics, the proton manifests not as a simple non-relativistic bound state, but as a highly relativistic, strongly coupled quantum many-body system. Deep inelastic scattering demonstrates that valence quarks carry only a fraction ($\langle x \rangle_{\text{valence}} \approx 0.4$) of the proton's total momentum, with the remainder distributed among fluctuating sea quarks and gluons interacting at characteristic scales of $\Lambda_{\text{QCD}} \sim 200$ MeV. Consequently, the proton's mass, spin, and electromagnetic form factors are fundamentally emergent properties arising from the collective behavior of these confined degrees of freedom. Decoding how this complex internal structure stems directly from the foundational phenomena of confinement and the Yang-Mills mass gap remains a central challenge in hadron physics.

Presumably, the future EICs will provide experimental insights into how the mass gap and confinement manifest in the structure of hadrons. For instance, precision measurements of the proton's gravitational form factors via DVCS and DVMP at the EicC will directly probe the pressure distribution and mechanical properties of the proton, offering a window into the dynamical origin of confinement. Furthermore, measurements of quarkonium production near threshold may provide constraints on the trace anomaly contribution to the proton mass, shedding light on the role of gluon dynamics in mass generation. These measurements will not only test our theoretical understanding but may also suggest new approaches to the mathematical formulation of the problem. The interplay between rigorous mathematical theory, lattice QCD simulations, and precision experiments represents a promising path toward a complete understanding of mass generation in QCD. In the following subsections, we will develop the theoretical framework for decomposing the proton mass and describe how gravitational form factors provide experimental access to the individual contributions from quarks, gluons, and the trace anomaly.

\subsection{Proton mass decomposition}

Understanding the origin of the observed proton mass requires connecting fundamental QCD to measurable quantities. This is where Einstein's insight $m= E/c^2$ becomes crucial: the proton's inertial mass reflects its total energy content beyond its constituent quark masses, including the kinetic energies of quarks and gluons, and the energy stored in the gluon field configuration.

The challenge is to develop a framework that allows us to decompose this total energy into physically meaningful components that can be probed experimentally. This is precisely what Ji's mass decomposition achieves through the energy-momentum tensor formalism. A systematic theoretical framework for decomposing the nucleon mass was developed by Ji~\cite{Ji:1994av,Ji:1995sv,Ji:2021mtz} through analyzing the matrix elements of the QCD energy-momentum tensor $T^{\mu\nu}$. The energy-momentum tensor, which encodes the distribution of energy, momentum, and stress within a system, plays a central role in understanding the mechanical properties and mass structure of the nucleon.

In QCD, the energy-momentum tensor can be decomposed into quark and gluon contributions:
\begin{equation}
T^{\mu\nu} = T^{\mu\nu}_q + T^{\mu\nu}_g,
\end{equation}
where the quark contribution is given by
\begin{equation}
T^{\mu\nu}_q = \frac{i}{2}\sum_f \bar{q}_f \gamma^{(\mu}\overleftrightarrow{D}^{\nu)} q_f,
\end{equation}
with $D^\mu = \partial^\mu - igA^\mu$ being the covariant derivative, and the gluon contribution is
\begin{equation}
T^{\mu\nu}_g = F^{\mu\alpha}F^{\nu}_{~\alpha} - \frac{1}{4}g^{\mu\nu}F^{\alpha\beta}F_{\alpha\beta},
\end{equation}
where $F^{\mu\nu} = \partial^\mu A^\nu - \partial^\nu A^\mu + ig[A^\mu, A^\nu]$ is the gluon field strength tensor. The parentheses in $\gamma^{(\mu}\overleftrightarrow{D}^{\nu)}$ denote symmetrization, and the sum runs over all active quark flavors $f$.

The nucleon mass can be obtained by taking the matrix element of the trace of the energy-momentum tensor between nucleon states. For a nucleon at rest with four-momentum $P^\mu = (M_p, \vec{0})$, the mass sum rule reads:
\begin{equation}
M_p = \frac{\langle P | T^{00} | P \rangle}{\langle P |  P \rangle} .
\end{equation}
In a gauge-invariant and frame-independent decomposition, Ji showed that the proton mass can be partitioned into four distinct contributions~\cite{Ji:1995sv,Ji:2021mtz}:
\begin{equation}
M_p = M_m + M_q + M_g + M_a,
\label{eq:mass_decomp}
\end{equation}
where:
\begin{itemize}
\item $M_m = \sum_f \left(1+\frac{1}{4}\gamma_m  \right) m_f \langle P | \bar{q}_f q_f | P \rangle$, where $\gamma_m$ is the anomalous quark mass dimension, represents the contribution from the current quark masses;

\item $M_q$ is the contribution from quark kinetic and potential energies;

\item $M_g$ is the contribution from gluon field kinetic energy;

\item $M_a$ is the trace anomaly contribution arising from the breaking of scale invariance in QCD through quantum effects.
\end{itemize}

This decomposition reveals that the proton's mass is largely generated by the dynamics of quarks and gluons, rather than the Higgs mechanism, highlighting the central role that gluons play in the proton's mass structure.

The trace anomaly term, which has no classical analog, can be expressed as:
\begin{equation}
M_a = \langle P | \frac{1}{4}\frac{\beta(g)}{2g} F^{\alpha\beta}F_{\alpha\beta} | P \rangle,
\end{equation}
where $\beta(g) = -\frac{11N_c - 2N_f}{48\pi^2}g^3 + O(g^5)$ is the QCD beta function, with $N_c = 3$ being the number of colors and $N_f$ the number of active quark flavors. 

Currently, it is debated whether the decomposition in Eq.~(\ref{eq:mass_decomp}) is unique or whether it depends on the choice of renormalization scheme and the definition of the energy-momentum tensor. Alternative decompositions, such as those proposed by other authors \cite{Rodini:2020pis,Lorce:2017xzd,Hatta:2018sqd,Tanaka:2018nae,Metz:2020vxd}, may redistribute contributions differently among the terms.

\subsection{Gravitational form factors: experimental access to mass structure}

The experimental extraction of the mass components in Eq.~(\ref{eq:mass_decomp}) can be achieved through measurements of gravitational form factors (GFFs), which parametrize the matrix elements of the energy-momentum tensor between nucleon states with different momenta. The gravitational form factors of the proton provide crucial insights into its internal structure and mass distribution. These form factors, particularly $A(t)$, $J(t)$ and $D(t)$, encode information about the proton's mass and spin distributions, pressure, and shear forces, analogous to how electromagnetic form factors reveal the distribution of electric charge.

For a nucleon with initial momentum $P$ and final momentum $P'$, with momentum transfer $\Delta = P' - P$ and $t = \Delta^2$, the matrix elements of the symmetric energy-momentum tensor $T^{\mu\nu}$ can be parametrized in terms of four independent form factors. For the vector component ($\mu, \nu = 0, 1, 2, 3$), the decomposition reads~\cite{Ji:1996ek}:
\begin{equation}
    \langle \mathbf{p}', s' | T^{\mu\nu} | \mathbf{P}, s \rangle
= \frac{1}{M_p} \bar{u}(\mathbf{p}', s')
\Bigg[
P^{\mu} P^{\nu} \, A(t)
+ i P^{\{\mu} \sigma^{\nu\}\rho} \Delta_{\rho}\, J(t)
+ \frac{1}{4} \left( \Delta^{\mu} \Delta^{\nu} - g^{\mu\nu} \Delta^2 \right)\, D(t)
+M_p^2 g^{\mu\nu} \bar{C}(t)\Bigg] u(\mathbf{p}, s),
\end{equation}
where $a^{\{\mu} b^{\nu\}} = (a^{\mu} b^{\nu} + a^{\nu} b^{\mu})/2$, $u(p, s)$ and $\bar{u}(p', s')$ are nucleon spinors and spin eigenvalue $s = \pm \tfrac{1}{2}$, $P = (p + p')/2$, $\Delta = p' - p$, $t = \Delta^2$, and $\sigma^{\mu\nu} = \frac{i}{2}[\gamma^{\mu}, \gamma^{\nu}]$, where $\gamma^{\mu}$ are the Dirac matrices. The form factors $A(t)$, $J(t)$ ($A(t)+B(t)=2J(t)$), $D(t)$ (or equivalently $C(t)$ in other conventions), and $\bar{C}(t)$ encode different aspects of the nucleon's internal structure:
\begin{itemize}
\item $A(t)$ encodes the information about the mass/momentum distribution. It is related to the momentum fraction carried by partons and reduces at $t = 0$ to the momentum sum rule: $A(0) = \int_0^1 dx \, x[q(x) + \bar{q}(x) + g(x)] = 1$;

\item $J(t)$ ($B(t)$) carries the information about the spin distribution of the proton, and it is related to GPDs through the famous Ji sum rule;

\item $D(t)$ is the $D$-term form factor (also known as the $C(t)$ form factor), 
and it encodes how mechanical forces are distributed inside the proton. The condition $D(0) < 0$ is a requirement of 
mechanical stability: for the proton to exist as a bound state, the internal 
forces must be net confining, analogous to the condition that the inward 
gravitational pressure of a stable star must overcome the outward radiation 
pressure. A positive $D(0)$ would signal that the proton is mechanically 
unstable;

\item $\bar{C}(t)$ carries no independent dynamical information for the complete 
system: the total sum over all partons vanishes identically, $\bar{C}(t) = 0$ 
for all $t$, as a consequence of EMT conservation, $\partial_\mu T^{\mu\nu} = 0$, 
which requires 
$\Delta_\mu \langle p'| T^{\mu\nu}(0)|p\rangle = 0$. The individual quark and 
gluon EMTs are, however, not separately conserved (they exchange four-momentum 
through the QCD interaction), so the per-parton form factors $\bar{C}^q(t)$ 
and $\bar{C}^g(t)$ are individually nonzero and encode physical content related 
to the partial mechanical pressures and the QCD trace anomaly. Only their sum 
is constrained: $\bar{C}^q(t) + \bar{C}^g(t) = 0$. Each piece depends on the 
renormalization scale $\mu$, but their cancellation is scale-independent, as 
befits a consequence of an exact symmetry;

\item It is believed that quark and gluon energy contributions $M_q$ and $M_g$ can be constrained by PDFs through $A_q(0)$ and $A_g(0)$, $M_m$ can be accessed via $\pi$N low energy scatterings, and possibly probe $M_a$ through heavy quarkonia threshold productions. 
\end{itemize}

At zero momentum transfer ($t = 0$), these form factors reduce to moments of parton distributions and directly encode the mass contributions. In particular, the quark and gluon contributions to $A(0)$ give the momentum fractions:
\begin{equation}
A_q(0) = \int_0^1 dx \, x[q(x) + \bar{q}(x)], \quad A_g(0) = \int_0^1 dx \, x g(x),
\end{equation}
with $A_q(0) + A_g(0) = 1$. These momentum fractions are related to the mass contributions through the energy-momentum tensor matrix elements. $J(0)=1/2$ comes from the fact that the contributions of quarks and gluons to the spin of the nucleon add up to $1/2$. 

The GFFs can be separated into quark and gluon contributions: $A(t) = A_q(t) + A_g(t)$, and similarly for $B(t)$, $D(t)$ and $\bar{C}(t)$. The gluonic GFFs, $A_g(t)$, $B_g(t)$, $D_g(t)$, and $\bar{C}_g(t)$, are particularly challenging to access experimentally because gluons do not couple directly to electromagnetic probes. However, they can be probed through processes sensitive to gluon dynamics, such as deeply virtual Compton scattering (DVCS) at high energies and exclusive heavy quarkonium production near threshold~\cite{Strakovsky:2021vyk}. Recent experimental advancements have made it possible to probe these form factors through threshold $J/\psi$ photoproduction measurements.

\begin{figure}[t]
\centering
\includegraphics[width=.4\textwidth]{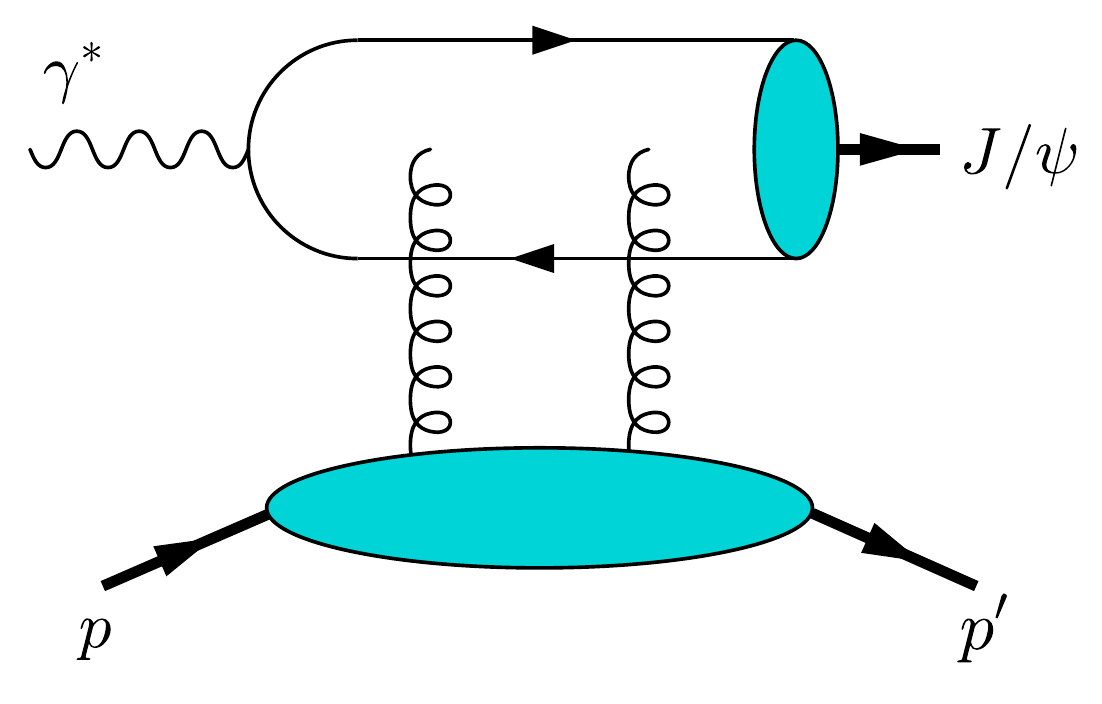}
\caption{The leading order Feynman diagram contributing to exclusive $J/\psi$ production.
}
\label{Jpsi}
\end{figure}
Near-threshold production of heavy quarkonia, particularly $J/\psi$ photoproduction, provides a clean and powerful probe of the gluonic GFFs. The process
\begin{equation}
\gamma + p \to J/\psi + p
\end{equation}
is dominated by gluon exchange at leading order in QCD as depicted in Fig.~\ref{Jpsi}, making it sensitive to the gluon content of the proton. Near the production threshold ($E_\gamma \approx 8.2$ GeV in the laboratory frame), the momentum transfer $t$ is relatively large ($|t| \sim 1$ GeV$^2$), and the skewness parameter $\xi$ (which measures the longitudinal momentum asymmetry between the initial and final states) approaches unity in the heavy quark limit.

In the GPD framework, the $J/\psi$ photoproduction amplitude can be factorized into a hard perturbative part (the photon-gluon fusion subprocess $\gamma + g \to J/\psi$) and a soft non-perturbative part encoded in gluon GPDs. The differential cross section can be written as~\cite{Guo:2023qgu}:
\begin{equation}
\frac{d\sigma}{dt} = \frac{\alpha_{em} e_c^2}{4(W^2 - M_p^2)^2} \frac{(16\pi\alpha_s)^2}{3M_{J/\psi}^3} |\psi_{NR}(0)|^2 |G(t, \xi)|^2,
\end{equation}
where $W$ is the center-of-mass energy, $e_c = 2/3$ is the charm quark charge, $M_{J/\psi}$ is the $J/\psi$ mass, $\psi_{NR}(0)$ is the $J/\psi$ wave function at the origin in non-relativistic QCD, and $G(t, \xi)$ is the hadronic matrix element related to gluon GPDs.

The key feature of near-threshold kinematics is that $\xi \to 1$, which simplifies the GPD structure. In this limit, the amplitude is dominated by the lowest moments of gluon GPDs, which are directly related to the gluonic GFFs. This leading-moment dominance allows extraction of gluonic GFFs through measurements of $J/\psi$ photoproduction cross sections as a function of $t$ near threshold.

Significant experimental progress has been made in recent years in measuring $J/\psi$ photoproduction near threshold, providing the first constraints on gluonic GFFs.
Concurrently, theoretical advances have refined our understanding of how to interpret these observations.

\textbf{Jefferson lab measurements:} A recent interesting measurement conducted by Duran et al.~\cite{Duran:2022xag} utilized a small color dipole via threshold $J/\psi$ photoproduction to probe the gluonic GFFs. This pioneering experiment determined, for the first time, the proton's gluonic mass radius, revealing that it is notably smaller than the electric charge radius ($r_{g} = 0.52 \pm 0.03$ fm compared to the charge radius $r_E = 0.84$ fm). This finding suggests the presence of a dense gluon core at the center of the proton, where gluons dominate the mass distribution. The experimental setup involved measuring the differential cross sections of $J/\psi$ photoproduction near the kinematic threshold, providing access to the gluonic gravitational form factors $A_g(t)$ and $C_g(t)$. The analysis revealed that the extracted mass radius from the gluonic GFFs is significantly smaller than the charge radius, suggesting a compact region of gluonic energy density within the proton.

The GlueX collaboration at Jefferson Lab has made significant contributions to this field through their comprehensive measurements of $J/\psi$ photoproduction cross sections. In their recent study~\cite{GlueX:2023pev}, they reported both total and differential cross sections over the full kinematic range of momentum transfer for photon energies from the threshold at 8.2 GeV up to 11.44 GeV. These measurements revealed possible structures in the total cross section around 9.1 GeV, potentially indicating contributions from open-charm intermediate states or other resonance effects. The precise near-threshold measurements provide crucial input for theoretical models studying the gluon structure of the proton, including gluon GPDs, proton mass radius, and trace anomaly contributions.

Most recently, Pybus et al.~\cite{Pybus:2024ifi} reported the first measurement of $J/\psi$ photoproduction from nuclei ($^2$H, $^4$He, $^{12}$C) near and below the threshold energy of 8.2 GeV. Using Jefferson Lab's GlueX detector, researchers observed $J/\psi$ production in the subthreshold region for the first time, with cross sections suggesting possible enhancement compared to theoretical predictions. The data allowed extraction of a gluonic radius for bound protons of $0.85 \pm 0.14$ fm, providing the first glimpse into how the gluonic structure of nucleons is modified in the nuclear environment, and hints at modified gluon structure in deeply bound nucleons.

Despite these experimental advancements, current datasets are insufficient to fully constrain the gluonic GFFs. Guo et al.~\cite{Guo:2023qgu} studied the exclusive near-threshold photoproduction of heavy quarkonium in the framework of GPD factorization. They found that near-threshold $J/\psi$ production is related to gluon GPDs at large skewness $\xi$, which is distinct from common high-energy kinematics where skewness is typically small. They proposed methods to extract the first few moments of gluon GPDs from these amplitudes, with the leading moments corresponding to the gluonic GFFs. However, their analysis of recent $J/\psi$ production measurements revealed that while the $\xi$-scaling of the measured differential cross sections is consistent with asymptotic behavior predicted by GPD factorization, current data are not yet accurate enough for a complete and model-independent determination of all gluonic GFFs.

\textbf{Weizs\"acker-Williams method:} In parallel with experimental efforts, theoretical developments have provided complementary approaches to studying gluonic GFFs. Ref.~\cite{Hagiwara:2024wqz} proposed using the Weizs\"acker-Williams (WW) method to compute gluonic GFFs in the high-energy limit. This method, which treats the fast-moving hadron as a source of a quasi-real gluon field, establishes a novel relationship between the gluonic momentum (A-type) GFF and the Laplacian of the dipole scattering amplitude in the small-$x$ framework:
\begin{equation}
A_g(t) = \frac{2N_c}{\alpha_s} \int_0^1 dx\int \frac{d^2\mathbf{b}_\perp}{4\pi^2} \, e^{-i\boldsymbol{\Delta}_\perp \cdot \mathbf{b}_\perp} \nabla_{\mathbf{r}_\perp}^2 \mathcal{N}_x(\mathbf{r}_\perp,\mathbf{b}_\perp )|_{\mathbf{r}_\perp =0},
\end{equation}
where $\mathcal{N}_x(\mathbf{r}_\perp,\mathbf{b}_\perp )$ is the dipole scattering amplitude, $\boldsymbol{\Delta}_\perp$ is the transverse momentum transfer, and $\nabla_{\mathbf{r}_\perp}^2$ is the transverse Laplacian operator. This formulation enables the extraction of gluonic GFFs through dipole scattering measurements and provides direct access to the gluon mean square radius $\langle r^2 \rangle_g = -6 \frac{dA_g(t)}{dt}\big|_{t=0}$, characterizing the spatial distribution of gluon energy-momentum inside hadrons.

Applying this framework to HERA data on diffractive vector meson production, Ref.~\cite{Hagiwara:2024wqz} extracted a gluon transverse radius of $\sqrt{\langle b^2_\perp \rangle_g} = 0.56 \pm 0.03$ fm for the proton, corresponding to $\sqrt{\langle r^2\rangle_g} = 0.61$ fm. This result is notably smaller than the recent determination from near-threshold $J/\psi$ photoproduction at Jefferson Lab off nuclei~\cite{Pybus:2024ifi}, which reported $\sqrt{\langle r^2 \rangle_g} = 0.85 \pm 0.14$ fm. The smaller radius from the WW method is more consistent with the $J/\psi$-007 analysis~\cite{Duran:2022xag}, which extracted $\sqrt{\langle r^2_m \rangle_g} = 0.52 \pm 0.03$ fm (dipole fit) or $0.755 \pm 0.035$ fm (holographic QCD, in agreement with lattice QCD~\cite{Pefkou:2021fni}), both notably smaller than the proton charge radius. The apparent discrepancies, ranging from approximately $0.6$ to $0.85$ fm, arise from differences in kinematic regimes (high-energy diffractive versus near-threshold), theoretical frameworks (dipole formalism versus GPD/holographic approaches), model dependence in extracting gravitational form factors, and the role of open-charm contributions near threshold. Future EIC measurements covering a broad kinematic range will be crucial for resolving these tensions and establishing a consistent picture of the proton's gluonic structure.

\textbf{Bayesian analysis and constraints on GFFs:} A breakthrough came with the recent work by Guo, Yuan, and Zhao~\cite{Guo:2025jiz}, who employed advanced Bayesian inference methods to extract nucleon gravitational form factors from near-threshold $J/\psi$ photoproduction data. This study represents a significant methodological advancement in the field by applying the GPD framework at next-to-leading order (NLO) in QCD and demonstrating stable expansion for near-threshold kinematics. The Bayesian approach provides a statistically rigorous and interpretable extraction that is particularly valuable since both quark and gluon GFFs enter at NLO and are strongly correlated in the analysis. (For a brief introduction to Bayesian inference, see Sec.~\ref{sec:ai_applications}.) The key findings from this Bayesian analysis include:
  Strong preference for negative values of $C_q(t)$ and $C_g(t)$, which is in excellent agreement with state-of-the-art lattice QCD simulations;
 Demonstration that the experimental constraints from the $J/\psi$-007 experiment (E12-16-007 experiment at JLab) and GlueX Collaboration are consistent with lattice predictions;
 Establishment of a robust framework for extracting both quark and gluon GFFs simultaneously at NLO, which was not possible in previous leading-order analyses;
 Validation of the conformal expansion approach for Compton form factors (CFFs) in the large $\xi$ kinematics, showing better convergence properties.

This work highlights the great potential to extract precise GFFs from future high-precision experiments at facilities like the US-EIC and EicC. The Bayesian inference methodology provides a pathway to combine experimental data with lattice QCD constraints, offering the most accurate picture of gluon GFFs to date.

\textbf{Perturbative QCD at large momentum transfer:} However, theoretical tensions remain. Sun et al.~\cite{Sun:2021pyw} applied perturbative QCD to investigate near-threshold heavy quarkonium photoproduction at large momentum transfer. Their analysis considered contributions from the leading three-quark Fock states of the nucleon and found that the dominant contribution comes from states with one unit of quark OAM, while contributions from zero quark OAM are suppressed at threshold. Importantly, they concluded that there is no direct connection between near-threshold heavy quarkonium photoproduction and the gluonic gravitational form factors of the nucleon. To extract or determine the gluon GFFs from the GPD formalism, additional assumptions and corrections must be made. As demonstrated in a recent analysis~\cite{Guo:2025jiz}, this requires expanding the amplitude at large skewness $\xi$. Crucially, this expansion breaks down at small $\xi$. While a direct extraction of GFFs from near-threshold quarkonium photoproduction may not be straightforward, a model-assisted indirect determination remains possible under well-defined kinematic approximations (large $\xi > 0.5$), provided the associated 
theoretical uncertainties are carefully accounted for. This highlights that while GFFs can be indirectly constrained under very specific kinematic approximations, the relationship is highly dependent on the  kinematic regime. In this regard, the EicC offers a distinct advantage over the EIC, as it is easier to access the large $\xi$ regime required for this expansion.

\textbf{Dispersive relations and emergent hadron mass:} An important progress has been made in the model-independent determination of GFFs through non-perturbative frameworks. Cao, et al.~\cite{Cao:2024zlf} utilized a dispersive analysis to provide a rigorous determination of nucleon GFFs by relating them to $\pi\pi \to N\bar{N}$ scattering amplitudes. Their results confirm the mechanical stability of the nucleon through a negative $D$-term and suggest that the mass radius of the nucleon is larger than its charge radius, providing a benchmark for upcoming EIC measurements.

First-principles lattice QCD calculations have made significant progress in computing the nucleon mass decomposition and GFFs directly from the fundamental theory. Recent lattice simulations~\cite{Yang:2018nqn,Alexandrou:2020sml} have provided estimates of the individual contributions to the proton mass:
\begin{itemize}
\item Quark mass contribution: $M_m \approx 90$ MeV (about 9\%);
\item Quark energy: $M_q \approx 330$ MeV (about 31\%);
\item Gluon energy: $M_g \approx 370$ MeV (about 37\%);
\item Trace anomaly: $M_a \approx 230$ MeV (about 23\%).
\end{itemize}

These lattice results indicate that gluons contribute roughly 37\% of the proton's mass through their kinetic energy ($M_g$), while the trace anomaly ($M_a$), which arises from quantum corrections and the breaking of scale invariance, contributes about 23\%. The quark energy ($M_q$), which includes both kinetic and potential energy contributions, accounts for about 31\%.

Lattice calculations have also computed the $t$-dependence of GFFs, providing predictions for the spatial distribution of mass and pressure inside the nucleon. These calculations predict that the pressure at the center of the proton is positive and extremely large (on the order of $10^{34}$ - $10^{35}$ Pa), comparable to the pressure inside neutron stars, while the pressure becomes negative at larger radii, providing the confining force that holds quarks together.

The agreement between lattice QCD predictions and experimental extractions from $J/\psi$ photoproduction, particularly for the sign and approximate magnitude of $C_g(t)$, provides strong validation of both approaches and demonstrates the power of combining complementary theoretical and experimental methods.

\subsection{Pion gravitational form factors via the Sullivan process}

While most efforts have focused on nucleon gravitational form factors, the pion GFFs are of special interest due to the pion's unique nature as the Nambu-Goldstone boson of spontaneously broken chiral symmetry(see a review on this topic~\cite{Ding:2022ows}). The soft pion theorem imposes stringent constraints on pion GFFs and GPDs~\cite{Polyakov:1998ze,Chavez:2021llq}.

Hatta and Schoenleber~\cite{Hatta:2025ryj} proposed a novel method to access pion GFFs through near-threshold vector meson production in the Sullivan process. The Sullivan process~\cite{Sullivan:1971kd} refers to electron-proton scattering where the proton transitions to a neutron by emitting a virtual pion, $p(p) \to \pi^+(p_\pi) + n(p_n)$ with virtuality $p_\pi^2 = t < 0$, which then interacts with the virtual photon from the electron. This process has been instrumental in extracting the pion electromagnetic form factor~\cite{Horn:2006tm,Huber:2008id,Qin:2017lcd}.

The key innovation is to study exclusive vector meson production $\gamma^*(q) + \pi^{+*}(p_\pi) \to V(q') + \pi^+(p'_\pi)$ in the threshold region $s_\pi \gtrsim (m_\pi + m_V)^2$, focusing on $\rho$, $\omega$, and $\phi$ electroproduction (where $Q^2 \gg \Lambda_{\text{QCD}}^2$ provides the hard scale) and $J/\psi$ photoproduction (where the heavy quark mass serves as the hard scale). In this region, the skewness variable becomes order unity,
\begin{equation}
\xi \approx \frac{x_B}{2x_\pi - x_B \left(1 - \frac{t_\pi}{Q^2}\right)} \sim \mathcal{O}(1),
\end{equation}
where $x_\pi = (p_\pi \cdot l)/(p \cdot l)$ is the momentum fraction of the proton carried by the pion, $x_B = Q^2/(2p_\pi \cdot q)$ is the Bjorken variable for the $\gamma^* + \pi^{+*}$ subprocess, and $t_\pi = (p_\pi - p'_\pi)^2$ is the pion momentum transfer. This large $\xi$ kinematics enables the threshold approximation~\cite{Hatta:2021can,Guo:2021ibg,Guo:2022upw,Guo:2023ahv} where the convolution integrals of GPDs,
\begin{equation}
\mathcal{H}^a(\xi; t; \mu^2) = \int_{-1}^1 dx \frac{1}{\xi - x - i\epsilon} 
\begin{cases}
\frac{1}{2} H^{q(+)}(x, \xi, t, \mu^2) \\
\frac{1}{x} H^g(x, \xi, t, \mu^2)
\end{cases},
\end{equation}
can be well approximated by their truncated versions containing only the gravitational form factors,
\begin{equation}
\mathcal{H}^a_{\text{trunc}}(\xi; t; \mu^2) = \frac{2}{\xi^2} \frac{5}{4} [A^a(t, \mu^2) + \xi^2 D^a(t, \mu^2)].
\end{equation}
A critical advantage is that the light pion mass allows $\xi$ to be tuned closer to unity than for nucleon targets, improving theoretical control.

In a comprehensive NLO analysis~\cite{Hatta:2025ryj} using the algebraic pion GPD model~\cite{Chavez:2021llq}, a remarkable finding shows that the threshold approximation exhibits dramatically different convergence for different parton flavors. For valence quarks ($u$, $d$), the approximation fails unless $\xi \gtrsim 0.7$ because the GPDs are well-supported in the DGLAP region $\xi < |x| < 1$. In contrast, for gluon and strange quarks, the approximation works excellently with truncation errors below 10\% for $\xi > 0.2$, since these GPDs are dominantly supported in the central region $|x| < \xi$. Consequently, $\phi$ and $J/\psi$ production provide clean access to gluon and strangeness GFFs.

Concrete NLO predictions are provided for the US-EIC and Jefferson Lab. The GFF parametrization adopts monopole and dipole forms,
\begin{equation}
A^a_\pi(t, \mu^2) = \frac{A^a_\pi(0, \mu^2)}{1 - t/m_A^2}, \quad D^a_\pi(t, \mu^2) = \frac{D^a_\pi(0, \mu^2)}{(1 - t/m_D^2)^2},
\end{equation}
with $m_A = 1.6$ GeV and $m_D = 1.1$ GeV. The soft pion theorem constraint $D^a_\pi(0, \mu^2) = -A^a_\pi(0, \mu^2)$ is imposed, though deviations would be experimentally interesting. For $\phi$ electroproduction at the US-EIC with $s_{ep} = 800$ GeV$^2$, the longitudinal cross section at $(x_\pi, x_B) = (0.3, 0.2)$ with $Q^2 = 5$ to 10 GeV$^2$ shows reduced scale uncertainties at NLO and is well within measurable range. For $J/\psi$ photoproduction at the proposed 22 GeV Jefferson Lab upgrade with $s_{\gamma p} = 40$ GeV$^2$, cross sections at $x_\pi = 0.45$ are comparable to nucleon production~\cite{Duran:2022xag,GlueX:2023pev}. The EicC, operating at a lower energy compared to the EIC, may allow one to better perform the longitudinal/transverse separation for the virtual photon. This work establishes the Sullivan process near threshold as a promising tool for constraining pion GFFs, particularly testing chiral symmetry predictions.

\subsection{Opportunities for mass studies at the EicC}

Future experimental facilities, such as US-EIC~\cite{Accardi:2012qut,AbdulKhalek:2021gbh} and EicC~\cite{CAO:2024fdz,Anderle:2021wcy}, will play crucial roles in measuring these mass contributions with greater precision.  With its specialized kinematic coverage and access to polarized proton, deuteron, and ${}^{3}\mathrm{He}$ beams, as well as unpolarized ion beams ranging from carbon to uranium~\cite{Anderle:2021wcy}, the EicC will offer unique opportunities to deepen our understanding of the nucleon mass structure through precision measurements of GFFs.

\textbf{Near-threshold heavy quarkonium production:} The EicC's energy range (center-of-mass energies of 15 to 20 GeV) is ideally suited for studying near-threshold $J/\psi$ production. The high luminosity of (2 to 4)$\times 10^{33}$ cm$^{-2}$s$^{-1}$ will enable precision measurements of differential cross sections as a function of $t$ with unprecedented statistical accuracy. These measurements will allow for:
\begin{itemize}
\item Constraints on the $t$-dependence of gluonic GFFs from exclusive near-threshold $J/\psi$ production~\cite{Duran:2022xag,Guo:2023qgu};
\item Determination of the gluonic mass radius and its comparison with the charge radius;
\item Mapping of the spatial distribution of gluon energy density inside the proton;
\item Constraints on the trace anomaly contribution to the proton mass.
\end{itemize}

\textbf{Flavor separation and quark contributions:} The EicC will enable measurements on different nuclear targets (protons, deuterons, $^3$He, and heavy nuclei), allowing for flavor separation of quark contributions to the mass. By comparing measurements on proton and neutron targets (using deuterium and $^3$He), the EicC can separately constrain the contributions of quarks of different flavors to GFFs. The quark energy ($M_q$) and gluon energy ($M_g$) components can be constrained through the second moment of the unpolarized PDFs, while the gluon trace anomaly contribution ($M_a$) may potentially be probed via color dipole interactions. 

\textbf{Nuclear modifications of mass structure:} Measurements across a broad range of ion species at the EicC will systematically map these modifications as a function of the nuclear mass number $A$ and provide crucial input for understanding CNM effects on the gluonic structure.
 The recent observation of subthreshold $J/\psi$ production in nuclei suggests possible enhancement effects that could arise from modified gluon distributions or collective nuclear effects. The EicC's comprehensive program on nuclear targets will systematically map these modifications as a function of nuclear mass number $A$ and provide crucial input for understanding CNM effects on the gluonic structure.

\textbf{Complementarity with DVCS measurements:} While $J/\psi$ production probes gluonic GFFs, deeply virtual Compton scattering (DVCS) provides access to quark GFFs. The EicC will perform precision DVCS measurements, enabling a complete separation of quark and gluon contributions to the nucleon mass. The combination of DVCS and exclusive meson production data will allow for a comprehensive determination of all GFFs and a complete experimental validation of Ji's mass decomposition.

	\newpage
	
\section{Quantum information at an electron-ion collider}

The intersection of quantum information theory (see 
Refs.~\cite{Benenti:2019non,Nielsen:2012yss,Horodecki:2009zz} 
and references therein for an introduction) with nuclear 
and high-energy physics phenomenology has emerged as a vibrant 
frontier~\cite{Barr:2024djo,Afik:2025ejh,Robin:2026lqp}, 
offering new perspectives on fundamental aspects of quantum 
mechanics, including entanglement, quantum complexity, and 
Bell nonlocality, at energy scales far beyond those of 
conventional atomic, molecular, and optical physics. Electron-ion colliders provide unique opportunities to study quantum entanglement and Bell nonlocality in the production of quark antiquark pairs, where the spin degrees of freedom of the produced particles serve as natural qubits. 

In particular, since 2022, the ATLAS and CMS collaborations have successfully measured quantum entanglement in top quark pair production at the LHC~\cite{ATLAS:2023fsd,CMS:2024pts}, demonstrating that quantum correlations survive at the highest accessible energy scales. In early 2026, the STAR collaboration reported the first observation of spin correlations between hyperon pairs at the Relativistic Heavy Ion Collider~\cite{STAR:2025njp}. Through proton-proton collisions at 200~GeV, the experiment measured spin correlations of strange quark-antiquark pairs during hadronization, finding an 18\% relative polarization signal for short-range hyperon pairs that vanishes when pairs are widely separated in angle, consistent with quantum decoherence. It is believed that hyperons carry spin information through the entire hadronization process, offering unique insights into how quantum correlations evolve during the transition from partonic to hadronic degrees of freedom and providing new experimental approaches to study QCD confinement and chiral symmetry breaking.

A self-contained introduction to the relevant concepts from quantum information theory is provided in Appendix~\ref{appendix}. In this section, we review recent theoretical developments that establish electron-ion colliders as promising platforms for quantum information studies. The EicC's kinematic coverage in the moderate-$x$ regime and its dedicated program of diffractive and exclusive measurements 
can provide distinctive contributions to the study of quantum entanglement and correlations in hadronic systems. For a comprehensive overview of related activities at the LHC and other facilities, see Ref.~\cite{Barr:2024djo,Afik:2025ejh} and references therein.

\subsection{Deep inelastic scattering as a probe of entanglement entropy}

Although the proton as a whole exists in a pure quantum state, its internal partonic structure is described by PDFs that represent classical probability distributions rather than quantum probability amplitudes. This classical nature arises because inclusive deep inelastic scattering (DIS) measurements detect only partons within a specific kinematic region while tracing over all other constituents. The hard momentum transfer in the probe effectively decoheres the quantum superposition of different partonic configurations. To illustrate this mechanism, consider the Bell state \(|\Psi^-\rangle = \frac{1}{\sqrt{2}}(\ket{\uparrow\downarrow} - \ket{\downarrow\uparrow})\): while the two-particle system is in a pure entangled state, measuring only one particle and tracing over the other yields a maximally mixed state with density matrix \(\rho = \frac{1}{2}\mathbb{I}_2\). The von Neumann entropy \(S = -\text{Tr}(\rho \ln \rho) = \ln 2\) quantifies the information loss from ignoring the unmeasured particle~\cite{Semenoff:2019dqe}. This entropy is maximal because both possible states \(\ket{\uparrow}\) and \(\ket{\downarrow}\) occur with equal probability \(1/2\), meaning we have complete ignorance about which state the measured particle is in—the quantum correlations with the unmeasured particle have been completely lost.
\begin{figure}[htb]
\begin{center}
    \begin{tikzpicture}

\fill[red!30] (0,0) circle (2cm);

\fill[blue!40] (0,0) circle (1cm);

\draw[thick, blue] (0,0) circle (1cm);
\draw[thick, red] (0,0) circle (2cm);

\node at (0.5,-0.4) {\textbf{A}};
\node at (-0.2,0) {$r=1/Q$};
\node at (1,1) {\textbf{B}};

\draw[->, thick, cyan, decorate, decoration={snake, amplitude=1.5mm, segment length=3mm}] 
    (-3,1) -- (0,0.3) node[at start, above, black] {$\gamma^\ast$ $(q^2=-Q^2)$};

\fill[cyan] (0,0.3) circle (2pt);

\end{tikzpicture}
\end{center}
\caption{Cartoon for the entropy production in DIS.}
\label{fig:entropy}
\end{figure}
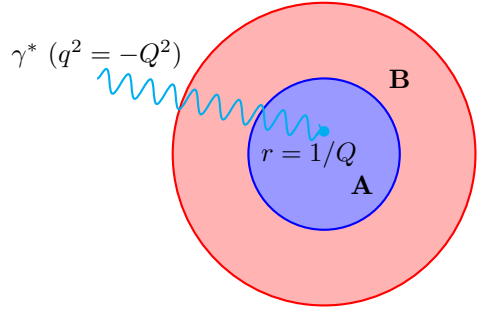

Similarly, in DIS, we probe only a subset of the proton's partons while tracing over the rest, converting the pure proton state into a mixed state characterized by PDFs. Kharzeev and Levin~\cite{Kharzeev:2017qzs} formalized this picture by computing the von Neumann entropy of the reduced density matrix obtained by tracing the pure proton state over the spatial region outside the DIS probe. In the proton's rest frame as shown in Fig.~\ref{fig:entropy}, DIS probes a spatial region \(A\) localized within a tube of radius \(\sim 1/Q\) and length \(\sim 1/(mx)\), where \(m\) is the proton mass and \(x\) the momentum fraction of the struck quark. Inclusive measurements sum over the unobserved part of the wave function in the complementary region \(B\), giving access only to the reduced density matrix \(\rho_A = \text{Tr}_B \rho\) rather than the full density matrix \(\rho = |\psi\rangle\langle\psi|\). The von Neumann entropy arising from the quantum entanglement between regions \(A\) and \(B\) is \(S_A = -\text{Tr}(\rho_A \ln \rho_A)\), and because region \(B\) is complementary to \(A\), the entanglement entropy \(S_B\) associated with it must equal \(S_A\). At small Bjorken \(x\), where gluons dominate, the entanglement entropy takes the remarkably simple form
\begin{equation}
S(x) = \ln[xG(x)].
\end{equation}
Since the gluon distribution grows as \(xG(x) \sim 1/x^\Delta\) at small \(x\) (where \(\Delta \approx 0.2\text{--}0.3\) from experimental data), the entropy becomes \(S \approx \Delta Y\), where \(Y = \ln(1/x)\) is the rapidity. This means the proton is composed of \(\exp(\Delta Y)\) partonic micro-states, each occurring with equal probability \(\exp(-\Delta Y)\). The fact that all micro-states have equal probabilities is the defining feature of maximal entanglement: just as in the Bell state where we have complete ignorance about the measured particle's state, in the small-\(x\) proton we have complete ignorance about which specific partonic configuration is realized. The reduced density matrix \(\rho_A\) is proportional to the identity matrix in the space of partonic configurations, meaning no particular configuration is preferred. This is the quantum analog of a thermal state at high temperature limit, where all accessible states are equally populated. Furthermore, Ref.~\cite{Hagiwara:2017uaz} introduces the semiclassical Wehrl entropy to quantify multiparton complexity in the nucleon's phase space, and evaluates it at small $x$ in comparison with the von Neumann entropy. In addition, after including the QCD evolution of the entanglement entropy \cite{Hentschinski:2024gaa}, a strong agreement between the rapidity dependence of von Neumann entropy and the corresponding experimental data on hadron entropy is then revealed. Recently, the entropy production due to soft gluon emissions in deep inelastic scattering using Monte Carlo simulations has been studied in Ref.~\cite{Hentschinski:2025pyq,Hentschinski:2026otq}. Hatta \emph{et al.}~\cite{Hatta:2024lbw} generalize the observation that quarks and gluons at small $x$ are maximally entangled Bell states to all $x$ values, describing gluons as maximally entangled qubit-qudit states and computing the conditional probability distribution of orbital angular momentum given helicity.

This situation presents a modern incarnation of the Einstein-Podolsky-Rosen paradox at subnucleonic scales: in the infinite-momentum frame, the parton probed by a virtual photon with virtuality \(Q^2\) appears causally disconnected from the rest of the nucleon during the hard interaction, yet color confinement requires that the parton and the remainder of the nucleon form a color-singlet state and thus must be in strongly correlated quantum states. The resolution of this paradox lies in quantum entanglement, which indicates that the partons are not truly independent but rather maximally entangled, allowing them to appear ``quasi-free'' to local measurements while maintaining global quantum correlations that ensure color confinement.

Tu, Kharzeev, and Ullrich~\cite{Tu:2019ouv} provided experimental verification of this entanglement picture using data from proton-proton collisions at the LHC. They devised a test based on the principle that the entanglement entropy \(S_A\) arising from tracing over the unobserved region should equal the entropy \(S_B\) of that region, which can be independently measured through the final-state hadron multiplicity distribution \(P(N)\)—the probability of producing \(N\) hadrons per event. The key insight is that if the initial quantum entanglement entropy is converted into classical Boltzmann entropy during hadronization, then the final-state entropy \(S_{\text{hadron}} = -\sum_N P(N) \ln P(N)\) should equal the initial entanglement entropy. Their analysis of LHC data confirmed the relation \(S_A = S_B = S_{\text{hadron}}\), providing strong direct evidence for quantum entanglement at subnucleonic scales and demonstrating that the parton model's ``quasi-free'' partons are in fact maximally entangled quantum states. This experimental confirmation also validates the interpretation that PDFs encode not just classical probabilities but the quantum entanglement structure of the proton's wave function.

Ref.~\cite{Hentschinski:2023izh} investigates the onset of maximal quantum entanglement inside the proton using Diffractive Deep Inelastic Scattering (DDIS). Employing a dipole cascade model and diffractive parton distribution functions, the authors show that the rapidity gap inherent to diffractive processes delays the QCD evolution and reduces the partonic Hilbert space, thereby postponing the onset of maximal entanglement. It has been shown that in the inclusive DIS case, the proton is found 
to be maximally entangled at sufficiently small $x$ \cite{Kharzeev:2017qzs}. In diffractive DIS, however, the presence of the 
rapidity gap delays the QCD evolution and thus the onset of maximal 
entanglement, which is only reached at small values of $\beta$ with $\beta$ defined as the Pomeron’s momentum fraction carried by the quark interacting with the virtual photon. The phrase 
``onset of maximal entanglement'' therefore refers to this kinematic 
transition specific to the diffractive process. By relating the entanglement entropy directly to the entropy of final-state hadrons---without invoking fragmentation functions---they find good agreement with H1 Collaboration data at HERA using both the exact entropy formula and its asymptotic expansion, the latter indicating a nearly maximally entangled state at small $\beta$. At larger $\beta$, configurations with few partons dominate and the entropy falls below its maximal value, revealing the transition region. Future opportunities at the US-EIC are discussed, where extended detector coverage will enable more precise studies of this transition in diffractive scattering.

The connection between entanglement and inclusive measurements extends beyond QCD: in quantum electrodynamics and perturbative quantum gravity, infrared divergences in the S-matrix arise from tracing over unobserved soft photons and gravitons~\cite{Semenoff:2019dqe}. When computing transition probabilities, one must sum over all possible soft radiation configurations that escape detection, effectively tracing over these unobserved degrees of freedom. This leads to decoherence and information loss that can be understood through the lens of entanglement entropy. These developments suggest that PDFs and fragmentation functions, traditionally viewed as purely classical probability distributions, actually encode quantum entanglement between observed and unobserved degrees of freedom in high-energy scattering processes, with the classical probabilities emerging from the loss of quantum coherence when we trace over what we do not measure.

\subsection{The entanglement of quark-antiquark pairs in DIS}
\textbf{Entanglement in inclusive DIS}: In a recent study~\cite{Qi:2025onf}, quantum entanglement (concurrence) and Bell nonlocality in quark-antiquark pair production via photon-gluon fusion at an EIC are computed for the first time. The process $\gamma^* g \to q\bar{q}$ provides a clean environment for studying quantum spin correlations, as the initial state is well-defined and the produced quark-antiquark pair carries the spin degrees of freedom that can be viewed as a two-qubit system. By computing the scattering amplitude without summing over the spins of the produced $q\bar{q}$ system, one can obtain the spin density matrix as follows:
\begin{equation}
\rho_{\alpha \alpha^\prime, \beta \beta^\prime} = \frac{R_{\alpha \alpha^\prime, \beta \beta^\prime}}{\text{tr}R} \quad \text{with} \quad R_{\alpha \alpha^\prime, \beta \beta^\prime} = \sum \mathcal{M}^\ast_{\alpha \alpha^\prime} \mathcal{M}_{\beta \beta^\prime},
\end{equation}
where $\alpha$ and $\beta$ ($\alpha^\prime$ and $\beta^\prime$) are the spin indices of the quark (antiquark) in the complex conjugate amplitude and amplitude, respectively.

\textbf{Longitudinal photons:} For longitudinally polarized virtual photons ($\gamma^*_L$), the calculation reveals vanishing polarizations and a correlation matrix:
\begin{equation}
C_{ij} = \begin{pmatrix} 
1 & 0 & 0 \\
0 & -\chi_1 & -\chi_2 \\
0 & -\chi_2 & \chi_1
\end{pmatrix}
\quad \text{with} \quad \chi_1 = \frac{1-2z^2+z^2 \beta^2}{1-z^2 \beta^2}, \quad \chi_2 = \sqrt{1-\chi_1^2},
\end{equation}
where $\beta$ is the velocity of the produced quark in the center-of-mass frame and $z = \cos\theta$ with $\theta$ being the angle between the beam direction and the quark outgoing direction. This result reveals a striking conclusion: the two-qubit system produced by the longitudinal photon is always in a \textbf{maximally entangled pure state}. More specifically, $\rho_L$ is given by $\rho_L = |\Psi\rangle\langle\Psi|$, with:
\begin{equation}
|\Psi\rangle = \frac{1}{2}\left(\sqrt{1+\chi_1}, \mathrm{i}\sqrt{1-\chi_1}, \mathrm{i}\sqrt{1-\chi_1}, \sqrt{1+\chi_1}\right),
\end{equation}
in the spin basis $\ket{\uparrow\uparrow}, \ket{\uparrow\downarrow}, \ket{\downarrow\uparrow}, \ket{\downarrow\downarrow}$. For longitudinally polarized virtual photons ($\gamma^*_L$), the produced $q\bar{q}$ pairs exhibit maximal entanglement at leading order in QCD, with concurrence $\mathcal{C} = 1$ independent of kinematic variables. This remarkable result persists throughout the accessible phase space, making EICs ideal environments for studying quantum correlations at high energies. The maximal entanglement arises from the helicity structure of the $\gamma^*_L g \to q\bar{q}$ amplitude, where angular momentum conservation constrains the final-state spin configuration to a maximally entangled Bell state. In fact, one can mathematically prove that any density matrix with $\mathcal{C} = 1$ must have rank $1$, and thus be given by a pure state. Using the Horodecki criterion with the correlation matrix which gives $C^TC= \mathbb{I}$, the state saturates the Tsirelson bound with $|\mathcal{B}|_{\max} = 2\sqrt{2}$, guaranteeing maximal violation of the CHSH inequality.

\begin{figure}[htbp!]
\centering
\includegraphics[width=0.7\textwidth]{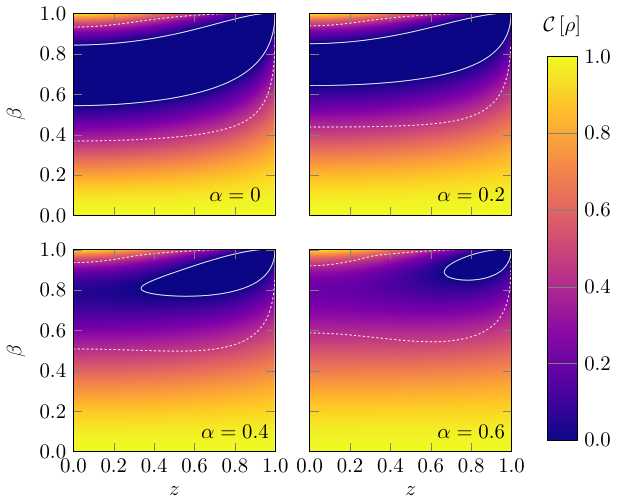}
\caption{Density plots of the concurrence $\mathcal{C}[\rho_T]$ associated with 
the transverse photon polarization state at EICs~\cite{Qi:2025onf}, presented as 
functions of the quark velocity $\beta$ and $z=\cos \theta$ for representative values of the virtuality parameter $\alpha$. 
The solid and dashed contour lines indicate the onset of quantum 
entanglement ($\mathcal{C}[\rho_T] = 0$) and Bell nonlocality 
($\mathcal{N}[\rho_T] = 0$), respectively. Figure replotted from Ref.~\cite{Qi:2025onf}.}
\label{Fig:concurrence}
\end{figure}

\textbf{Transverse photons:} For transversely polarized photons ($\gamma^*_T$), the correlation matrix takes a different form, and significant entanglement as shown in Fig.~\ref{Fig:concurrence} emerges primarily in two kinematic regimes: near the production threshold where the quark mass $m_q$ is comparable to the partonic collisional energy scale, and in the ultra-relativistic limit where the quarks are highly boosted. 
\begin{itemize}
    \item In the threshold limit ($\beta \to 0$), the state approaches the singlet Bell state $|\Psi^-\rangle = \frac{1}{\sqrt{2}}(\ket{\uparrow\downarrow} - \ket{\downarrow\uparrow})$ with correlation matrix $C_{ij} = -\delta_{ij}$, achieving maximal entanglement with $\mathcal{C} = 1$ and $|\mathcal{B}|_{\max} = 2\sqrt{2}$. 
    \item In the ultra-relativistic limit ($\beta \to 1$), the state approaches $|\Phi^-\rangle = \frac{1}{\sqrt{2}}(\ket{\uparrow\uparrow} - \ket{\downarrow\downarrow})$ with correlation matrix $C_{ij} = \text{diag}(-1, 1, 1)$, also achieving maximal entanglement.
    \item Between these limits, the concurrence depends on the ratio $\alpha=Q^2/\hat{s}$ with $\hat{s}$ being the partonic center of mass energy, the velocity $\beta$ and the scattering angle, interpolating between separable and maximally entangled configurations as the kinematics vary. Also, the Horodecki criterion~\cite{Horodecki:1995nsk} can be applied to determine the CHSH violation potential: computing the eigenvalues $\mu_1 \geq \mu_2 \geq \mu_3$ of $C^T C$ gives $|\mathcal{B}|_{\max} = 2\sqrt{\mu_1 + \mu_2}$, showing that Bell nonlocality is accessible in kinematic regions where $\mu_1 + \mu_2 > 1$.
\end{itemize}

The clean experimental environment of electron-ion colliders, compared to hadron colliders where initial-state effects and underlying event activity complicate the interpretation, provides superior conditions for measuring these quantum correlations through the angular distributions of the produced quarks or their decay products. Recent studies indicated that EICs are promising platforms for testing fundamental quantum mechanics at unprecedented energy scales, where the interplay between QCD dynamics and quantum information phenomena can be systematically explored. Moreover, there have been significant efforts~\cite{Fucilla:2025kit, Hatta:2025obw, Cheng:2025zaw, Fucilla:2026mkg} to establish theoretical frameworks that use quantum information techniques to study nuclear physics, thereby bridging quantum information science and hadronic physics in the EIC era. In particular, such approaches enable the study of many QCD effects, including the transversity, the gluon saturation effect, and the linearly polarized gluon distribution in inclusive and exclusive DIS at EICs.

\textbf{Diffractive production and Pomeron exchange}: In a recent work~\cite{Fucilla:2025kit} by Fucilla and Hatta, the spin-spin entanglement in diffractive heavy quark production in deep inelastic scattering (DIS) and ultraperipheral collisions (UPCs) is studied in detail, focusing on the role of Pomeron exchange. Diffractive processes, characterized by a large rapidity gap and an intact final-state proton, offer a particularly clean environment for quantum information studies because the color-singlet exchange mechanism provides additional theoretical control, and the intact proton serves as an experimental tag to reduce backgrounds. Working within the dipole framework of high-energy QCD, where the virtual photon fluctuates into a $q\bar{q}$ dipole that subsequently scatters off the target via Pomeron exchange, the authors calculated the spin density matrix for diffractively produced $b\bar{b}$, $c\bar{c}$, and $s\bar{s}$ pairs. In high-energy QCD, the Pomeron provides an effective description of the exchange of two or more gluons in a color-singlet configuration, with the dipole-proton scattering amplitude encoding the density information of the target.

First, for longitudinally polarized virtual photons, the Pomeron exchange preserves maximal entanglement with $\mathcal{C} = 1$ and maximal violation of the Bell-CHSH inequality with $|\mathcal{B}| = 2\sqrt{2}$. This result coincides with the inclusive case, but with distinct dynamical origins rooted in the color-singlet nature of the diffractive exchange.

Second, for transversely polarized photons, the produced $q\bar{q}$ pairs are always entangled and Bell-violating across the accessible kinematic range, with the degree of quantum correlation depending on the transverse momentum $k_\perp$ and the longitudinal momentum fraction $z$ of the produced quark. A particularly striking prediction is that maximal entanglement ($\mathcal{C} = 1$) and maximal Bell violation ($|\mathcal{B}| = 2\sqrt{2}$) occur simultaneously when the transverse momentum approximately equals the quark mass, $k_\perp \approx m_q$, in the heavy quark mass limit. This provides a clear experimental signature that can be targeted in dedicated measurements at EICs. For charm quarks with $m_c \approx 1.3$ GeV and bottom quarks with $m_b \approx 4.2$ GeV, the optimal kinematic configurations are well within the reach of both US-EIC and EicC. More interestingly, for the $q\bar{q}$ pairs produced in this channel via pomeron exchanges, the requirements for quantum entanglement and Bell nonlocality become the same. In other words, the entangled $q\bar{q}$ pairs always violate the Bell inequality in this channel, which is peculiar since Bell-nonlocal states usually form a subset of entangled states.

At last, the framework also incorporates saturation effects through the dipole scattering amplitude, allowing exploration of how gluon saturation in the target modifies the quantum correlations of the produced pairs. This connection between saturation physics and quantum information provides a novel perspective on the dense gluon regime.

\textbf{Quantum entanglement for GPD studies}: Furthermore, Hatta and Schoenleber studied quantum entanglement in exclusive quark-antiquark pair production at EICs using GPDs~\cite{Hatta:2025obw}. GPDs provide a unified description of the longitudinal momentum and transverse spatial structure of partons inside hadrons, and their framework connects quantum information observables to the three-dimensional tomographic structure of the nucleon.

The calculation of the spin density matrix within collinear factorization reveals a rich pattern of entanglement arising from both real and imaginary parts of GPD amplitudes. The interference between different GPD contributions—including both quark-antiquark and two-gluon exchange mechanisms—generates non-trivial quantum correlations that depend sensitively on the kinematic variables. The authors map out kinematical regions where $q\bar{q}$ pairs of various flavors ($u\bar{u}$, $d\bar{d}$, $s\bar{s}$, $c\bar{c}$, $b\bar{b}$) exhibit entanglement, Bell nonlocality~\cite{Bell:1964kc}, and non-stabilizerness (magic)~\cite{Bravyi:2004isx,Emerson:2013zse}.

A particularly striking prediction is that massive quarks and antiquarks produced in these exclusive processes are transversely polarized, analogous to the well-known transverse hyperon polarization observed in unpolarized hadron-hadron collisions. For strange, charm, and bottom quark production, the transverse polarization can reach 50--80\% in certain kinematic regions accessible in low-energy US-EIC runs and possibly at EicC. This large polarization arises from the interference between different helicity amplitudes and provides a clear experimental signature.

The GPD framework bridges the low-energy regime, where perturbative QCD may not be fully applicable, and the high-energy Regge limit, where Pomeron exchange dominates. By incorporating both the $H$ and $E$ type GPDs along with their skewness dependence, the formalism captures the full complexity of color-singlet exchange in QCD. The presence of both real and imaginary parts of the Compton form factors, related to GPDs through dispersion relations, introduces interference effects that are essential for generating the observed entanglement patterns.

Exclusive processes provide a cleaner environment for quantum information studies compared to inclusive production, as the requirement of an intact final-state proton constrains the kinematics and significantly reduces backgrounds from inelastic events. The connection between entanglement measures and GPDs suggests that quantum information observables could provide novel constraints on these distributions, complementing traditional extractions from deeply virtual Compton scattering and meson production.

\subsection{Quantum information with polarizations}
\textbf{Quantum Information with Polarized beams:} Cheng, Han, and Trifinopoulos~\cite{Cheng:2025zaw} investigated quantum information theoretic observables in electron-proton scattering at EICs, focusing on both entanglement and magic, which are two complementary indicators of non-classicality. Quantum magic~\cite{Bravyi:2004isx,Emerson:2013zse} is a concept related to quantum computing, and it reveals the non-classical resources required to make quantum computers more powerful than the classical ones. Their analysis systematically explores how the polarization state of the initial beams determines the quantum information properties of the final state.

First, it is found that unpolarized and longitudinally polarized beams yield unentangled, separable outcomes in the final state in electron-proton scatterings and $t$-channel electron-quark scatterings. In the helicity basis, which is the natural basis for electromagnetic interactions, the scattering matrix is block-diagonal: different helicity configurations scatter independently without creating quantum coherence between them. As a result, the final state remains separable, which is a product state or classical mixture of product states, with no entanglement.

In contrast, transverse polarization prepares the initial beams in coherent superpositions of helicity eigenstates, such as $ \ket{\rightarrow_x} = \tfrac{(\ket{ \uparrow_z} + \ket{\downarrow_z})}{\sqrt{2}}$.
The interference between different helicity amplitudes then generates off-diagonal terms in the density matrix, producing genuine quantum entanglement in the final electron-proton state. When both electron and proton initial beams are transversely polarized along the transverse direction, their spin density matrices are given by
\begin{equation}
    \rho^{e/p} = \frac{\mathbb{I}_2 + B^{e/p}_\perp \cdot \sigma_{\perp}}{2},
\end{equation}
where $\sigma_\perp =\{\sigma_x, \sigma_y\}$.

In the deep inelastic scattering regime, the degree of quantum correlation is governed by transversity PDFs $h_1(x)$, which describe the distribution of transversely polarized quarks inside a transversely polarized nucleon. This connection provides a novel perspective on spin dynamics within QCD: the transversity PDFs, which are among the least well-constrained of the leading-twist distributions, directly control the entanglement properties of the scattered particles. Measurements of quantum correlations could therefore provide complementary constraints on transversity, supplementing traditional extractions from semi-inclusive DIS and Drell-Yan processes.

Beyond entanglement, the authors analyze quantum magic (also known as non-stabilizerness), which quantifies the departure of a quantum state from the set of stabilizer states that can be efficiently simulated on classical computers~\cite{Bravyi:2004isx,Emerson:2013zse}. Magic is quantified using stabilizer R\'enyi entropies, which measure the resources required for quantum computation beyond Clifford gates. The analysis shows that transverse polarization not only generates entanglement but also produces states with nonzero magic, representing genuine computational resources for quantum information processing. In addition, quantum entanglement, discord, steering, and magic in the DIS final state have been shown to carry nontrivial sensitivity to the nucleon's transversity PDFs and tensor 
charges~\cite{Bloss:2026yrf}, providing a tomographically 
clean probe that is complementary to conventional global 
QCD analyses.

The capability of EICs to prepare both electron and proton (or ion) beams with varying degrees of transverse polarization makes them ideal facilities for initializing scattering processes in well-defined quantum states. By systematically varying the beam polarizations, experimenters can explore the full landscape of quantum correlations accessible in high-energy scattering, from separable states to maximally entangled configurations.

\textbf{Polarization, Maximal Concurrence and Pure States:} Recently, an exact analytical upper bound on the concurrence of a general two-qubit state subject to fixed local polarization magnitudes $a = \|\vec{B}^{+}\|$ and 
$b = \|\vec{B}^{-}\|$ was proposed in Ref.~\cite{Liu:2026dzv}. The bound takes the form
\begin{equation}
    \mathcal{C}(\rho)
    \;\leq\;
    \mathcal{C}_{\max}(a,b)
    \;=\;
    \sqrt{\bigl(1-\max\{a,b\}\bigr)\bigl(1+\min\{a,b\}\bigr)},
    \label{eq:main_result}
\end{equation}
and is shown to be tight: it is saturated by an explicit family of 
$X$-state density matrices. This result represents an improvement over the previously 
established bound of Refs.~\cite{Zhang:2008ebu, Barr:2024djo},
\begin{equation}
    \mathcal{C}(\rho) \;\leq\; \sqrt{1 - \max\{a^2, b^2\}},
\end{equation}
which depends solely on the larger of the two polarization magnitudes.
In contrast, Eq.~\eqref{eq:main_result} involves \emph{both} 
polarizations simultaneously and yields a strictly tighter constraint 
whenever $a \neq b$, thereby providing a more accurate characterization 
of the interplay between local spin information and bipartite entanglement.

Two physically motivated special cases merit particular attention. 
First, when the total polarization $B = a^2 + b^2$ is held fixed, 
or second, when the two subsystems are equally polarized ($a = b$), 
the maximal concurrence simplifies to
\begin{equation}
    \mathcal{C}_{\max} 
    \;=\; \sqrt{1 - a^2} 
    \;=\; \sqrt{1 - b^2}
    \;=\; \sqrt{1 - \tfrac{B}{2}},
\end{equation}
and the optimal state is necessarily \emph{pure}. This observation 
establishes a fundamental connection between polarization, entanglement, 
and quantum purity: under these constraints, the density matrix that 
maximizes entanglement is a rank-one projector, revealing that the 
trade-off between local (polarization) and nonlocal (entanglement) quantum correlations is most efficiently realized in pure states.

To illustrate the physical implications of these results, one can consider the parity-violating process 
$e^+e^- \to Z^0 \to q\bar{q}$, in which the chiral structure of the 
weak interaction induces a net polarization in the 
final-state quark-antiquark pair. One finds that the maximal concurrence is determined by $c_A^q$ and $c_V^q$, the axial-vector and vector couplings of quark $q$ to the $Z^0$ boson, respectively, and is attained in the ultra-relativistic limit ($u = 1$) at perpendicular scattering angle ($\theta = \pi/2$), where it takes the value
\begin{equation}
    \mathcal{C}_{\max}
    \;=\;
    \frac{(c_A^q)^2 - (c_V^q)^2}{(c_A^q)^2 + (c_V^q)^2}
    \;\simeq\;
    \begin{cases}
        0.744 & \text{(up-type quarks)}, \\
        0.353 & \text{(down-type quarks)}.
    \end{cases}
\end{equation}
These values are significantly suppressed relative to the unpolarized 
maximum $\mathcal{C}_{\max} = 1$ obtained in Ref.~\cite{Qi:2025onf}, 
demonstrating quantitatively that parity-violating polarization 
fundamentally limits the achievable quantum entanglement. 
The corresponding optimal density matrix is a pure state, in agreement with the general 
analysis. Taken together, these findings establish a comprehensive, 
process-independent framework relating local polarization, maximal 
entanglement, and the role of pure states in high-energy collisions, 
with direct implications for experimental tests of quantum correlations 
at collider facilities.

\textbf{Outlook}: Recent theoretical developments can lead to a rich program for quantum information studies at EICs in the years to come. For EicC, operating at lower center-of-mass energies compared to the US-EIC, several complementary advantages emerge. The lower energy regime provides better access to the threshold region for heavy quark pair production, where the reduced phase space enhances sensitivity to production mechanisms and spin correlations. This kinematic advantage extends to hyperon pair production, enabling systematic studies of strange quark hadronization and providing crucial insights into the origin of $\Lambda$ polarization and related spin phenomena in QCD. The reduced boost of produced heavy quarks and hyperons facilitates spin measurements through the angular distributions of their decay products, as the decay kinematics are less compressed in the laboratory frame. The enhanced role of valence quark contributions in the accessible kinematic regime provides sensitivity to different aspects of nucleon structure compared to the sea-quark and gluon-dominated regime at higher energies.

The connection between quantum information observables and parton distributions—including GPDs, transversity PDFs, and the correlation matrix elements accessible through polarized beams—suggests that precision measurements of entanglement and Bell nonlocality could provide novel constraints on nucleon structure. Conversely, improved knowledge of these distributions from global QCD analyses would sharpen the predictions for quantum correlations, enabling more stringent tests of quantum mechanics at high energies. The interplay between threshold production dynamics, spin correlations, and quantum entanglement at EicC energies thus opens new avenues for exploring the quantum nature of strong interactions in a regime where perturbative and nonperturbative effects are both significant.

\newpage
   \section{Artificial intelligence and machine learning for EICs}
\label{sec:ai_applications}

\subsection{Introduction: AI/ML in modern collider physics}

The rapid advancement of artificial intelligence (AI) and machine 
learning (ML) technologies has ushered in a transformative era across 
virtually all domains of scientific research. From drug discovery and 
climate modeling to materials science and fundamental physics, AI/ML 
methods have demonstrated remarkable capabilities in pattern recognition, 
optimization, and predictive modeling. In high energy and nuclear physics, 
the explosion of data volumes and the increasing complexity of experimental 
apparatus have made AI/ML techniques not merely advantageous but essential 
for extracting meaningful physics from modern 
experiments~\cite{Guest:2018yhq,Carleo:2019ptp,Shanahan:2022ifi,Alexandrou:2026cnj}. The application of AI, and in particular ML, in particle physics has experienced remarkable growth in 
recent years, with an ever-expanding scope of topics and use cases, 
as systematically indexed by the Living Review of Machine Learning for Particle Physics~\cite{Feickert:2021ajf} (\href{https://iml-wg.github.io/HEPML-LivingReview/}{HEP ML Living Review}).

The convergence of unprecedented computational resources, sophisticated 
algorithm development, and the availability of large-scale datasets has 
positioned AI/ML as a cornerstone technology for the next generation of 
particle and nuclear physics experiments. These techniques offer 
transformative potential across the entire experimental workflow, from 
accelerator operations and detector design to real-time data processing, 
event reconstruction, and physics analysis. In particular, AI tools 
provide significantly more efficient approaches for both processing 
high-dimensional detector data and accelerating Monte Carlo simulations.

For many years, high energy and nuclear physicists have been employing 
techniques known as multivariate analysis, which can be viewed as 
early applications of ML methods. Since the emergence of deep learning 
in 2012, which enabled the training of large neural networks, jet physics, 
tracking, and simulation at the LHC have been revolutionized. Neural 
network-based PDFs~\cite{Forte:2002fg,NNPDF:2014otw}, first introduced 
in 2002, remain one of the most sophisticated applications of ML in QCD 
phenomenology. Given the rapid pace of development in this field and the 
growing role of AI/ML in EIC and EicC research, this chapter is devoted 
to highlighting these emerging tools, new ideas, and new directions that 
are shaping the methodological landscape of the EicC physics program.

Let us first introduce some key concepts commonly used in AI and ML 
studies and their applications.

\textbf{Fundamentals of AI/ML for physicists}: \textit{What is AI?}---AI refers to computational systems designed to perform tasks that typically require human intelligence, such as learning, reasoning, problem-solving, perception, language understanding, pattern recognition, decision-making, prediction, and optimization. In the context of physics, AI encompasses a broad spectrum of techniques that enable computers to learn from data, identify complex patterns, and make informed predictions without being explicitly programmed for every specific scenario. For physicists approaching AI/ML from a traditional computational physics background, the following concepts form the essential foundation.

\textbf{Machine Learning (ML)} constitutes the primary subset of AI used in physics research, focusing on algorithms that learn patterns from data to improve accuracy without explicit programming for specific tasks. 

\textbf{Neural Networks (NN)} are computational models inspired by biological neural systems, consisting of interconnected layers of artificial neurons (nodes). Each neuron applies a weighted sum followed by an activation function:
\begin{equation}
y = f\left(\sum_i w_i x_i + b\right),
\end{equation}
where $x_i$ are inputs, $w_i$ are learnable weights, $b$ is a bias term, and $f$ is an activation function. Networks are trained by adjusting weights through backpropagation to minimize a loss function that quantifies prediction errors.

\textbf{Deep Learning (DL)} is a specialized branch of ML employing neural networks with multiple hidden layers (deep architectures). Deep networks can learn hierarchical representations, with early layers capturing simple features and deeper layers combining them into complex abstractions. The ``deep learning revolution'' of 2012 demonstrated that sufficiently deep networks, trained on large datasets with modern computational resources, could dramatically outperform traditional methods across diverse tasks.

\textbf{Bayesian Analysis}, or Bayesian Inference, provides a statistical framework for extracting physics parameters by systematically combining experimental data with theoretical knowledge. The method uses Bayes' theorem to update our prior understanding:
\begin{equation}
P(\theta|D) = \frac{P(D|\theta)P(\theta)}{P(D)},
\end{equation}
where $P(A|B)$ is a conditional probability: the probability of event $A$ occurring given that $B$ is true. The left-hand side $P(\theta|D)$ is the output of this analysis, showing how likely different parameter values $\theta$ are given the observed data $D$. On the right-hand side: $P(D|\theta)$ is the likelihood function that quantifies how probable the observed data would be if the parameters had value $\theta$ (higher values indicate parameters that better predict the observations); $P(\theta)$ (the prior probability) incorporates what we knew beforehand from theory or previous experiments; and $P(D)$ is the total probability of observing the data (averaging over all possible parameter values), which serves as a normalization constant. This approach provides complete probability distributions for parameters rather than single values with uncertainties, naturally handling correlations and systematically including information from multiple measurements and theoretical constraints.

Taking the Higgs boson mass measurement as an example: the \textit{input data} $D$ are the reconstructed invariant masses from decay products (e.g., diphoton events); the \textit{likelihood} $P(D|\theta)$ describes the expected mass distribution for each hypothetical Higgs mass, including detector resolution effects; the \textit{prior} $P(\theta)$ might encode theoretical expectations (e.g., $m_H \sim 125$ GeV from electroweak fits) or previous LHC measurements; and the \textit{output posterior} $P(\theta|D)$ gives the full probability distribution for the Higgs mass after including the input data, showing not just the best estimate but the complete uncertainty structure and correlations with other parameters. For a more detailed application to jet mass calculations, see Ref.~\cite{Gao:2025fch}.

\textbf{Generative  Models} are AI systems, mostly based on deep learning, that learn the underlying probability distribution of data and can generate new samples resembling the training data. This capability is particularly important for collider physics applications such as fast simulation and data augmentation.

The hierarchy among these concepts reflects both their logical structure and historical development: AI encompasses all intelligent computational systems, with ML forming the dominant subset used in physics. DL is a powerful specialization of ML using multi-layer neural networks, enabled by modern computational resources and large datasets. Bayesian methods provide a complementary statistical framework that can be combined with ML (e.g., Bayesian neural networks, simulation-based inference) or used independently for rigorous uncertainty quantification. Generative models, including large language models (LLMs), represent state-of-the-art AI techniques that leverage DL architectures to learn and sample from high-dimensional probability distributions, with physics applications ranging from fast simulation to data augmentation in collider physics.

The application of AI/ML techniques to collider physics addresses several fundamental challenges in modern experiments and phenomenological studies. DL excels at processing massive data volumes with real-time event filtering and pattern recognition across millions of detector channels, while handling complex high-dimensional data through specialized architectures. Generative AI models accelerate computationally expensive Monte Carlo simulations, enabling rapid exploration of systematic uncertainties. ML-enhanced Bayesian analysis efficiently explores high-dimensional parameter spaces in QCD phenomenology, providing rigorous uncertainty quantification for simultaneous fitting of multiple observables. Key physics applications~\cite{Astrand:2025sij} include event selection and classification with superior discrimination power compared to traditional cut-based methods, jet physics~\cite{Larkoski:2017jix,Larkoski:2024uoc} and flavor tagging~\cite{Mondal:2024nsa} using jet substructure, particle identification~\cite{Graczykowski:2022zae,Karwowska:2023dhl,Kasak:2023hhr,Karwowska:2024xqy,CMS:2026znb} integrating information from multiple detector subsystems, and model-independent anomaly detection~\cite{ATLAS:2023ixc,CMS:2024nsz} for beyond Standard Model searches.

\subsection{AI initiatives at EICs}

The US-EIC, currently under construction at Brookhaven National Laboratory, is emerging as a pioneering facility in the systematic integration of AI/ML technologies into all aspects of collider science~\cite{Allaire:2023fgp, Alexandrou:2026cnj}. The AI4EIC collaboration has established a comprehensive framework for applying artificial intelligence across five major areas: (1) simulation, (2) reconstruction, (3) particle identification and detector design, (4) physics analysis, and (5) accelerator design and operations. This effort has produced numerous publications and fostered a vibrant community dedicated to advancing AI applications in electron-ion physics. These publications span a broad range of topics, including fast 
simulation with generative models~\cite{Giroux:2025mit,Araz:2024bom}, 
event reconstruction and track finding with graph neural 
networks~\cite{Gardner:2024ihn, Gavalian:2024icb}, 
particle identification and AI-assisted detector 
design~\cite{Matousek:2024vpa, Fanelli:2022rdm, AID2E:2024gyl, Fanelli:2022kro, Kelleher:2025yem}, 
and accelerator control and 
operations~\cite{Jeske:2022nws, Jeske:2024arh, Britton:2024pdy}. The EicC, as a complementary facility operating at lower center-of-mass energies, stands to benefit significantly from these developments. While the AI4EIC initiative has made substantial progress, there is a growing interest in developing AI/ML applications specifically tailored for the EicC. This part serves as a comprehensive review of AI/ML techniques developed for electron-ion collider science, with particular emphasis on their potential adaptation and implementation for the EicC program.

Modern electron-ion collider experiments generate vast quantities of data that require sophisticated computing infrastructure for storage, processing, and analysis. This necessitates hardware investment, software development including generative pretrained transformers, as well as the deployment of AI agents for research purposes, along with a team of technicians. Furthermore, ML algorithms can optimize data placement and job scheduling across computing resources, significantly improving throughput and reducing latency in data processing pipelines. For EicC, establishing a robust AI-integrated computing infrastructure from the outset would provide substantial advantages in managing the expected data volumes and ensuring efficient utilization of computational resources across collaborating institutions.

\textbf{Fast simulation with generative models}: Monte Carlo simulations are essential for detector design, physics analysis, and systematic uncertainty estimation, yet they can impose enormous computational burdens. Generative AI models offer a promising solution by learning to produce realistic detector responses orders of magnitude faster than traditional simulation methods~\cite{Allaire:2023fgp}. Recent developments have demonstrated the application of generative adversarial networks (GANs), variational autoencoders (VAEs), and normalizing flows for fast simulation of calorimeter showers and tracking detector responses. 

Of particular relevance to the EicC, generative models have been specifically developed for simulating Cherenkov detectors at EICs, effectively addressing the computationally intensive optical photon propagation that dominates simulation time in ring imaging Cherenkov (RICH) detectors~\cite{Giroux:2025mit}. These models can accurately reproduce complex photon hit patterns while achieving speedups of several orders of magnitude compared to traditional Geant4 simulations. Integrating these fast simulation techniques will be invaluable for optimizing EicC detectors and conducting large-scale physics projections, enabling rapid exploration of design parameter spaces that would otherwise be computationally prohibitive.

Furthermore, score-based diffusion models using point cloud 
representations have been developed specifically for EIC events, 
demonstrating the ability to generate full collider events with 
complete kinematic information for all particle species while 
accurately preserving event-wide conservation laws~\cite{Araz:2024bom}. 
This work also explores the adaptation of the \textsc{OmniLearn} 
foundation model for event generation, suggesting a broader shift 
toward fine-tuning general-purpose models for collider physics tasks 
rather than training dedicated models from scratch.

\textbf{Reconstruction}: Event reconstruction, encompassing track finding, vertex reconstruction, and particle flow algorithms, represents one of the most computationally demanding aspects of collider data processing. Machine learning approaches have demonstrated significant improvements in both reconstruction quality and processing speed compared to traditional algorithmic methods~\cite{Allaire:2023fgp}. For example, graph neural networks (GNNs) have emerged as particularly powerful tools for track reconstruction, naturally representing the sparse hit patterns in tracking detectors as graph structures where nodes correspond to detector hits and edges encode potential track connections~\cite{Gardner:2024ihn}. In the extreme occupancy conditions of quasi-real photon tagging, where bremsstrahlung processes produce a very high flux of electrons near the beamline, object condensation methods based on GNNs have demonstrated exceptional performance, achieving track finding efficiencies exceeding 95\% with purities above 90\% even in the presence of noise and hit detection inefficiencies. These algorithms are particularly well-suited for tracking scattered electrons at small angles relative to the beamline, where the high flux of tagged almost-real photons enables photo-production measurements but creates substantial backgrounds that challenge traditional tracking approaches.

Real-time track reconstruction using ML has been successfully implemented for the CLAS12 detector at Jefferson Lab, demonstrating the viability of AI-based reconstruction in production environments~\cite{Gavalian:2024icb}. In this approach, machine learning algorithms reconstruct tracks, including their momentum and direction, with high accuracy directly from raw hits of the CLAS12 drift chambers at the rate of data acquisition, enabling real-time identification of event topologies. This capability can significantly improve nuclear physics data processing by allowing experimental data to be identified and categorized on the fly, leading to significant reductions in offline processing requirements and enabling more efficient streaming readout applications. For EicC, where precise reconstruction of scattered electrons and hadronic final states is crucial for accessing the physics program, the adoption of ML-based reconstruction algorithms could substantially enhance physics reach while managing computational costs.

\textbf{Particle identification and detector design}: Particle identification (PID) is fundamental to the physics program of EICs, requiring the integration of information from multiple detector subsystems including tracking, calorimetry, and Cherenkov detectors. Machine learning methods excel at combining heterogeneous detector signals to achieve optimal PID performance, surpassing traditional likelihood based approaches~\cite{Allaire:2023fgp,Graczykowski:2022zae}. Gradient boosted decision trees and deep neural networks have been successfully deployed for photon classification and hadron identification in existing experiments such as CLAS12~\cite{Matousek:2024vpa}. Beyond analysis applications, AI techniques have revolutionized the detector design process itself through multi objective optimization frameworks that can simultaneously optimize multiple detector parameters against competing physics requirements~\cite{Fanelli:2022rdm}. The AI-assisted detector design for US-EIC, known as AID2E, employs surrogate models and Bayesian optimization to efficiently explore high dimensional design spaces~\cite{AID2E:2024gyl}. The US-EIC Comprehensive Chromodynamics Experiment (ECCE) tracking system optimization demonstrated the power of AI-assisted design by simultaneously optimizing tracking resolution, mechanical constraints, and cost considerations~\cite{Fanelli:2022kro}. These methodologies are directly transferable to EicC detector development, where similar multi-objective optimization challenges exist in balancing physics performance, technical feasibility, and resource constraints.

\textbf{Accelerator design and operations}: The operation of modern particle accelerators involves managing thousands of parameters in real time to maintain beam quality, luminosity, and machine protection. AI/ML techniques offer transformative capabilities for accelerator control, from predictive modeling of beam dynamics to autonomous tuning and fault detection~\cite{Allaire:2023fgp,Jeske:2022nws}. Machine learning models can predict beam behavior and preemptively adjust accelerator parameters to maintain optimal performance, reducing the need for manual intervention and improving integrated luminosity. At Jefferson Lab, AI for experimental controls has been implemented to automate routine operations and provide intelligent monitoring of accelerator systems~\cite{Jeske:2022nws}. Computer vision techniques have been developed for data quality monitoring, enabling automated detection of anomalies in detector performance during data taking~\cite{Jeske:2024arh}. ML-based calibration and control systems have been successfully deployed for drift chamber operations, demonstrating improved stability and reduced systematic uncertainties~\cite{Britton:2024pdy}. For EicC, incorporating AI driven accelerator controls from the design phase would enable more efficient commissioning, higher operational reliability, and optimized luminosity delivery throughout the experimental program.

\subsection{AI applications to related inverse problems}
\label{AI-inverse-problem}

In high-energy QCD, we often have to tackle many difficult inverse problems. An \textbf{inverse problem} is a fundamental challenge in science and engineering where we attempt to infer the underlying causes or parameters from observed effects, essentially working backwards from measurements to the hidden structure that produced them. This concept is in contrast to the \textbf{forward problem}, where we predict observable outcomes from known inputs. For example, in hadronic structure studies, the forward problem involves calculating measurable cross-sections $\sigma$ given known PDFs $f_i(x, Q^2)$ and the corresponding partonic cross-sections, while the inverse problem requires extracting these PDFs from experimental cross-section data. Schematically, this can be expressed as:
\begin{equation}
\text{Data} = \mathcal{F}[\text{Parameters}] + \text{noise},
\end{equation}
where $\mathcal{F}$ represents the forward model (such as convolutions with partonic cross-sections and DGLAP evolution in QCD). The central challenge of inverse problems is that they are often \textbf{ill-posed} in the Hadamard sense, meaning they may lack existence (no solution fits the data), uniqueness (multiple solutions fit equally well), or stability (small measurement errors lead to large variations in inferred parameters). The difficulty of inverse problems arises from limited data coverage, noise amplification, and the mathematical structure of the problem itself, such as the convolution integrals in PDF extraction that smear out information. Addressing these challenges requires sophisticated techniques, including Bayesian priors, machine learning with appropriate constraints, and physics-driven approaches that embed theoretical knowledge directly into the inference framework~\cite{Aarts:2025gyp}. 

Typical examples of inverse problems in QCD include extracting PDFs from deep inelastic scattering and hadron collider data, determining fragmentation functions from hadron multiplicities in $e^+e^-$ annihilation, inferring GPDs from DVCS measurements and TMDs from SIDIS measurements, and obtaining real-time spectral functions from Euclidean correlation functions in lattice QCD. The fundamental differences among traditional, Bayesian, and machine learning approaches to PDF extraction lie in how they address the ill-posedness of the inverse problem and quantify uncertainties:
\begin{itemize}
   \item \textbf{Traditional methods} (such as CTEQ-TEA (CT18)~\cite{Hou:2019efy}) rely on predetermined functional forms to parametrize PDFs at an initial scale $Q_0$, with modern implementations like CT18 employing sophisticated parametrizations including Bernstein polynomials combined with power-law terms of the form $x^{\alpha}(1-x)^{\beta}$ to enhance flexibility while maintaining stability. The small-$x$ behavior ($x^{\alpha}$) encodes Regge-theory expectations and the rise of parton densities at low momentum fractions, while the large-$x$ behavior ($\sim(1-x)^{\beta}$) reflects quark counting rules and the suppression of PDFs as quarks carry most of the proton's momentum. QCD constraints (momentum and valence sum rules, positivity) are enforced through explicit parametrization choices and Lagrange multipliers, with uncertainties estimated through the Hessian method based on a dynamical tolerance criterion. While this approach has evolved to minimize parametrization bias through more flexible functional forms and systematic exploration of alternative parametrizations, it still requires careful validation that the chosen basis functions can adequately represent the true PDF shapes, particularly in kinematic regions with sparse data coverage.

    \item \textbf{Bayesian methods} (such as the technique used by the JAM Collaboration~\cite{Cocuzza:2021cbi,Cocuzza:2025qvf}) treat PDFs as probability distributions rather than fixed functions, using physically motivated priors to encode expectations about smoothness, positivity, and asymptotic behavior, then employing the Bayesian Monte Carlo framework to explore the posterior distribution $P(\text{PDFs}|\text{Data}) \propto P(\text{Data}|\text{PDFs}) \cdot P(\text{PDFs})$. This provides rigorous uncertainty quantification through credible intervals while naturally regularizing the ill-posed problem through prior information, though results remain somewhat dependent on the subjective choice of priors.
    
    \item \textbf{Machine learning approaches} (pioneered by NNPDF~\cite{Forte:2002fg}) can, in principle, significantly reduce parametrization bias by using neural networks as highly flexible universal function approximators with reduced functional-form bias, generating Monte Carlo replicas of the data to train an ensemble of networks, and quantifying uncertainties through the spread across replicas. This maximizes flexibility and minimizes theoretical assumptions but can lead to unphysical behavior in extrapolation regions, may violate sum rules or positivity constraints, and requires large datasets for reliable results. \textbf{Physics-driven machine learning} represents an emerging hybrid approach that combines the flexibility of neural networks with hard QCD constraints (sum rules, symmetries) embedded directly into the architecture or loss function. This method maintains theoretical consistency while avoiding parametrization bias and provides reliable extrapolation to unmeasured kinematic regions, offering the best balance between data-driven flexibility and physics compliance.
\end{itemize}

The field is evolving toward physics-driven approaches that leverage machine 
learning's representational power while respecting the fundamental structure 
of QCD theory. In the following, we will review some recent applications of 
these AI techniques to these problems in detail.

\textbf{Neural Network extraction of parton distributions}: The determination 
of PDFs, TMDs, and GPDs from experimental data represents one of the most 
successful applications of machine learning in hadron physics. Traditional PDF 
extractions rely on parametrizing the $x$ dependence of distributions at an 
initial scale using functional forms with a limited number of free parameters, 
which are then evolved to experimental scales using DGLAP equations and fitted 
to data. However, such parametrizations inevitably introduce theoretical bias, 
as the chosen functional form constrains the space of allowed solutions. The 
NNPDF collaboration pioneered the use of neural networks to provide unbiased 
parametrizations for PDF extraction, replacing rigid functional forms with 
flexible neural network architectures capable of learning arbitrary functional 
relationships from data~\cite{Ball:2012cx,Forte:2002fg,NNPDF:2014otw,NNPDF:2002,
Ball:2008by,NNPDF:2017mvq,NNPDF:2021njg}.

\begin{figure}[htbp!]
\centering
\includegraphics[width=0.8\textwidth]{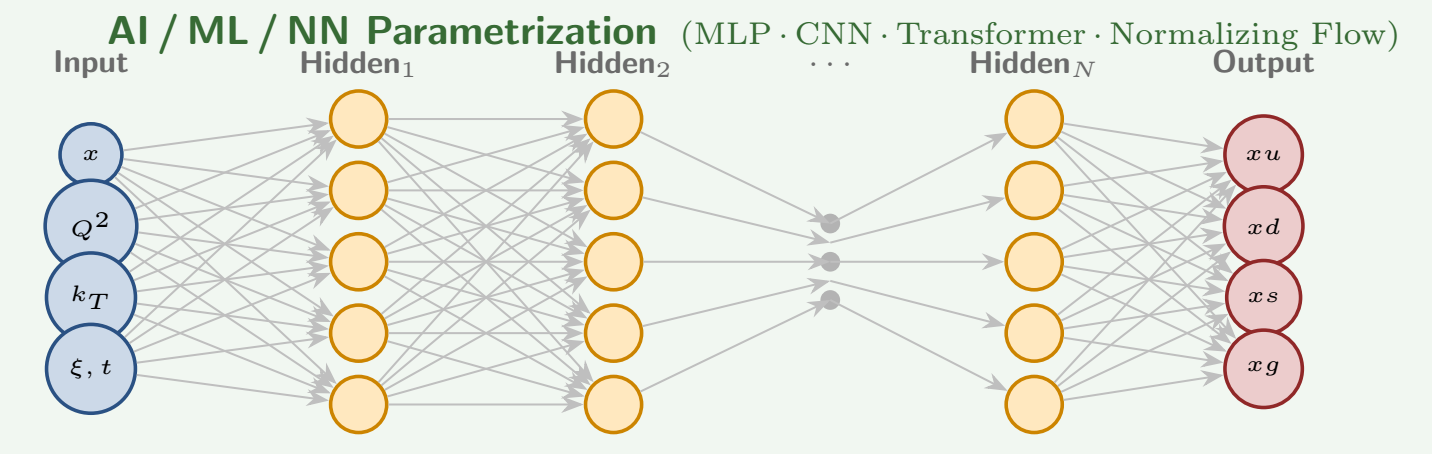}
\caption{Illustration of the application of NN on the extraction of PDFs.}
\label{Fig:nnpdf}
\end{figure}

The general AI/ML-based extraction pipeline proceeds as follows. Raw 
experimental measurements, including DIS, SIDIS, Drell-Yan, and 
proton-proton collision data, are first subjected to preprocessing, where 
kinematic cuts are applied, normalization and systematic uncertainties are 
accounted for, and the full experimental covariance matrix is constructed. 
As illustrated in Fig.~\ref{Fig:nnpdf}, the parton distributions are parametrized at an initial scale $Q_0$ by a neural network (NN), which serves as a flexible parametrization with reduced functional-form bias. The inputs are the relevant kinematic variables $(x,\, Q^2,\, k_T,\, \xi,\, t)$, and the outputs are the distribution functions for each parton flavor. This 
parametrization, realized through architectures such as Multi-Layer Perceptron (MLP), convolutional NNs, transformers, or normalizing flows, replaces the model-dependent polynomial 
ans\"{a}tze used in conventional global analyses. In this approach, each PDF 
flavor $f_i(x, Q_0^2)$ at the initial scale is represented by a NN that 
takes Bjorken $x$ as input and outputs the corresponding parton density. The 
network output is then evolved from $Q_0$ to the experimental scale $Q$ via 
perturbative QCD evolution kernels, namely DGLAP evolution at NLO/NNLO for 
collinear PDFs and CSS evolution for TMDs. The evolved distributions are 
subsequently convoluted with the appropriate hard-scattering coefficient 
functions to produce theoretical predictions for the measured observables. 
The network parameters are optimized by minimizing a loss function that 
quantifies the discrepancy between these theoretical predictions and 
experimental measurements across a global dataset encompassing DIS, 
Drell-Yan, and hadron collider data, using gradient-based optimizers such as 
Adam or L-BFGS, or alternatively via genetic algorithms, as employed in the 
NNPDF framework.

A crucial aspect of neural network PDF extraction is the rigorous propagation 
of experimental uncertainties to the final parton distributions. The NNPDF 
methodology employs a Monte Carlo replica approach, where $N_{\rm rep}$ 
artificial replicas of the experimental data are generated by fluctuating 
central values according to their uncertainties, assuming a multivariate 
Gaussian distribution characterized by the experimental covariance 
matrix~\cite{NNPDF:2014otw,NNPDF:2017mvq}. An independent neural network is 
trained on each replica, yielding an ensemble of $N_{\rm rep}$ PDF sets whose 
spread encodes the propagated experimental uncertainty. For any observable 
$\mathcal{O}$ that depends on PDFs, the central value and uncertainty can be 
computed as ensemble averages:
\begin{equation}
\langle \mathcal{O} \rangle = \frac{1}{N_{\rm rep}} \sum_{k=1}^{N_{\rm rep}} 
\mathcal{O}[f^{(k)}], \quad 
\sigma_{\mathcal{O}}^2 = \frac{1}{N_{\rm rep}} \sum_{k=1}^{N_{\rm rep}} 
\left( \mathcal{O}[f^{(k)}] - \langle \mathcal{O} \rangle \right)^2,
\end{equation}
where $f^{(k)}$ denotes the PDF set obtained from the $k$-th replica. 
This Monte Carlo approach provides a faithful representation of PDF 
uncertainties without requiring assumptions about the functional form of the 
parton densities. Alternatively, uncertainty quantification can be carried 
out through the Hessian eigenvector method or through fully Bayesian 
inference, the latter of which naturally yields a posterior distribution 
over the space of parton densities. Recent developments have further enhanced 
the NNPDF framework through the incorporation of Bayesian reweighting 
techniques, which allow efficient updating of PDF ensembles when new 
experimental data become available without complete 
refitting~\cite{Ball:2010gb,Ball:2011gg}. The weight for each replica given 
new data $D_{\rm new}$ is computed as $w_k \propto \exp(-\chi^2_k/2)$, where 
$\chi^2_k$ measures the agreement between replica $k$ and the new data. This 
enables rapid assessment of the impact of new data from future measurements 
(including EICs) on global PDF constraints by updating the probability 
density distribution in the fit.

A physically motivated extension of the standard NN framework is the 
Physics-Informed Neural Network (PINN) approach~\cite{Aarts:2025gyp}, in which the loss function 
is augmented by an additional penalty term encoding known QCD 
constraints,
\begin{equation}
\mathcal{L} \;=\; \mathcal{L}_{\rm data} \;+\; \lambda\,\mathcal{L}_{\rm phys},
\end{equation}
where $\mathcal{L}_{\rm data}$ is the standard $\chi^2$ data loss and 
$\mathcal{L}_{\rm phys}$ penalizes violations of physical constraints --- 
such as the momentum and valence quark sum rules, positivity bounds on the 
distributions, and the residual of the DGLAP evolution equations --- directly 
during the training process, rather than enforcing them only as post-fit 
checks. The key distinction between a standard NN and a PINN therefore lies 
in where the physics enters: in the standard approach the network is a 
free-form interpolator and physical constraints are verified after training, 
whereas in the PINN approach the network is continuously steered toward 
physically consistent solutions throughout optimization. This encourages physical consistency of the extracted distributions during training, even in kinematic regions where experimental data are sparse, which is particularly relevant for GPD extractions from the limited DVCS datasets currently available, and for TMD extractions where perturbative matching conditions at large transverse momentum (or hard scale $Q$) must be satisfied. The extracted distributions are finally 
validated against physical sum rules, positivity constraints, closure tests 
on pseudo-data, and comparisons with lattice QCD moments, before being used 
to produce phenomenological predictions for existing and future experiments 
at facilities such as the Electron-Ion Collider at BNL (US-EIC), EicC, and JLab.

\textbf{Extraction of GPDs}: Beyond collinear PDFs, neural network methods have been extended to the extraction of GPDs, which encode the three-dimensional structure of hadrons by correlating longitudinal momentum and transverse spatial distributions of partons~\cite{Mueller:1998fv,Ji:1996ek,Radyushkin:1997ki}. GPDs depend on three kinematic variables: the average longitudinal momentum fraction $x$, the longitudinal momentum transfer $\xi$ (skewness), and the squared momentum transfer $t$.

The application of neural networks to GPD phenomenology has evolved significantly over the past decade. The pioneering work in 2011 introduced neural network parametrizations of CFFs from DVCS data~\cite{Kumericki:2011rz}, demonstrating the feasibility of model-independent CFF extraction. The CFFs, which are convolutions of GPDs with hard scattering kernels, can be parametrized using neural networks and fitted to experimental cross sections and asymmetries. For the dominant GPD $H$, the CFF $\mathcal{H}$ is related to the GPD through:
\begin{equation}
\mathcal{H}(\xi, t, Q^2) = \sum_q e_q^2 \int_{-1}^{1} dx \left[ \frac{1}{x - \xi - i\epsilon} + \frac{1}{\xi + x - i\epsilon} \right] H^q(x, \xi, t, Q^2), \label{gpdk}
\end{equation}
where the sum runs over quark flavors with charges $e_q$. The neural network approach is particularly valuable for GPD extraction because the inverse problem of deconvolving GPDs from CFFs is inherently ill-posed, and the flexibility of neural networks allows exploration of the full space of solutions consistent with data.

This initial framework was subsequently refined and extended to comprehensive global analyses. In 2016, a systematic fitting procedure incorporating DVCS and DVMP data was developed~\cite{Kumericki:2016ehc}, establishing the foundation for modern GPD phenomenology. The methodology was further advanced in 2018 with the introduction of border and skewness functions that parametrize the $x$ and $\xi$ dependence of GPDs~\cite{Moutarde:2018kwr}, providing more flexible and model-independent constraints on nucleon structure. In 2022, the artificial neural network modeling was used to directly parametrize GPDs rather than just CFFs~\cite{Dutrieux:2021wll}, and it enables more direct access to the underlying parton distributions.

Recent years have witnessed rapid progress in deep learning applications to GPD extraction. The EXCLusives with AI and Machine learning (EXCLAIM) collaboration is set to develop a framework to implement AI and machine learning techniques and extract CFFs and GPDs from existing data and future measurements in exclusive scattering processes~\cite{Liuti:2024zkc}. In 2025, global deep neural network modeling of CFFs constrained from local $\chi^2$ maps was introduced~\cite{CaleroDiaz:2025luc}, combining the strengths of traditional fitting methods with neural network flexibility. Most recently, direct deep neural network extraction of GPDs from CFFs has been achieved~\cite{Watkins:2025apc}, representing a nonparametric approach to solving the inverse problem of deconvolving GPDs from their convolutions with hard scattering kernels (Eq.~\ref{gpdk}), without assuming specific functional forms for the $x$, $\xi$, and $t$ dependences. Complementary to DVCS measurements, Bayesian inference methods have been applied to extract GFFs (the Mellin moments of GPDs) from near-threshold $J/\psi$ photoproduction data~\cite{Guo:2025jiz}, where the GPD framework at next-to-leading order provides access to both quark and gluon contributions. The combined analysis of experimental data from J/$\psi$-007 and GlueX at JLab with lattice QCD constraints demonstrates the power of advanced statistical methods in constraining the momentum and spatial distributions of partons.

For EicC, these NN and Bayesian inference techniques will be 
essential for extracting GPDs and TMDs from the facility's 
dedicated measurements of DVCS and DVMP processes. Operating 
at center-of-mass energies of $\sqrt{s} \approx 15$--$20$~GeV, 
EicC will provide high-statistics data in the valence and 
sea quark transition region ($0.01 \lesssim x \lesssim 0.3$), 
where current experimental constraints remain limited. In 
particular, the precise determination of the gluon GPD, 
which governs the spatial distribution of gluons inside the 
nucleon, requires the kind of multidimensional neural network 
fits that can simultaneously handle the $x$, $\xi$, and $t$ 
dependences of the DVMP cross section. The high luminosity 
and well-controlled systematic uncertainties anticipated at 
EicC will directly reduce the dominant sources of uncertainty 
in these extractions, enabling more reliable tomographic 
imaging of the nucleon's gluon content than is currently 
possible.

\textbf{Neural Network extraction of TMDs}: The application of deep neural networks has also been extended to TMDs, which describe the three dimensional momentum structure of partons inside hadrons. Studies in Refs.~\cite{Fernando:2023obn, Bacchetta:2025ara, Fernando:2025xzv} pioneered the NN extraction of TMDs by applying deep learning techniques to unpolarized and Sivers quark distributions from experimental Drell-Yan data, using NNs to parametrize their nonperturbative parts. It was shown that NNs can outperform traditional parametrizations providing a more accurate description of data by demonstrating the feasibility and advantage of using NNs to explore the multi-dimensional partonic structure of hadrons.

In particular, Fernando and Keller applied the DL techniques to the Sivers function, one of the eight leading twist TMDs for polarized nucleons~\cite{Fernando:2023obn}. The Sivers function $f_{1T}^{\perp q}(x, k_\perp^2)$ describes the correlation between the transverse spin of the nucleon and the transverse momentum of unpolarized quarks, with a nonzero Sivers function indicating a contribution from quark OAM to the nucleon spin. The NN extraction constructs a minimally biased model by training on COMPASS and HERMES SIDIS data, with the Sivers asymmetry expressed as:
\begin{equation}
A_{UT}^{\sin(\phi_h - \phi_S)} = \frac{\sum_q e_q^2 \, f_{1T}^{\perp q}(x, k_\perp^2) \otimes D_h^q(z, p_\perp^2)}{\sum_q e_q^2 \, f_1^q(x, k_\perp^2) \otimes D_h^q(z, p_\perp^2)},
\end{equation}
where $D_h^q$ denotes the unpolarized fragmentation function and $\otimes$ represents the appropriate TMD convolution. The trained NN successfully reproduces the measured asymmetries and provides predictions for Drell-Yan kinematics. Building upon this foundation, Fernando and Keller subsequently developed a momentum space, physics-informed deep learning framework for extracting unpolarized TMDs directly from Drell-Yan data~\cite{Fernando:2025xzv}. Unlike conventional approaches that transform to impact parameter ($b_T$) space to simplify the convolution structure, this method operates directly in transverse momentum ($k_\perp$) space, avoiding potential biases from the Fourier transform. Applied to Fermilab E288 and E605 fixed target Drell Yan data, the method reproduces measured $q_T$ spectra across different invariant masses and yields TMDs that exhibit the expected $Q$ dependent broadening from TMD evolution. Uncertainties from experimental errors, PDF inputs, and methodological choices are propagated through Monte Carlo replicas, providing rigorous uncertainty quantification. 

The MAP Collaboration recently presented a comprehensive neural-network-based extraction of unpolarized quark TMD parton distribution functions from Drell-Yan data~\cite{Bacchetta:2025ara}. Neural networks are employed to parametrize the nonperturbative component of TMD PDFs, while the perturbative part is calculated at next-to-next-to-next-to-leading logarithmic (N$^3$LL) accuracy. The analysis includes Drell-Yan data from fixed-target experiments (Fermilab E605, E288, E772), colliders at Tevatron (CDF, D0) and RHIC (STAR), and the LHC (LHCb, CMS, ATLAS). The neural network parametrization demonstrates superior fit quality compared to traditional functional forms, particularly for high-precision LHC measurements. This proof-of-concept study establishes the feasibility of machine-learning techniques for extracting the multi-dimensional partonic structure of hadrons, with future extensions planned to incorporate semi-inclusive deep-inelastic scattering data and TMD flavor dependence.

A recent study~\cite{Fernando:2026vqf} explores the potential of Quantum Deep Neural Networks (QDNNs) for addressing challenges in hadronic tomography, particularly the extraction of CFFs and related distributions from sparse and noisy experimental data. The work proposes a hybrid quantum-classical architecture that combines a quantum simulator, which encodes physics-informed priors based on QCD structure, with a classical neural network that models experimental effects. The study concludes that hadronic tomography offers a testable program for selective quantum advantage, where the near-term approach uses quantum circuits to generate states and structured priors while classical layers handle nuisance parameters and perform uncertainty-aware inference. These QDNN technological advances in hadronic tomography could have a significant impact on nuclear physics experiments at facilities such as the US-EIC and EicC.

\textbf{Probing gluon saturation with AI/ML techniques}: One of the important goals of EICs is the exploration of the gluon saturation regime, where the density of gluons inside hadrons and nuclei becomes so large that nonlinear QCD dynamics become dominant. At small values of Bjorken $x$, the gluon distribution function $xg(x, Q^2)$ grows rapidly as $x$ decreases, following the DGLAP and BFKL evolution equations~\cite{Gribov:1972ri,Altarelli:1977zs,Dokshitzer:1977sg,Balitsky:1978ic} in collinear and $k_T$ factorizations, respectively. However, this growth cannot continue indefinitely without violating unitarity bounds. The resolution lies in gluon recombination processes ($gg \rightarrow g$), which tame the growth at sufficiently high densities. This gives rise to the saturation scale $Q_s(x)$, which marks the boundary between the dilute regime governed by linear QCD evolution and the dense regime where nonlinear effects dominate. The saturation scale can be parametrized as $Q_s^2(x) \sim Q_0^2 (x_0/x)^\lambda$, where $\lambda \approx 0.2$--$0.3$ characterizes the speed of evolution~\cite{Golec-Biernat:1998zce,Golec-Biernat:1999qor}. For nuclear targets, the saturation scale is enhanced by a factor proportional to $A^{1/3}$, making heavy nuclei particularly attractive for saturation searches at EICs. Geometric scaling of cross sections where observables depend on the single variable $Q^2/Q_s^2(x)$ rather than on $Q^2$ and $x$ independently~\cite{Stasto:2000er} is considered one of the most compelling signatures of saturation effect. The US-EIC facilities provide an ideal environment for saturation studies through clean electromagnetic probes with precise kinematic control and the ability to vary both center-of-mass energy and nuclear target species. However, identifying saturation effects is complicated by subtle modifications to distributions and the need to disentangle saturation from other nuclear effects such as shadowing, energy loss, and CNM modifications.

The theoretical framework for describing the saturated gluon regime is the CGC effective field theory~\cite{Gelis:2010nm,Albacete:2014fwa}. In this framework, small-$x$ gluons are treated as classical color fields generated by fast-moving color sources at larger $x$ values, with dynamics governed by the JIMWLK evolution equation~\cite{Jalilian-Marian:1997qno, Jalilian-Marian:1997jhx, Iancu:2000hn, Ferreiro:2001qy} or its alternative formulation, the Balitsky-Kovchegov (BK) equation~\cite{Balitsky:1995ub,Kovchegov:1999yj}. The BK equation describes the evolution of the dipole scattering amplitude $N(r, Y)$ with rapidity $Y = \ln(1/x)$, incorporating both linear BFKL growth and nonlinear saturation effects. A crucial development has been the formulation of transverse momentum dependent factorization in the saturation regime. This allows one to build the the theoretical foundation for dijet correlations in deep inelastic scattering off nuclei and proton-nucleus collisions~\cite{Marquet:2007vb,Dominguez:2010xd, Dominguez:2011wm}. These developments provide the foundation for precision tests of gluon saturation at EICs, where the suppression of back-to-back dihadron correlations serves as a golden signature of the saturated gluon regime.

Artificial intelligence and machine learning techniques, particularly those based on Bayesian inference, offer powerful new approaches to address various challenges in probing saturation effects in high-energy collisions. In saturation physics, Bayesian methods enable systematic extraction of saturation scale parameters and their uncertainties from measured cross sections while properly accounting for theoretical and experimental uncertainties, and thus provide not only best-fit values but also rigorous uncertainty quantification essential for claiming discovery of saturation effects. ML classifiers trained on theoretical predictions from CGC calculations and conventional linear evolution frameworks can distinguish between these scenarios, identifying combinations of observables with maximal discriminating power~\cite{Allaire:2023fgp}. Neural networks can serve as powerful surrogate models for computationally intensive CGC calculations, enabling rapid evaluation of theoretical predictions across large parameter spaces required for Bayesian inference. Furthermore, extracting fundamental properties of the saturated gluon distribution from measured final-state observables constitutes an inverse problem naturally suited to ML techniques. Simulation-based inference methods, which combine neural density estimation with forward simulations, are particularly promising for saturation studies where the likelihood function is intractable but samples from the theoretical model can be readily generated. For EicC, which operates at lower center-of-mass energies compared to the US-EIC, accessing the saturation regime relies more heavily on nuclear enhancement effects in electron-nucleus collisions, making precision AI-assisted analysis techniques essential for maximizing sensitivity to saturation phenomena in the available kinematic range.

A significant recent advance in applying AI techniques to saturation physics is the development of physics-informed neural networks (PINNs) for extracting the universal dipole amplitude. Dai et al.~\cite{Dai:2026nzp} first introduced a PINN-based global analysis that determines the dipole scattering amplitude $N(r,x_B)$ as a differentiable function without imposing an a priori parametrization of the initial condition. In this framework, the collinearly improved BK evolution equation is embedded directly into the loss function as a physics constraint, while the network is simultaneously trained on DIS reduced cross sections, charm-production data, and exclusive $J/\psi$ photoproduction measurements, together with positivity constraints on the momentum-space dipole amplitude. More recently, the same framework was extended to nuclei in Ref.~\cite{Dai:2026rgl}, where the impact-parameter-averaged $^{208}\mathrm{Pb}$ dipole amplitude was extracted from forward-hadron nuclear-modification-factor and coherent $J/\psi$ photoproduction data without assuming any parametric form for the nuclear initial condition. The analysis yielded a saturation-scale ratio $Q_{s0,\mathrm{Pb}}^2/Q_{s0,p}^2 \simeq 3.2$, consistent with geometric expectations, and found that the extracted nuclear initial condition is well described by a McLerran-Venugopalan model~\cite{McLerran:1993ni,McLerran:1993ka} form. Together, these studies demonstrate how machine learning can go beyond conventional fitting strategies by incorporating nonlinear QCD evolution and physical constraints directly into the training procedure, thereby providing smooth, positive-definite, and phenomenologically robust dipole amplitudes for CGC applications. Such methods are expected to play an important role in extracting saturation physics from the precision measurements anticipated at future EIC facilities.

While EicC operates at lower center-of-mass energies ($15-20$ GeV) compared to the US-EIC, it provides crucial complementary insights into CGC physics by probing the initial conditions of gluon saturation. Specifically, EicC's measurements of gluon distributions in heavy nuclei at intermediate Bjorken-$x$ values around $x \sim 0.01$ serve as essential boundary conditions for understanding the evolution toward the saturated regime at smaller $x$, enabling precise constraints on the initial state of dense gluonic matter that evolves into the CGC at higher energies and smaller $x$ values accessible at the US-EIC.

While many of the AI/ML techniques discussed above were developed in the context of the US EIC program, several of them find particularly natural and well-motivated applications at EicC owing to its distinct physics program and kinematic regime. EicC operates at center-of-mass energies of 15 to 20 GeV with high-luminosity polarized beams, placing it in a regime where TMD and GPD measurements demand exceptional control of systematic uncertainties, a challenge that ML-based global QCD analyses are uniquely suited to address, enabling unbiased extraction of multi-dimensional parton distributions with rigorous uncertainty quantification. The EicC detector's reliance on RICH technology for $\pi/K/p$ separation makes fast generative simulation of Cherenkov photon propagation not merely convenient but essential for large-scale detector optimization and physics projections, where full Geant4 simulation would be computationally prohibitive. Furthermore, EicC's focus on exclusive processes such as DVCS and DVMP requires precise reconstruction of final states with low multiplicity but stringent kinematic constraints, making GNN-based track finding and vertex reconstruction particularly well-suited to the signal topology. Finally, while EicC does not access the deeply small-$x$ regime, its measurements of electron-nucleus collisions can provide crucial constraints on the initial conditions for CGC evolution, a role that is further reinforced by recent PINN-based studies that directly solve the BK evolution equation and for which EicC data would serve as valuable boundary inputs.

Looking further ahead, foundation models are large pretrained transformers trained in a self-supervised or generative manner on broad physics datasets, and they offer a promising paradigm for EicC analyses. Analogous to OmniJet-$\alpha$~\cite{Birk:2024knn}, which demonstrated transfer learning from jet generation to jet tagging, a foundation model pretrained on the full body of EicC-relevant data encompassing simulated events, existing HERA and JLab measurements, and lattice QCD inputs could be fine-tuned for individual downstream tasks such as Sivers function extraction, CFF fitting, or nuclear modification studies with minimal additional training data. Given the rapid pace of development in this area, we expect that further dedicated applications will emerge and mature alongside the EicC technical design.

\newpage
 \section{Summary and outlook}\label{sec:sum}

The EicC represents a transformative facility in the global landscape of nuclear and hadronic physics. Anchored at the HIAF in Huizhou, and operating at a center-of-mass energy of $\sqrt{s} = 15$--$20$ GeV with luminosity $(2$--$4)\times 10^{33}$ cm$^{-2}$s$^{-1}$, the EicC is uniquely positioned to address some of the most profound open questions in QCD through precision measurements in the intermediate-to-large $x$ regime ($0.005 < x < 0.3$).

This review has presented the broad and interconnected physics program of the EicC across several frontier topics.

\textbf{One-dimensional spin structure.} The EicC will deliver precision measurements of helicity distributions $\Delta f(x, Q^2)$ for quarks and gluons through doubly polarized SIDIS with $e$-$p$ and $e$-$^3$He beams. With 50 fb$^{-1}$ of integrated luminosity, uncertainties in sea quark helicity distributions will be reduced by nearly an order of magnitude for $x > 0.01$, providing the first reliable constraints on the polarized strange sea and the first precise determination of $\Delta g/g$ in the large-$x$ region via the photon-gluon fusion process. These measurements will directly confront the long-standing proton spin puzzle, where quark spins are known to account for only $\sim 30\%$ of the total proton spin.

\textbf{Three-dimensional tomography.} Through DVCS, DVMP, and SIDIS, the EicC will map the full three-dimensional landscape of the proton in both position and momentum space. The kinematic range $1~\text{GeV}^2 < Q^2 < 30~\text{GeV}^2$ makes EicC well suited for studying GPDs and TMDs in the sea quark region, where current knowledge is most limited. Precise measurements of the transverse spin asymmetry $A_{UT}$ in DVCS will substantially reduce uncertainties in the Compton form factors $H$ and $E$, enabling a first reliable extraction of quark orbital angular momentum via the Ji sum rule. The EicC can also provide the first measurements of sea quark worm-gear functions and chiral-odd GPDs through $\pi^0$ DVMP asymmetries, completing the leading-twist TMD and GPD landscapes.

\textbf{Origin of nucleon mass.} By measuring gravitational form factors through threshold $J/\psi$ photoproduction and studying the trace anomaly contribution via the energy-momentum tensor, the EicC will shed new light on the dynamical origin of the proton mass, which is only $\sim 1\%$ attributable to the Higgs mechanism. The gluon mass radius and its relation to the charge radius will be precisely determined, testing the emerging picture of a dense gluonic core at the center of the proton.

\textbf{Cold nuclear matter and exotic states.} With a flexible program of electron-nucleus collisions spanning from deuterium to uranium, the EicC will map nuclear modifications of parton distributions across the EMC, shadowing, and anti-shadowing regions, and constrain the jet transport parameter $\hat{q}$ for CNM in the intermediate-$x$ regime. The EicC's energy range is also well suited for searches for hybrid mesons and other exotic QCD states in exclusive production channels.

\textbf{Quantum information and novel observables.} The EicC will serve as a new laboratory for quantum information science at high energies. The spin degrees of freedom of produced quark-antiquark pairs in 
$\gamma^* g \to q\bar{q}$ and $\gamma^* \text{Pomeron} \to q\bar{q}$ (diffractive 
scatterings) provide natural qubits for testing Bell nonlocality and measuring 
quantum entanglement. Longitudinally polarized virtual photons produce maximally 
entangled quark pairs ($\mathcal{C} = 1$) regardless of kinematics, while for 
transversely polarized photons the degree of entanglement varies across kinematic 
regions, being maximal near the production threshold and in the ultra-relativistic 
limit, allowing the EicC to map the entanglement structure of QCD over a broad and 
continuous kinematic landscape. Furthermore, the entanglement entropy arising from 
information loss when the unobserved nuclear remnant is traced out can be studied 
in detail across different $x$ and $Q^2$ regimes, connecting the non-perturbative 
partonic structure of the proton to the fundamental concept of quantum information loss 
in high-energy scattering. These novel observables open a new interdisciplinary 
frontier connecting QCD and quantum information theory.

\textbf{AI and machine learning.} The EicC program will be deeply integrated with modern AI and ML techniques, from fast detector simulation using generative models to physics-informed neural networks for solving the ill-posed inverse problems of extracting PDFs, GPDs, and TMDs from experimental data. These tools will be essential for handling the high-dimensional, high-luminosity data environment and for providing rigorous uncertainty quantification.

\subsection*{Outlook}

The EicC is designed as a facility complementary to, rather than duplicating, the US-EIC at Brookhaven. While the US-EIC will explore the high-energy frontier of gluon saturation and the CGC, the EicC will deliver the most precise three-dimensional images of the sea quark and valence-to-sea transition region, filling a critical gap in the global QCD program. Together, these two facilities will provide a comprehensive, multi-scale understanding of hadron structure from first principles.

Looking forward, the completion of the EicC Conceptual Design Report in 2026 marks a pivotal milestone. The next steps toward construction will require continued advances in polarized beam technology, detector design, and theoretical frameworks that connect measurements across different energy regimes. In particular, bridging the gap among collinear factorization, GPD and TMD formalisms, and small-$x$ formalisms will be essential for a unified interpretation of the rich dataset that the EicC will produce.

Beyond refining existing knowledge, the EicC will open new windows onto the non-perturbative dynamics of QCD, the quantum mechanical nature of the strong interaction, and the emergent properties of CNM. By addressing the fundamental questions of how mass, spin, and spatial structure arise from quarks and gluons, the EicC will make lasting contributions to our understanding of the visible universe.

	\section*{Acknowledgements}  
	We thank Zuo-Tang Liang, Jie Liu, Yu-Xuan Liu, Jian-Ping Ma, Wei Qi, Enke Wang, Hongxi Xing, Nu Xu, and Kai Zhou for comments and discussions. We are grateful to Prof. Mei Huang for suggesting this review article. This work was supported in part by the Ministry of Science and Technology of China under Grant No. 2024YFA1611004, by the Natural Science Foundation of Guangdong Province under Grant No.~2026A1515011242, by the CUHK (Shenzhen) University Development Fund under Grant No.~UDF01001859, by the CAS Project for Young Scientists in Basic Research under Grant No. YSBR-117, and by National Science Foundations of China under Grant No.~12175118, and No.~12321005.

	\bibliography{mybibfile}
	
	


	
	\newpage
	\appendix
	\renewcommand*{\thesection}{\Alph{section}}
	\section{Fundamentals of quantum information theory}\label{appendix}

    This appendix provides a self-contained introduction to the key concepts
of quantum information theory that underpin the discussions in the main
text, with particular emphasis on their application to high-energy and
nuclear physics. We begin with the historical and conceptual foundations
of the field, tracing the debate over the interpretation of quantum
mechanics from the Einstein-Podolsky-Rosen (EPR) paradox~\cite{Einstein:1935rr}
and its spin-based reformulation by Bohm (EPRB), through Bell's
theorem~\cite{Bell:1964kc} and the Clauser-Horne-Shimony-Holt (CHSH)
inequality~\cite{Clauser:1969ny}, to the experimental confirmation of
quantum nonlocality. We then introduce the quantum information formalism
required for the analysis of spin correlations in particle collisions:
the qubit and its Bloch sphere representation, the density matrix
framework for both pure and mixed states, and
the von Neumann entropy as a measure of quantum entanglement. For
two-qubit systems, we introduce the correlation matrix parametrization
of the spin density matrix, the Peres-Horodecki separability
criterion~\cite{Peres:1996dw,Horodecki:1996nc}, and the concurrence as
an analytically tractable entanglement measure~\cite{Wootters:1997id,
Hill:1997pfa}, elucidating its deep connection to time-reversal symmetry.
The CHSH inequality and the Horodecki condition~\cite{Horodecki:1995nsk}
for its maximal violation are then derived, with the Tsirelson
bound~\cite{Cirelson:1980ry} identified as the quantum-mechanical
ceiling. Finally, we discuss decoherence in the context of high-energy
collisions, where renormalization group evolution acts as a quantum
channel that drives the irreversible loss of quantum coherence in the
observable sector~\cite{Semenoff:2019dqe}.

\textbf{The EPR paradox and interpretations of quantum mechanics:} The debate over the interpretation of quantum mechanics dates back to the
earliest days of the theory, with the famous exchanges between Bohr and
Einstein at the Solvay Conferences of 1927 and 1930, and was brought into
sharp focus by the EPR paradox in 1935~\cite{Einstein:1935rr}, which has since played a pivotal role in shaping our understanding of quantum mechanics.

\textbf{Three philosophical positions on quantum reality:}
The interpretation of quantum mechanics has historically been divided into three main philosophical positions:
\begin{enumerate}
    \item \textit{Realism:} This view, advocated by Einstein, holds that physical properties of objects exist objectively and independently of observation. Einstein famously encapsulated this position with the rhetorical question: \textit{``Is the moon there when nobody looks?''}~\cite{Mermin:1985ks}. According to this perspective, the moon possesses definite properties, such as position, momentum, and other physical attributes, whether or not any observer measures them. Einstein argued that quantum mechanics, with its inherently probabilistic predictions, must be incomplete. He proposed that additional ``hidden variables'' should exist to determine measurement outcomes, restoring determinism and local realism to physics.
    \item \textit{Copenhagen interpretation:} In stark contrast, the orthodox Copenhagen interpretation asserts that physical properties do not have definite values until they are measured. The act of measurement causes the wave function to ``collapse'' from a superposition of possibilities to a single definite outcome. From this viewpoint, quantum mechanics cannot provide a definite description of a physical system prior to measurement. If the ``moon'' were a microscopic object, it would not, in a sense, have a definite position until someone looks. 
    \item \textit{Agnostic/Pragmatic view:} A third position refuses to answer and avoids metaphysical questions about what exists when unobserved. This pragmatic approach focuses solely on the predictive power of quantum mechanics, treating the formalism as a tool for computing observable quantities without committing to any particular position about the nature of reality in the absence of measurement.
\end{enumerate}

\textbf{The EPRB thought experiment:} The original EPR argument was later reformulated by David Bohm into a more concrete form known as the EPRB experiment~\cite{Bohm:1957zz}. Consider the ``gedanken'' decay of a neutral pion at rest into an electron-positron pair:
\begin{equation}
\pi^0 \to e^- + e^+.
\end{equation}
Since the pion has zero spin, conservation of angular momentum requires the electron-positron pair to be produced in a spin-singlet state:
\begin{equation}
|\Psi^-\rangle = \frac{1}{\sqrt{2}}\left(\ket{\uparrow\downarrow} - \ket{\downarrow\uparrow}\right).
\end{equation}
This state predicts perfect anti-correlation: if the electron's spin is measured along any axis and found to be up, the positron's spin along the same axis is guaranteed to be down, and vice versa.

Suppose the electron and positron are later separated to a macroscopic distance, and the electron's spin is measured along a chosen axis. Once the result is obtained, the outcome of a spin measurement on the distant positron along the same axis becomes certain. From a realist viewpoint, this presents no conceptual difficulty: the spins of both particles had definite values from the moment of their creation, and measurement merely reveals these pre-existing properties. In this sense, quantum mechanics is incomplete because the wave function does not encode all elements of physical reality.

The Copenhagen interpretation takes a different stance. It holds that neither particle has a definite spin value before measurement. Measuring the electron collapses the joint wave function and thereby instantaneously determines the spin of the positron, even when the particles are spacelike separated. 

The EPR paradox thus exposes a fundamental tension: either quantum mechanics is incomplete (requiring hidden variables), or it involves nonlocal correlations that appear to violate the spirit of special relativity, although communication faster than the speed of light is not possible. Einstein found the latter option deeply troubling, famously referring to it as ``spooky action at a distance''.

\textbf{Bell's theorem: Making the debate testable}: For nearly three decades, the EPR paradox remained a philosophical debate with no clear experimental resolution. The decisive breakthrough came in 1964 when John Bell demonstrated that any local hidden variable (LHV) theory satisfying Einstein's locality principle makes predictions that are incompatible with quantum mechanics~\cite{Bell:1964kc}. Bell's theorem transformed the EPR paradox from a philosophical argument into an experimentally testable proposition.

Bell showed that LHV theories must satisfy certain inequalities, which is now known as Bell inequalities, that constrain the correlations between measurements on spatially separated systems. Quantum mechanics, however, predicts violations of these inequalities for entangled states. Subsequent experiments have consistently confirmed the predictions of quantum mechanics and ruled out local realism. The 2022 Nobel Prize in Physics was awarded to Alain Aspect, John Clauser, and Anton Zeilinger for their pioneering experimental tests~\cite{Aspect:1982fx,Clauser:1969ny,Weihs:1998gy} of Bell inequalities.

The violation of Bell inequalities demonstrates that nature exhibits genuine quantum nonlocality: correlations between entangled particles cannot be explained by any theory in which measurement outcomes are predetermined by local hidden variables. This does not, however, permit faster-than-light communication, as the outcome of the measurements is probabilistic and the correlations become apparent only when measurement results from both locations are compared, which requires classical communication.

\textbf{Qubits and single-particle states}: In quantum information theory (see Refs.~\cite{Benenti:2019non,Nielsen:2012yss,Horodecki:2009zz} for an introduction), the fundamental unit of quantum information is the qubit, a two-level quantum system that generalizes the classical bit. While a classical bit can only take values $0$ or $1$, a qubit can exist in a coherent superposition of both basis states simultaneously. Mathematically, a general single-qubit pure state is written as:
\begin{equation}
|\psi\rangle = \alpha|0\rangle + \beta|1\rangle,
\end{equation}
where $|0\rangle$ and $|1\rangle$ form an orthonormal basis called the computational basis, and $\alpha, \beta \in \mathbb{C}$ are complex amplitudes satisfying the normalization condition $|\alpha|^2 + |\beta|^2 = 1$. Upon measurement in the computational basis, the qubit collapses to $|0\rangle$ with probability $|\alpha|^2$ or to $|1\rangle$ with probability $|\beta|^2$.

In high-energy and nuclear physics, qubits can be naturally realized through the spin degrees of freedom of spin-$1/2$ particles such as electrons, quarks, and their antiparticles. The spin states along a chosen quantization axis (conventionally the $z$-axis) serve as the computational basis:
\begin{equation}
|0\rangle \equiv \ket{\uparrow} = \begin{pmatrix} 1 \\ 0 \end{pmatrix}, \qquad |1\rangle \equiv \ket{\downarrow} = \begin{pmatrix} 0 \\ 1 \end{pmatrix}.
\end{equation}
A general spin state of a spin-$1/2$ particle is usually parametrized using the Bloch sphere representation:
\begin{equation}
|\psi\rangle = \cos\frac{\theta}{2}|0\rangle + e^{i\phi}\sin\frac{\theta}{2}|1\rangle,
\end{equation}
where $\theta \in [0, \pi]$ and $\phi \in [0, 2\pi)$ are the polar and azimuthal angles on the Bloch sphere, respectively. This geometric representation makes manifest that the space of pure single-qubit states forms a two-dimensional sphere $S^2$.

\textbf{Density matrix formalism}: While pure states provide a complete description of isolated quantum systems with maximal knowledge, physical systems often involve statistical mixtures or entanglement with external degrees of freedom that cannot be described by a single wave function. A mixed state represents either (1) a classical statistical ensemble where the system is in one of several possible pure states but we lack complete information about which one, or (2) a subsystem that is entangled with an environment, where tracing out the environmental degrees of freedom necessarily produces a mixed state even if the total system is pure. In both cases, the state cannot be written as $|\psi\rangle$ for any single wave function $|\psi\rangle$, but rather requires a probabilistic combination of multiple pure states. The density matrix formalism provides a unified framework for describing both pure and mixed quantum states (see, for example, Ref.~\cite{Sakurai:2011zz}).

For a pure state $ |\psi \rangle$, the density matrix is defined as the projector $\rho = |\psi\rangle\langle\psi|$, and the expectation value of any observable $\hat{A}$ can be computed as:
\begin{equation}
\langle A \rangle = \text{Tr}(\rho \hat{A}).
\end{equation}
A pure-state density matrix satisfies the idempotency condition $\rho^2 = \rho$, or equivalently $\text{Tr}(\rho^2) = 1$.

More generally, a mixed state arises when the system is described by a statistical ensemble of pure states $\{|\psi_i\rangle\}$ with classical probabilities $\{p_i\}$. The corresponding density matrix is written as a convex combination as follows
\begin{equation}
\rho = \sum_i p_i |\psi_i\rangle\langle\psi_i|,
\end{equation}
where $p_i \geq 0$ and $\sum_i p_i = 1$. A valid density matrix must satisfy three properties: (i) Hermiticity: $\rho = \rho^\dagger$; (ii) positive semi-definiteness: $\langle\phi|\rho|\phi\rangle \geq 0$ for all $|\phi\rangle$; and (iii) unit trace: $\text{Tr}(\rho) = 1$. The purity $\text{Tr}(\rho^2)$ quantifies the degree of mixedness, ranging from $1$ for pure states to $1/d$ for maximally mixed states in a $d$-dimensional Hilbert space.

For a single qubit, the density matrix can be cast into the Bloch representation:
\begin{equation}
\rho = \frac{1}{2}\left(\mathbb{I} + \vec{r} \cdot \vec{\sigma}\right) = \frac{1}{2}\begin{pmatrix} 1 + r_z & r_x - ir_y \\ r_x + ir_y & 1 - r_z \end{pmatrix},
\end{equation}
where $\vec{\sigma} = (\sigma_x, \sigma_y, \sigma_z)$ are the Pauli matrices:
\begin{equation}
\sigma_x = \begin{pmatrix} 0 & 1 \\ 1 & 0 \end{pmatrix}, \quad \sigma_y = \begin{pmatrix} 0 & -i \\ i & 0 \end{pmatrix}, \quad \sigma_z = \begin{pmatrix} 1 & 0 \\ 0 & -1 \end{pmatrix},
\end{equation}
and $\vec{r} = (r_x, r_y, r_z)$ is the Bloch vector satisfying $|\vec{r}| \leq 1$. Pure states correspond to $|\vec{r}| = 1$ (points on the Bloch sphere surface with unit normalization), while mixed states have $|\vec{r}| < 1$ (points inside the Bloch sphere). The components of the Bloch vector are related to the spin polarization of the spinor: $r_i = \text{Tr}(\rho \sigma_i) = \langle \sigma_i \rangle$.

\textbf{Von Neumann entropy and entanglement}: The von Neumann entropy provides a fundamental measure of quantum entanglement and information content in quantum systems. For a quantum system described by a density matrix \(\rho\), the von Neumann entropy is defined as
\begin{equation}
S = -\text{Tr}(\rho \ln \rho).
\end{equation}
Since the trace is basis-independent, this can be equivalently expressed in 
the eigenbasis of $\rho$ in terms of its eigenvalues $\{\lambda_i\}$ as 
$S = -\sum_i \lambda_i \ln \lambda_i$, where the sum runs over all eigenvalues. 
For a pure state $|\psi\rangle$ with $\rho = |\psi\rangle\langle\psi|$, the 
von Neumann entropy vanishes, $S = 0$, reflecting the fact that one has 
complete information about the system; the density matrix of a pure state has 
a single nonzero eigenvalue $\lambda_1 = 1$, with all others vanishing. 
This serves as a baseline reference for the subsequent discussion of entanglement entropy and quantum information content.

When a pure quantum system is divided into subsystems \(A\) and \(B\), and we have access only to subsystem \(A\), we must describe it using the reduced density matrix obtained by tracing over the degrees of freedom in \(B\), yielding $\rho_A = \text{Tr}_B \rho$. The von Neumann entropy of this reduced density matrix,
\begin{equation}
S_A = -\text{Tr}(\rho_A \ln \rho_A),
\end{equation}
quantifies the entanglement between subsystems \(A\) and \(B\): it measures how much information about \(A\) is stored in the correlations with \(B\), or equivalently, how much information is lost when we ignore \(B\). For a bipartite pure state, the entanglement entropy is symmetric:
\begin{equation}
S_A = S_B.
\end{equation}

To illustrate maximal entropy, consider the single-qubit Bloch sphere representation. The maximally mixed state corresponds to the center of the Bloch sphere with $\vec{r} = 0$, giving the density matrix
\begin{equation}
\rho_{\text{max}} = \frac{1}{2}\mathbb{I} = \frac{1}{2}\begin{pmatrix} 1 & 0 \\ 0 & 1 \end{pmatrix}.
\end{equation}
This density matrix has two equal eigenvalues $\lambda_1 = \lambda_2 = 1/2$, corresponding to the two basis states $|0\rangle$ and $|1\rangle$ occurring with equal probability. The von Neumann entropy is
\begin{equation}
S_{\text{max}} = -\left(\frac{1}{2}\ln\frac{1}{2} + \frac{1}{2}\ln\frac{1}{2}\right) = \ln 2,
\end{equation}
which is the maximum possible entropy for a two-dimensional system. More generally, for a $d$-dimensional system, the maximally mixed state 
$\rho_{\text{max}} = \frac{1}{d}\mathbb{I}$ has maximal von Neumann entropy 
$S_{\text{max}} = \ln d$: all accessible states occur with equal probability, 
no measurement outcome can be predicted with certainty, and the reduced density 
matrix is proportional to the identity. In the bipartite context, this 
corresponds to maximal entanglement, where the information about subsystem $A$ 
is not absent but entirely encoded in its correlations with $B$, so that $A$ 
considered in isolation appears maximally disordered.

\textbf{Two-qubit systems and the correlation matrix}: For a bipartite system consisting of two qubits A and B, the Hilbert space corresponds to the tensor product $\mathcal{H} = \mathcal{H}_A \otimes \mathcal{H}_B$, which is four-dimensional with the computational basis $\{|00\rangle, |01\rangle, |10\rangle, |11\rangle\}$. A general two-qubit density matrix can be parametrized as:
\begin{equation}
\rho = \frac{1}{4}\left(\mathbb{I}_4 + \sum_{i=1}^{3} B^{+}_i \sigma_i \otimes \mathbb{I}_2 + \sum_{j=1}^{3} B^{-}_j \mathbb{I}_2 \otimes \sigma_j + \sum_{i,j=1}^{3} C_{ij} \sigma_i \otimes \sigma_j\right),
\label{eq:two_qubit_density}
\end{equation}
where $\mathbb{I}_2$ and $\mathbb{I}_4$ denote the $2 \times 2$ and $4 \times 4$ identity matrices, respectively. The fifteen real physical quantities appearing in this decomposition are:

\begin{itemize}
\item $B^{+}_i = \text{Tr}[\rho(\sigma_i \otimes \mathbb{I}_2)]$: the $i$-th component of the Bloch vector for subsystem A, representing the expectation value of spin along direction $i$ for particle A.

\item $B^{-}_j = \text{Tr}[\rho(\mathbb{I}_2 \otimes \sigma_j)]$: the $j$-th component of the Bloch vector for subsystem B, representing the expectation value of spin along direction $j$ for particle B.

\item $C_{ij} = \text{Tr}[\rho(\sigma_i \otimes \sigma_j)]$: the spin correlation matrix, describing the spin correlation, namely, the joint expectation value of spin component $i$ measured on particle A and spin component $j$ measured on particle B.
\end{itemize}
The coefficients $B^{\pm}_i$ characterize the local properties of each particle individually, while the $3 \times 3$ correlation matrix $C_{ij}$ encodes the spin-spin correlations between the two particles. Together, these 15 real parameters (3 + 3 + 9) provide a complete parametrization of an arbitrary two-qubit state, with the remaining degree of freedom fixed by the unit trace condition.

\textbf{Entanglement and separability}: Entanglement is a uniquely quantum mechanical phenomenon with no classical analog, describing correlations between subsystems that cannot be explained by any local realistic theory~\cite{Horodecki:2009zz}. A bipartite pure state $|\psi\rangle \in \mathcal{H}_A \otimes \mathcal{H}_B$ is called separable (or product) if it can be written as:
\begin{equation}
|\psi\rangle = |\phi_A\rangle \otimes |\phi_B\rangle,
\end{equation}
for some states $|\phi_A\rangle \in \mathcal{H}_A$ and $|\phi_B\rangle \in \mathcal{H}_B$. A pure state that cannot be expressed in this form is called entangled.

The quintessential examples of maximally entangled two-qubit states are the four Bell states:
\begin{align}
|\Phi^{\pm}\rangle &= \frac{1}{\sqrt{2}}\left(|00\rangle \pm |11\rangle\right) = \frac{1}{\sqrt{2}}\left(\ket{\uparrow\uparrow} \pm \ket{\downarrow\downarrow}\right), \\
|\Psi^{\pm}\rangle &= \frac{1}{\sqrt{2}}\left(|01\rangle \pm |10\rangle\right) = \frac{1}{\sqrt{2}}\left(\ket{\uparrow\downarrow} \pm \ket{\downarrow\uparrow}\right).
\end{align}
The singlet state $|\Psi^{-}\rangle = \frac{1}{\sqrt{2}}(\ket{\uparrow\downarrow} - \ket{\downarrow\uparrow})$, which appears in the spin-zero decay of a neutral pion $\pi^0 \to e^+ e^-$, exhibits perfect anti-correlation: if particle A is measured to have spin up along any axis, particle B is guaranteed to have spin down along the same axis, and vice versa. This is precisely the state that appears in the EPRB thought experiment discussed above.

For mixed states, the definition of separability is extended as follows~\cite{Werner:1989zz}: a density matrix $\rho$ on $\mathcal{H}_A \otimes \mathcal{H}_B$ is separable if it admits a convex decomposition:
\begin{equation}
\rho = \sum_k p_k \rho^A_k \otimes \rho^B_k,
\end{equation}
where $p_k \geq 0$, $\sum_k p_k = 1$, and $\rho^A_k$, $\rho^B_k$ are valid density matrices on $\mathcal{H}_A$ and $\mathcal{H}_B$, respectively. The partial transpose of $\rho$ with respect to subsystem B only is given by $\rho^{T_B}= \sum_k p_k \rho^A_k \otimes \left(\rho^B_k \right)^T$. A mixed state that cannot be written in this form is entangled. Determining whether a given mixed state is separable is generally a computationally hard problem (NP-hard in general~\cite{Gurvits:2003mcm}), but for two-qubit systems, the Peres-Horodecki criterion provides a necessary and sufficient condition: a two-qubit state $\rho$ is separable if and only if its partial transpose $\rho^{T_B}$ (transposition with respect to subsystem B only) has no negative eigenvalues~\cite{Peres:1996dw,Horodecki:1996nc}.

\textbf{Concurrence as an entanglement measure}: To quantify the degree of entanglement in two-qubit systems, the concurrence provides a widely used and analytically tractable measure~\cite{Wootters:1997id,Hill:1997pfa}. Unlike the von Neumann entropy which requires diagonalizing density matrices, the concurrence offers a direct algebraic formula. For a pure state $|\psi\rangle = \alpha\ket{\uparrow\uparrow} + \beta\ket{\uparrow\downarrow} + \gamma\ket{\downarrow\uparrow} + \delta\ket{\downarrow\downarrow}$, the concurrence is defined as:
\begin{equation}
\mathcal{C}(|\psi\rangle) = 2|\alpha\delta - \beta\gamma|.
\end{equation}
This simple expression has a deep physical interpretation rooted in time-reversal symmetry, making it particularly natural for physicists familiar with discrete symmetries.

\textit{Time-Reversal Interpretation:} The concurrence can be understood through the action of the time-reversal operator on spin-$1/2$ particles. As an angular momentum, spin must flip under time reversal. For a single spin-$1/2$ particle, the time-reversal operator acts as:
\begin{equation}
\hat{\Theta} \begin{pmatrix} a \\ b \end{pmatrix} = \begin{pmatrix} -b^* \\ a^* \end{pmatrix},
\end{equation}
which can be written as $\hat{\Theta} = -i\sigma_y \mathcal{K}$, where $\mathcal{K}$ denotes complex conjugation. This operator is anti-unitary and satisfies $\hat{\Theta}^2 = -1$ for spin-$1/2$ particles. Explicitly, we have:
\begin{equation}
\hat{\Theta}\ket{\uparrow} = -\ket{\downarrow}, \quad \hat{\Theta}\ket{\downarrow} = \ket{\uparrow}.
\end{equation}

For a two-qubit system, the time-reversal operator acts on both spins:
\begin{equation}
\hat{\Theta}^{(2)} = (\hat{\Theta} \otimes \hat{\Theta}) = (-i\sigma_y \mathcal{K}) \otimes (-i\sigma_y \mathcal{K}) = -(\sigma_y \otimes \sigma_y) \mathcal{K}.
\end{equation}
The overall minus sign is conventional and can be absorbed into the definition. The key insight is that the concurrence measures the overlap between a state and its time-reversed counterpart:
\begin{equation}
\mathcal{C}(|\psi\rangle) = |\langle\psi|\tilde{\psi}\rangle|, \quad \text{where} \quad |\tilde{\psi}\rangle = -(\sigma_y \otimes \sigma_y)|\psi^*\rangle.
\end{equation}

To see why this definition captures entanglement, let us examine the behavior of different classes of states under time reversal:
\begin{itemize}
\item \textit{Separable states have zero concurrence:} Consider a product state $|\psi\rangle = |\phi_A\rangle \otimes |\phi_B\rangle$ where $|\phi_A\rangle = a\ket{\uparrow} + b\ket{\downarrow}$ and $|\phi_B\rangle = c\ket{\uparrow} + d\ket{\downarrow}$. Under time reversal:
\begin{equation}
|\tilde{\psi}\rangle = \hat{\Theta}^{(2)}|\psi\rangle = (\hat{\Theta}|\phi_A\rangle) \otimes (\hat{\Theta}|\phi_B\rangle) = |\tilde{\phi}_A\rangle \otimes |\tilde{\phi}_B\rangle.
\end{equation}
For a single spin-$1/2$ particle, one can show that $\langle\phi|\tilde{\phi}\rangle = \langle\phi|\hat{\Theta}|\phi\rangle = 0$ for any pure state $|\phi\rangle$. This orthogonality follows from the anti-unitary nature of $\hat{\Theta}$: if $|\phi\rangle = a\ket{\uparrow} + b\ket{\downarrow}$, then $|\tilde{\phi}\rangle = -b^*\ket{\uparrow} + a^*\ket{\downarrow}$, giving $\langle \tilde{\phi} | \phi\rangle = (-b)a + (a)b =  0$. Therefore, $\langle\psi|\tilde{\psi}\rangle = \langle\phi_A|\tilde{\phi}_A\rangle \langle\phi_B|\tilde{\phi}_B\rangle = 0$, 
yielding $\mathcal{C} = 0$ for all separable states.

\item \textit{Bell states have maximal concurrence:} Consider the singlet state $|\Psi^-\rangle = \frac{1}{\sqrt{2}}(\ket{\uparrow\downarrow} - \ket{\downarrow\uparrow})$. Applying time reversal to each term:
\begin{align}
\hat{\Theta}^{(2)}\ket{\uparrow\downarrow} &= \hat{\Theta}\ket{\uparrow} \otimes \hat{\Theta}\ket{\downarrow} = (-\ket{\downarrow}) \otimes (\ket{\uparrow}) = -\ket{\downarrow\uparrow}, \\
\hat{\Theta}^{(2)}\ket{\downarrow\uparrow} &= \hat{\Theta}\ket{\downarrow} \otimes \hat{\Theta}\ket{\uparrow} = (\ket{\uparrow}) \otimes (-\ket{\downarrow}) = -\ket{\uparrow\downarrow}.
\end{align}
Therefore:
\begin{equation}
|\tilde{\Psi}^-\rangle = \frac{1}{\sqrt{2}}(-\ket{\downarrow\uparrow} - (-\ket{\uparrow\downarrow})) = \frac{1}{\sqrt{2}}(\ket{\uparrow\downarrow} - \ket{\downarrow\uparrow}) = |\Psi^-\rangle.
\end{equation}
The singlet is an eigenstate of time reversal with eigenvalue $+1$, giving $|\langle\Psi^-|\tilde{\Psi}^-\rangle| = 1$ and maximal concurrence $\mathcal{C} = 1$. Similar calculations show that all four Bell states achieve $\mathcal{C} = 1$.
\item \textit{General two-qubit states:} For the general state $|\psi\rangle = \alpha\ket{\uparrow\uparrow} + \beta\ket{\uparrow\downarrow} + \gamma\ket{\downarrow\uparrow} + \delta\ket{\downarrow\downarrow}$, applying time reversal gives: $|\tilde{\psi}\rangle = \delta^*\ket{\uparrow\uparrow} - \gamma^*\ket{\uparrow\downarrow} - \beta^*\ket{\downarrow\uparrow} + \alpha^*\ket{\downarrow\downarrow}$ and then the overlap is: $\langle\psi|\tilde{\psi}\rangle = \alpha\delta^* - \beta\gamma^* - \gamma\beta^* + \delta\alpha^* = 2(\alpha\delta - \beta\gamma)$.
Taking the absolute value gives $\mathcal{C} = 2|\alpha\delta - \beta\gamma|$, recovering the original definition.
\end{itemize}

\textit{Physical interpretation:} The time-reversal operator exhibits seemingly paradoxical behavior: it destroys single-particle coherence while preserving quantum correlations. For a single spin-$1/2$ state $|\phi\rangle = a\ket{\uparrow} + b\ket{\downarrow}$, time reversal gives $|\tilde{\phi}\rangle = -b^*\ket{\uparrow} + a^*\ket{\downarrow}$, yielding orthogonality: $\langle\phi|\tilde{\phi}\rangle = 0$. This is a consequence of $\hat{T}^2 = -1$ for fermions (Kramers degeneracy), which requires the existence of an orthogonal degenerate state under time-reversal. However, for the singlet state $|\Psi^-\rangle = \frac{1}{\sqrt{2}}(\ket{\uparrow\downarrow} - \ket{\downarrow\uparrow})$, time reversal maps it back to itself: $|\tilde{\Psi}^-\rangle = |\Psi^-\rangle$. The anti-correlation structure remains unchanged when both spins flip simultaneously. Correlations are relational properties between particles, not intrinsic to either particle alone, and time reversal preserves such relational structures. For separable states $|\psi\rangle = |\phi_A\rangle \otimes |\phi_B\rangle$, the overlap factorizes and it gives $\langle\psi|\tilde{\psi}\rangle  = 0$ due to single-particle orthogonality. The concurrence $\mathcal{C} = |\langle\psi|\tilde{\psi}\rangle|$ thus measures the degree to which quantum correlations survive time reversal: $\mathcal{C} = 0$ indicates independent single-particle Kramers doublets, while $\mathcal{C} = 1$ indicates maximal correlation preservation.

\textit{Connection to reduced density matrix:} An equivalent formulation uses the reduced density matrix. For a bipartite pure state $|\psi\rangle$, the reduced density matrix of subsystem A is obtained by tracing over subsystem B:
\begin{equation}
\rho_A = \text{Tr}_B(|\psi\rangle\langle\psi|).
\end{equation}
The concurrence can then be expressed as:
\begin{equation}
\mathcal{C}(|\psi\rangle) = \sqrt{2(1 - \text{Tr}(\rho_A^2))}.
\end{equation}
Due to the Schmidt decomposition, the same result is obtained if one instead considers $\rho_B = \text{Tr}_A(|\psi\rangle\langle\psi|)$ by tracing over subsystem A, since both reduced density matrices share identical eigenvalue spectra. For a general two-qubit pure state $|\psi\rangle = \alpha\ket{\uparrow\uparrow} + \beta\ket{\uparrow\downarrow} + \gamma\ket{\downarrow\uparrow} + \delta\ket{\downarrow\downarrow}$, this formula recovers the explicit expression $\mathcal{C} = 2|\alpha\delta - \beta\gamma|$. For a separable state, $\rho_A$ is pure with $\text{Tr}(\rho_A^2) = 1$, giving $\mathcal{C} = 0$. For a maximally entangled state, $\rho_A = \mathbb{I}_2/2$ is maximally mixed with $\text{Tr}(\rho_A^2) = 1/2$, giving $\mathcal{C} = 1$. This formula reveals that entanglement is intimately connected to the mixedness of the reduced state: the more entangled a pure bipartite state is, the more mixed its subsystems appear when considered individually.

This mixedness reflects information loss inherent in the tracing operation. When we trace over subsystem B to obtain $\rho_A$, we discard information about the quantum correlations between A and B, rendering subsystem B inaccessible. The reduced density matrix $\rho_A$ alone cannot capture these non-local correlations. This information loss distinguishes an entangled pure state from a genuinely mixed state: the full bipartite state $|\psi\rangle$ is pure with zero entropy, yet the reduced state $\rho_A$ exhibits non-zero von Neumann entropy $S(\rho_A) = -\text{Tr}(\rho_A \log \rho_A)$. This entropy quantifies precisely the information stored in correlations with subsystem B, information that becomes inaccessible once we trace over B and focus only on subsystem A.

\textit{Extension to mixed states:} For mixed states, Wootters derived a closed-form expression for the concurrence~\cite{Wootters:1997id} to quantify the level of quantum entanglement. Given a two-qubit density matrix $\rho$, the procedure is:

\begin{enumerate}
\item Compute the spin-flipped density matrix after the time reversal operation:
\begin{equation}
\tilde{\rho} = (\sigma_y \otimes \sigma_y) \rho^* (\sigma_y \otimes \sigma_y),
\end{equation}
where $\rho^*$ denotes complex conjugation of $\rho$. Taking into account the spin flip for each Pauli matrix, one finds 
\begin{equation}
\tilde{\rho} = \frac{1}{4}\left(\mathbb{I}_4 - \sum_{i=1}^{3} B^{+}_i \sigma_i \otimes \mathbb{I}_2 - \sum_{j=1}^{3} B^{-}_j \mathbb{I}_2 \otimes \sigma_j + \sum_{i,j=1}^{3} C_{ij} \sigma_i \otimes \sigma_j\right). 
\label{eq:two_qubit_density_tilde}
\end{equation}
\item Construct the auxiliary Hermitian matrix:
\begin{equation}
R = \sqrt{\sqrt{\rho} \tilde{\rho} \sqrt{\rho}}.
\end{equation}
It is important to note that $\tilde{\rho}$ becomes the same $\rho$, thus $R$ reduces to $\rho$ when the Bloch polarization vectors $B^\pm =0$ in $\rho$. This is expected since Bloch vector components change sign (reflecting the reversal of spin angular momentum), while the correlation matrix remains unchanged because both qubits transform with the same minus sign, giving $(-1) \times (-1) = +1$.
\item Compute the eigenvalues $\lambda_1 \geq \lambda_2 \geq \lambda_3 \geq \lambda_4 \geq 0$ of $R$ (or equivalently, the square roots of the eigenvalues of $\rho \tilde{\rho}$).

\item The concurrence is defined as follows:
\begin{equation}
\mathcal{C}(\rho) = \max(0, \lambda_1 - \lambda_2 - \lambda_3 - \lambda_4).
\label{eq:concurrence}
\end{equation}
\end{enumerate}

The concurrence ranges from $\mathcal{C} = 0$ for separable states to $\mathcal{C} = 1$ for maximally entangled states. The appearance of the spin-flip operation $\tilde{\rho}$ in this formula is a direct generalization of the time-reversal interpretation for pure states, extending the physical picture of time-reversal overlap to the space of mixed states.

\textbf{Bell inequalities and nonlocality}: Bell inequalities provide experimental tests to distinguish quantum mechanics from local hidden variable (LHV) theories, which assume that measurement outcomes are predetermined by hidden variables and that no superluminal influences exist between spacelike-separated measurements~\cite{Bell:1964kc}. The most widely used form is the Clauser-Horne-Shimony-Holt (CHSH) inequality~\cite{Clauser:1969ny}.

Consider a bipartite experiment where Alice measures one of two observables $A_1$ or $A_2$ on her particle, and Bob measures one of two observables $B_1$ or $B_2$ on his particle. Each measurement yields outcomes $\pm 1$. In an LHV theory, the outcomes are determined by a hidden variable $\lambda$ with probability distribution $\rho(\lambda)$:
\begin{equation}
A_i = A_i(\lambda), \quad B_j = B_j(\lambda), \quad \text{with} \quad A_i, B_j \in \{+1, -1\}.
\end{equation}
The correlation function is:
\begin{equation}
E(A_i, B_j) = \int d\lambda \, \rho(\lambda) \, A_i(\lambda) B_j(\lambda).
\end{equation}

The CHSH parameter is defined as:
\begin{equation}
\mathcal{B} = E(A_1, B_1) + E(A_1, B_2) + E(A_2, B_1) - E(A_2, B_2).
\end{equation}
For any LHV theory, one can show that:
\begin{equation}
|\mathcal{B}| \leq 2 \quad \text{(CHSH inequality)}.
\end{equation}

The derivation proceeds as follows. For each value of $\lambda$, define $S(\lambda) = A_1(B_1 + B_2) + A_2(B_1 - B_2)$. Since $B_1, B_2 \in \{+1, -1\}$, either $B_1 + B_2 = 0$ (when $B_1 = -B_2$) or $B_1 - B_2 = 0$ (when $B_1 = B_2$). In the former case, $S(\lambda) = 2A_2 B_1$, and in the latter, $S(\lambda) = 2A_1 B_1$. In both cases, $|S(\lambda)| = 2$. Therefore, $|S(\lambda)| \leq 2$ for all $\lambda$. Averaging over $\lambda$ gives:
\begin{equation}
|\mathcal{B}| = \left|\int d\lambda \, \rho(\lambda) \, S(\lambda)\right| \leq \int d\lambda \, \rho(\lambda) \, |S(\lambda)| \leq 2.
\end{equation}
This inequality holds for any LHV theory, regardless of the specific form of the hidden variable distribution $\rho(\lambda)$.

Meanwhile, in quantum mechanics, the correlation function for spin measurements along directions $\vec{a}$ and $\vec{b}$ on a two-qubit state $\rho$ is:
\begin{equation}
E(\vec{a}, \vec{b}) = \text{Tr}[\rho (\vec{a} \cdot \vec{\sigma}) \otimes (\vec{b} \cdot \vec{\sigma})] = \sum_{i,j} a_i b_j C_{ij},
\end{equation}
where $C_{ij}$ is the correlation matrix from Eq.~(\ref{eq:two_qubit_density}). For the singlet state $|\Psi^{-}\rangle$, the correlation matrix is $C_{ij} = -\delta_{ij}$, meaning the spins are always anti-parallel. This gives $E(\vec{a}, \vec{b}) = -\vec{a} \cdot \vec{b}$.

To find the quantum bound, we evaluate the CHSH combination:
\begin{equation}
|\mathcal{B}| = |E(\vec{a}, \vec{b}) - E(\vec{a}, \vec{b}') + E(\vec{a}', \vec{b}) + E(\vec{a}', \vec{b}')| = |\vec{a}' \cdot (\vec{b} + \vec{b}') + \vec{a} \cdot (\vec{b} - \vec{b}')|.
\end{equation}
Applying the triangle inequality and Cauchy-Schwarz inequality for unit vectors $\vec{a}$ and $\vec{a}'$, we obtain $|\mathcal{B}| \leq |\vec{b} + \vec{b}'| + |\vec{b} - \vec{b}'|$. For unit vectors $\vec{b}$ and $\vec{b}'$, this sum is maximized when $\vec{b} \perp \vec{b}'$, yielding:
\begin{equation}
|\vec{b} + \vec{b}'| + |\vec{b} - \vec{b}'| = \sqrt{2 + 2\vec{b} \cdot \vec{b}'} + \sqrt{2 - 2\vec{b} \cdot \vec{b}'} \leq 2\sqrt{2}.
\end{equation}
Thus, the quantum-mechanical bound is:
\begin{equation}
|\mathcal{B}| \leq 2\sqrt{2} \approx 2.828.
\end{equation}
This maximum value is known as the Tsirelson bound~\cite{Cirelson:1980ry}, which represents the maximum violation achievable by any quantum state. The bound can be reached by choosing the setup with $\vec{b} \perp \vec{b}'$, $\vec{a} \parallel (\vec{b} - \vec{b}')$, and $\vec{a}' \parallel (\vec{b} + \vec{b}')$.

For the other Bell states, the correlation matrices are $C_{ij} = \text{diag}(1, 1, -1)$ for $|\Psi^{+}\rangle$, $C_{ij} = \text{diag}(1, -1, 1)$ for $|\Phi^{+}\rangle$, and $C_{ij} = \text{diag}(-1, 1, 1)$ for $|\Phi^{-}\rangle$. The Horodecki condition~\cite{Horodecki:1995nsk} provides a powerful method to determine the maximum CHSH violation directly from the correlation matrix without optimizing over measurement settings. For a two-qubit state with correlation matrix $\mathbf{C}$, let $\mu_1 \geq \mu_2 \geq \mu_3$ be the eigenvalues of $\mathbf{C}^T \mathbf{C}$. The maximum CHSH parameter over all measurement choices is then
\begin{equation}
|\mathcal{B}|_{\max} = 2\sqrt{\mu_1 + \mu_2}.
\end{equation}
Bell nonlocality occurs when $|\mathcal{B}|_{\max} > 2$, which requires $\mu_1 + \mu_2 > 1$.

For all four Bell states, the correlation matrices are diagonal with entries $\pm 1$, giving $\mathbf{C}^T \mathbf{C} = \mathbf{I}$ (the identity matrix). Thus $\mu_1 = \mu_2 = \mu_3 = 1$, and $|\mathcal{B}|_{\max} = 2\sqrt{1 + 1} = 2\sqrt{2}$, which is precisely the Tsirelson bound~\cite{Cirelson:1980ry}. This demonstrates that all Bell states achieve maximal quantum violation of the CHSH inequality. The Horodecki approach reveals that this bound arises from the spectral properties of the correlation matrix: the Tsirelson bound corresponds to the case where the two largest eigenvalues of $\mathbf{C}^T \mathbf{C}$ are both unity. While all Bell states saturate this bound, the optimal measurement configurations differ due to the distinct signs in their correlation matrices. A convenient measure of Bell nonlocality is:
\begin{equation}
\mathcal{N} = \max(0, \mu_1 + \mu_2 - 1),
\end{equation}
which ranges from $\mathcal{N} = 0$ for states satisfying local realism to $\mathcal{N} = 1$ for maximally nonlocal states for states with maximal nonlocality, such as the Bell states.

The experimental confirmation of Bell inequality violations has profound implications. It demonstrates that Einstein's hope for a local realistic completion of quantum mechanics cannot be realized: nature does not admit a description in which measurement outcomes are predetermined by local hidden variables. The ``moon'', in Einstein's metaphor, does not have a definite position until someone looks. More precisely, the correlations between spatially separated measurements cannot be explained by any theory in which properties are determined locally and independently. This represents one of the most striking departures of quantum mechanics from classical intuition and establishes entanglement as a genuine physical resource with no classical counterpart.

\textbf{Decoherence and renormalization group flow}: High-energy collision processes are inherently open quantum systems: the particles produced at the hard scattering vertex inevitably interact with their environment through radiation before reaching the detector. In such open systems, entanglement with unobserved degrees of freedom leads to decoherence, the degradation of quantum coherence in the observable sector. A fundamental question then arises: how does the emission of soft and collinear radiation lead to information loss, and consequently to decoherence and the suppression of measurable entanglement?

For example, recent work using the open quantum system framework~\cite{Gu:2025ijz, Agrawal:2026zwa} shows that renormalization group (RG) evolution of the spin density matrix constitutes a quantum channel, a completely positive, trace-preserving map, where the RG flow parameter $t = \ln(Q/\mu)$ (the logarithm of the ratio between the hard production scale $Q$ and the soft measurement scale $\mu$) drives a Markovian loss of quantum information. This provides a novel physical interpretation: RG flow itself is the engine of decoherence in particle physics, analogous to temporal evolution in conventional open quantum systems.

For fermion pair production with QED-like final-state radiation, the framework reveals that decoherence proceeds through a phase-flip channel, where the concurrence $\mathcal{C}$ decreases exponentially with the RG parameter:
\begin{equation}
\mathcal{C} \propto e^{-\Gamma t},
\end{equation}
where $\Gamma$ depends on the detector resolution and coupling strength. This exponential suppression reflects the fundamental information loss mechanism: unresolved soft photon emission creates an environment that becomes entangled with the charged particle system. As emphasized in Ref.~\cite{Semenoff:2019dqe}, this infrared-induced decoherence represents a genuine loss of quantum information from the observable sector. The soft photons carry away correlations that cannot be recovered without detecting the complete final state, including arbitrarily soft radiation. The phase-flip nature of the channel means that collinear radiation preferentially destroys phase coherence (off-diagonal density matrix elements) while preserving populations (diagonal elements), making entanglement measures more fragile than simple spin polarizations.
	
\end{document}